\documentclass[twocolumn,tighten]{aastex61}
\usepackage{graphicx}
\usepackage[space]{grffile}
\usepackage{latexsym}
\usepackage{textcomp}
\usepackage{longtable}
\usepackage{multirow,booktabs}
\usepackage{amsfonts,amsmath,amssymb}
\newif\iflatexml\latexmlfalse
\DeclareGraphicsExtensions{.pdf,.PDF,.png,.PNG,.jpg,.JPG,.jpeg,.JPEG}

\usepackage[utf8]{inputenc}
\usepackage[english]{babel}

\newcommand{\amm}{NH$_3$}
\newcommand{\kms}{km\,s$^{-1}$}
\newcommand{\cc}{cm$^{-3}$}
\newcommand{\vlsr}{$v_{\mathrm{LSR}}$}
\newcommand{\sigv}{$\sigma_v$}
\newcommand{\tkin}{$T_\mathrm{K}$}
\newcommand{\tex}{$T_\mathrm{ex}$}

\newcommand{\nh}{$N(\mathrm{H}_2)$}
\newcommand{\namm}{N(NH$_3$)}

\newcommand{\av}{$A_\mathrm{V}$}


\providecommand{\sorthelp}[1]{}

\shorttitle{GAS: First Results}
\shortauthors{Friesen \& Pineda et al.}
\submitjournal{ApJS}
\accepted{March 28, 2017}

\begin{document}


\title{The Green Bank Ammonia Survey (GAS): First results of NH$_3$ mapping the Gould Belt}


\author{Rachel K. Friesen}
\affiliation{Dunlap Institute for Astronomy \& Astrophysics, University of Toronto, 50 St. George Street, Toronto, Ontario, Canada M5S 3H4} 

\author{Jaime E. Pineda}
\affiliation{Max-Planck-Institut f\"ur extraterrestrische Physik, Giessenbachstrasse 1, 85748 Garching, Germany}

\collaboration{(co-PIs)}

\author{Erik Rosolowsky}
\affiliation{Department of Physics, University of Alberta, Edmonton, AB, Canada} 

\author{Felipe Alves}
\affiliation{Max-Planck-Institut f\"ur extraterrestrische Physik, Giessenbachstrasse 1, 85748 Garching, Germany}

\author{Ana Chac\'on-Tanarro}
\affiliation{Max-Planck-Institut f\"ur extraterrestrische Physik, Giessenbachstrasse 1, 85748 Garching, Germany}

\author{Hope How-Huan Chen}
\affiliation{Harvard-Smithsonian Center for Astrophysics, 60 Garden St., Cambridge, MA 02138, USA}

\author{Michael Chun-Yuan Chen}
\affiliation{Department of Physics and Astronomy, University of Victoria, 3800 Finnerty Road, Victoria, BC, Canada V8P 5C2}

\author{James Di Francesco}
\affiliation{Department of Physics and Astronomy, University of Victoria, 3800 Finnerty Road, Victoria, BC, Canada V8P 5C2}
\affiliation{Herzberg Astronomy and Astrophysics, National Research Council of Canada, 5071 West Saanich Road, Victoria, BC, V9E 2E7, Canada}

\author{Jared Keown}
\affiliation{Department of Physics and Astronomy, University of Victoria, 3800 Finnerty Road, Victoria, BC, Canada V8P 5C2}

\author{Helen Kirk}
\affiliation{Herzberg Astronomy and Astrophysics, National Research Council of Canada, 5071 West Saanich Road, Victoria, BC, V9E 2E7, Canada}

\author{Anna Punanova}
\affiliation{Max-Planck-Institut f\"ur extraterrestrische Physik, Giessenbachstrasse 1, 85748 Garching, Germany}

\author{Youngmin Seo}
\affiliation{Jet Propulsion Laboratory, NASA, 4800 Oak Grove Dr, Pasadena, CA 91109, USA}

\author{Yancy Shirley}
\affiliation{Steward Observatory, 933 North Cherry Avenue, Tucson, AZ 85721, USA}

\author{Adam Ginsburg}
\affiliation{National Radio Astronomy Observatory, Socorro, NM 87801, USA}

\author{Christine Hall}
\affiliation{Department of Physics, Engineering Physics \& Astronomy, Queen's University, Kingston, Ontario, Canada K7L 3N6}

\author{Stella S. R. Offner}
\affiliation{Department of Astronomy, University of Massachusetts, Amherst, MA 01003, USA}

\author{Ayushi Singh}
\affiliation{Department of Astronomy \& Astrophysics, University of Toronto, 50 St. George Street, Toronto, Ontario, Canada M5S 3H4}

\author{H\'ector G. Arce}
\affiliation{Department of Astronomy, Yale University, P.O. Box 208101, New Haven, CT 06520-8101, USA}

\author{Paola Caselli}
\affiliation{Max-Planck-Institut f\"ur extraterrestrische Physik, Giessenbachstrasse 1, 85748 Garching, Germany}

\author{Alyssa A. Goodman}
\affiliation{Harvard-Smithsonian Center for Astrophysics, 60 Garden St., Cambridge, MA 02138, USA}

\author{Peter G. Martin}
\affiliation{Canadian Institute for Theoretical Astrophysics, University of Toronto, 60 St. George St., Toronto, Ontario, Canada, M5S 3H8}

\author{Christopher Matzner}
\affiliation{Department of Astronomy \& Astrophysics, University of Toronto, 50 St. George Street, Toronto, Ontario, Canada M5S 3H4}

\author{Philip C. Myers}
\affiliation{Harvard-Smithsonian Center for Astrophysics, 60 Garden St., Cambridge, MA 02138, USA}

\author{Elena Redaelli}
\affiliation{Max-Planck-Institut f\"ur extraterrestrische Physik, Giessenbachstrasse 1, 85748 Garching, Germany}
\affiliation{Dipartimento di Fisica \& Astronomia, Universita' degli Studi di Bologna, Viale Berti Pichat, 6/2, I - 40127 Bologna, Italy}

\collaboration{(The GAS collaboration)}

\begin{abstract}
We present an overview of the first data release (DR1) and first-look science from the Green Bank Ammonia Survey (GAS). 
GAS is a Large Program at the Green Bank Telescope to map all Gould Belt star-forming regions with $A_\mathrm{V} \gtrsim 7$~mag visible from the northern hemisphere in emission from \amm\ and other key molecular tracers. 
This first release includes the data for four regions in Gould Belt clouds: B18 in Taurus, NGC 1333 in Perseus, L1688 in Ophiuchus, and Orion A North in Orion. 
We compare the \amm\ emission to dust continuum emission from \textit{Herschel}, and find that the two tracers correspond closely. 
\amm\ is present in over 60\ \% of lines-of-sight with $A_\mathrm{V} \gtrsim 7$~mag in three of the four DR1 regions, in agreement with expectations from previous observations. 
The sole exception is B18, where \amm\ is detected toward $\sim 40$\ \% of lines-of-sight with $A_\mathrm{V} \gtrsim 7$~mag.
Moreover, we find that the \amm\ emission is generally extended beyond the typical 0.1\ pc length scales of dense cores.
We produce maps of the gas kinematics, temperature, and \amm\ column densities through forward modeling of the hyperfine structure of the \amm\ (1,1) and (2,2) lines. 
We show that the \amm\ velocity dispersion, \sigv, and gas kinetic temperature, \tkin, vary systematically between the regions included in this release, with an increase in both the mean value and spread of \sigv\ and \tkin\ with increasing star formation activity. 
The data presented in this paper are publicly available.
\end{abstract}


\section{Introduction}

The past several years have seen tremendous advancements in our ability to characterize the structure of nearby molecular clouds and the substructures in which dense star-forming cores are born. Within 500\ pc of the Sun, nearly all the ongoing, predominantly low-mass star formation is contained within  a ring of young stars and star-forming regions named the Gould Belt. These nearby star-forming clouds range in size and activity, from the nearly inactive Pipe Nebula, through the star-forming but largely quiescent Taurus molecular cloud, to stellar group and cluster-forming clouds like Perseus and Ophiuchus. The Orion molecular cloud is our nearest example of a high-mass star-forming region. This variation in star formation activity, coupled with their relatively nearby distances, make the Gould Belt clouds excellent survey targets to understand how star formation proceeds in different environments, and to constrain star formation theories and simulations through detailed observations.

Consequently, the Gould Belt star-forming regions have been surveyed by several Key Programs in the infrared and sub-millimeter continuum. 
The infrared surveys have characterized the embedded young stellar populations, identifying the loci of active star formation and constraining the lifetimes of the starless and embedded stages of star-forming cores
(the \textit{Spitzer} Gould Belt\added{, Spitzer Orion} and c2d surveys; \citealt{Enoch_2008,Harvey_2008,Kirk_2009}\added{,  \citealt{Megeath_2012},}
 and \citealt{Evans_II_2003,Evans_2009,Dunham_2014}). 
At longer wavelengths, the emission from cold dust within dense, star-forming cores, and the larger filaments and clumps that form the cores and young stellar objects (YSOs), are being surveyed extensively through Gould Belt Legacy surveys with the \textit{Herschel Space Observatory} \citep[\textit{Herschel} GBS;][]{andre10} and the James Clerk Maxwell Telescope \citep[JCMT GBLS; ][]{Ward_2007}. The \textit{Herschel} GBS has revealed the striking prevalence of filaments within star-forming regions as seen in the dust emission \citep{Andr__2010,K_nyves_2010}. While the presence of filaments in star-forming regions was not a new discovery \citep[e.g.,][]{Schneider_1979}, their ubiquity at both low and high column densities suggests that they are integral in the build-up of the required mass to form stars. 
Furthermore, the \textit{Herschel} GBS has found gravitationally unstable cores predominantly within (apparently) gravitationally unstable filaments. Previous studies have shown that dense molecular cores are mostly found above a visual extinction threshold of $A_\mathrm{V} > 5 \sim 7$\ mag (\nh $\sim 5 - 7 \times 10^{21}$\ cm$^{-2}$; e.g., \citealt{Onishi_1998,Johnstone_2004,Kirk_2006}). 
With a suggested characteristic filament width of $\sim 0.1$\ pc \citep[][ but see \citealt{Panopoulou_2016}]{Arzoumanian_2011}, this extinction threshold agrees within a factor of two with the critical mass per unit length of an isothermal, thermally-supported cylinder ($M_\mathrm{line,crit} \sim 16$\ M$_\odot$\ pc$^{-1}$ at 10 K; \citealt{Inutsuka_1997,Hocuk_2016}). 
Therefore, understanding filament instability might be central to understanding the formation of dense cores and, consequently, stars.

If the massive and long filaments found by the \textit{Herschel} GBS are an important piece of the star formation puzzle \citep{Andr__2014}, then we must understand what are their main formation mechanisms, evolutionary drivers, and fragmentation triggers. 
The power of the legacy surveys described above lies in the large areal coverage and consistency in observing strategies between nearby molecular clouds. A major gap in the present data is a comparable survey to characterize the dense gas properties. Continuum data show the dust column density structure, but do not provide information about the gas kinematics. Kinematics and gas temperatures are key to understanding the history and future fate of star-forming material.

Kinematic studies of filaments are relatively few \citep[e.g.,][]{Pineda_2011,Hacar_2011,Hacar_2013,Pineda_2015,Hacar_2016,Henshaw_2016a}.
The CARMA Large Area Star Formation Survey \citep{Storm_2014} targeted several regions within nearby Gould Belt clouds, at higher resolution but on a smaller scale relative to the JCMT and \textit{Herschel} continuum surveys. 
The JCMT GBLS has produced the widest survey thus far, using observations of CO (3-2) and its isotopologues to trace moderately dense gas ($10^3$\ \cc, e.g., \citealt{Graves_2010}). 
CO, however, is frequently very optically thick and the gas phase abundance can be severely affected by depletion onto dust grains (e.g., \citealt{Caselli_1999,Tafalla_2002,Christie_2012}) in star-forming filaments and cores, and therefore it is not a useful tracer of the dense gas. 

The lower \amm\ metastable inversion transitions primarily trace gas of density $n \geq 2\times 10^3$\ \cc (for gas at 10\,K, \citealt{Shirley_2015}). Furthermore, \amm\ does not typically suffer from depletion.
Since \amm\ observations allow us to identify actual high volume density features rather than column density peaks, \amm\ is then an ideal tracer of the hierarchical nature of star-forming regions, and a powerful probe of the kinematics of star-forming filaments and cores.

\floattable
\twocolumngrid
\begin{deluxetable}{cccc}
\tablecolumns{4}
\tablewidth{4cm}
\tablecaption{GAS DR1 region data \label{tab:DR1_regions}}
\tablehead{
\colhead{Cloud} & \colhead{Region} & \colhead{Distance} & \colhead{Number of} \\
\colhead{} & \colhead{} & \colhead{(pc)} & \colhead{footprints\tablenotemark{a}}
}
\startdata
Taurus & B18 & $135 \pm 20$\tablenotemark{b} & 11 \\
Ophiuchus & L1688 & \replaced{$119 \pm 6$\tablenotemark{c}}{$137.3 \pm 6$\tablenotemark{c}} & 10  \\
Perseus & NGC 1333 & $260 \pm 26$\tablenotemark{b} & 8  \\
Orion & Orion A-North & $414 \pm 7$\tablenotemark{d} & 12 \\
\enddata
\tablenotetext{a}{Each footprint is 10\arcmin $\times$ 10\arcmin.}
\tablenotetext{b}{\citet{schlafly14}}
\replaced{\tablenotetext{c}{\citet{lombardi08}}}{\tablenotetext{c}{\citet{OrtizLeon_2017}}}
\tablenotetext{d}{\citet{Menten_2007}}
\end{deluxetable}

\floattable
\begin{deluxetable}{cccc}
\tablecolumns{4}
\tablewidth{0pt}
\tablecaption{GAS VEGAS spectral line setup \label{tab:spectral_setup}}
\tablehead{
\colhead{Species} & \colhead{Transition} & \colhead{Rest frequency} & \colhead{Reference}\\
\colhead{} & \colhead{} & \colhead{(MHz)} & \colhead{}
}
\startdata
\amm\ & $(1,1)$ & 23694.4955 & \\
\amm\ & $(2,2)$ & 23722.6336 & \\
\amm\ & $(3,3)$ & 23870.1296 & \\
HC$_5$N & $9 - 8$ & 23963.9010 & LOVAS \\
HC$_7$N & $21 - 20$ & 23687.8974(6) & CDMS \\
HC$_7$N & $22 - 21$ & 24815.8772(6) & CDMS \\
C$_2$S\tablenotemark{a} & $2_1 - 1_0$ & 22344.030(1) & CDMS \\
\enddata
\tablenotetext{a}{Observed in central KFPA beam only}
\end{deluxetable}

Past \amm\ observations of dense gas were primarily pencil-beam pointings or small maps around interesting targets (e.g.,  \citealt{Jijina_1999,Rosolowsky_2008}).
The \amm-identified cores are generally cold and quiescent   
\citep{Goodman_1998}, either in early stages of collapse or near-critical equilibrium \citep{Benson_1983}, and often show significant velocity gradients 
\citep{Goodman_1993}. The local environment influences key core properties, such as the ratio of thermal vs. non-thermal motions and temperature \citep{Myers_1991,Jijina_1999,Foster_2009}, both important parameters for investigations of core stability and collapse. 
Larger, more sensitive maps are needed to relate the physical properties of the cores to their surrounding environment. These recent studies have highlighted the sharp transition between turbulent and thermal line widths in dense cores \citep{Pineda_2010,Seo_2015}, identified incongruities between structures identified via continuum and line emission \citep{Friesen_2009}, shown that cores in more active environments, such as Orion, tend to be more unstable, despite having similar internal motions \citep{Li_2013}, identified large-scale gravitational instability in a young filamentary cluster \citep{Friesen_2016}, and revealed evidence for filamentary accretion in Serpens South \citep[][also in N$_2$H$^+$ by \citealt{Kirk_2013}]{Friesen_2013}.

Here, we describe the Green Bank Ammonia Survey (GAS\added{, co-PIs: R. Friesen and J. E. Pineda}), an ambitious Large Project to survey all the major, nearby ($120 < d < 500$\ pc), northern Gould Belt star-forming regions with \replaced{optical}{visual} extinctions $A_\mathrm{V} \gtrsim 7$ (matching the extinction threshold for dense cores in continuum studies
\added{e.g., \citealt{Johnstone_2004,Enoch2007,Lada2010}}) in emission from \amm, as well as the carbon-chain molecules C$_2$S, HC$_5$N, and HC$_7$N. Within this distance range, the GBT beam at 23\ GHz (32\arcsec\ FWHM) subtends $0.02\ \mathrm{pc} - 0.08\ \mathrm{pc}$, able to resolve both the Jeans length (0.12 pc for $n = 10^{4}$\ \cc\ and $T = 10$\ K) and the $\sim 0.1$\ pc typical filament width suggested by \citet{Arzoumanian_2011}, and is furthermore well-matched to the resolution of \textit{Herschel} SPIRE at 500\ \micron\ \citep[$\theta_{\mathrm{beam}} = 36$\arcsec\ FWHM; ][]{Griffin_2010}. 
This coordinated, large-scale survey of \amm\  in the Gould Belt clouds will play a key role in understanding the evolution of dense gas in star-forming regions as a function of environment, expanding in a consistent way the \amm\ studies described above, and will complement and enhance the Gould Belt continuum surveys. 

In this paper, we present the first data release (hereafter referred to as DR1) of a subset of the survey targets, describing in detail the calibration, imaging, and analysis pipelines, and discussing some general trends in the data.  
The target regions presented are Barnard 18 in Taurus (hereafter B18), NGC 1333 in Perseus, L1688 in Ophiuchus, and Orion A (North).  
Future research will focus on the key scientific goals of GAS, including analyzing the structure and stability of the dense gas, the dissipation of turbulence from large to small scales, and the evolution of angular momentum in dense cores.

In \S \ \ref{sec:data}, we discuss the source selection, observations, data calibration and imaging. 
In \S \ \ref{sec:linefit}, we describe in detail the hyperfine modeling of the \amm\ emission and the determination of key physical parameters of the emitting gas. 
In \S \ \ref{sec:results}, we present the moment maps and noise properties of the regions included in this first data release, as well as the results from the \amm\ line modeling. 
We present a summary in \S\ \ref{sec:summary}. 

\section{Data}
\label{sec:data}

\subsection{Source selection and description}

\label{sources}

The goal of GAS was to map \amm\ emission toward all areas within the nearby ($d < 500$\ pc), GBT-visible, star-forming molecular clouds of the Gould Belt with extinctions $A_\mathrm{V} > 7$\ mag [N(H$_2$) $\gtrsim 6.7 \times 10^{21}$\ cm$^{-2}$, assuming $N(\mathrm{H}_2) = 9.4 \times 10^{20}$\ cm$^{-2}$ ($A_\mathrm{V}$ mag$^{-1}$)] 
\added{\citep{Johnstone_2004,Enoch2007,Heiderman2010,Lada2010}}. 
The initial map extents were identified in each cloud using either publicly released \textit{Herschel} GBS 500\ \micron\ continuum data, unreleased JCMT GBLS 850\ \micron\ continuum data, or extinction maps derived from 2MASS data \citep{dobashi05,Ridge_2006-COMPLETE}. Further refinement of mapping targets was done as new data were publicly released (e.g., \textit{Herschel} GBS). 

For our initial observing season we focused on four regions that span a range of physical environments and star formation activity: B18, NGC 1333, L1688, and Orion A (North). These regions comprise the first data release (DR1) of GAS. 

B18 is a filamentary structure located in the south-east of the Taurus molecular cloud, and runs roughly parallel in projection with the molecular gas structure containing Heiles Cloud 2 and the B213/L1495 filament. B18 has a projected length of $\sim 3\degr$ ($\sim 7$\ pc) with a total mass of $380 - 440$\ M$_\odot$  \citep{Heyer_1987,Mizuno_1995}, giving a mean gas number density $n \sim 300$\ cm$^{-3}$, and contains $\sim 9$ molecular clumps as traced by CO \citep{Heyer_1988}. Based on the numbers of associated young stars, the star formation efficiency (SFE), where $SFE = \mathrm{current} \ M_* / M_{\mathrm{gas}}$, is $\sim 1-2$\ \%\ \citep{Mizuno_1995}.

NGC 1333 is the most active star-forming region within the Perseus molecular cloud and contains approximately 450\ M$_\odot$ of molecular gas \citep{Warin_1996} and $\sim 150$ young stars \citep{Gutermuth_2008} in a region $\sim 1$\ pc$^{2}$. The mean gas density is therefore a few $\times 10^3$\ cm$^{-3}$. Rough estimates suggest the  star formation efficiency is $\sim 13-15$\ \%\ in the region over the past $10^6$ years \citep{Walsh_2006,Jorgensen_2008}.

L1688 is the centrally-concentrated, dense hub of the Ophiuchus molecular cloud and is $\sim 1-2$\ pc in radius and $\sim 2 \times 10^3$\ M$_\odot$ in mass \citep{Loren_1989}. Similarly to NGC 1333, the mean gas density is approximately a few $\times 10^3$\ cm$^{-3}$. L1688 contains regions of extremely high visual extinction, with $A_\mathrm{V} \sim 50 - 100$~mag \cite[e.g., ][]{Wilking_1983}, and over 300 young stellar objects \citep{Wilking_2008}. The star formation efficiency of the dense gas cores, identified in submillimeter continuum emission from dust, is $\sim 14$\ \%\ \citep{Jorgensen_2008}.

Orion A (North), an integral-shaped filamentary structure of dense gas that extends both north and south of the Orion Nebula and its associated young stellar cluster, including the Trapezium stars, is our nearest example of ongoing high-mass star formation. The filament is characterized by varying physical conditions along the $\sim 12$\ pc length mapped for DR1. A compact, narrow ridge north of the Orion Nebula, the filament becomes wider and less dense toward the south, and contains a total mass of $\sim 5 \times 10^3$\ M$_\odot$ in the area mapped here \citep[e.g., ][]{Bally_1987,Bally_2008}. Overall, the star formation efficiency in Orion A is $\sim 3-5$\ \%\ in gas with extinction $A_\mathrm{K} > 2$, with significantly higher values (up to several tens of percent) found toward young protostellar groups and clusters \citep{Megeath_2015}. 

Table \ref{tab:DR1_regions} lists the assumed distances and number of footprints (areal coverage of 10\arcmin\ $\times$ 10\arcmin) completed for the GAS DR1 regions.

\begin{figure}
\begin{center}
\includegraphics[width=0.9\columnwidth]{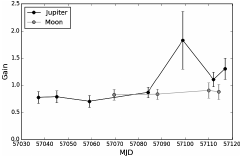}
\caption{Average gains over all beams and polarizations for the \amm\ (1,1) spectral window as a function of date derived from observations of Jupiter and the Moon. Error bars show the $1-\sigma$ variation. As discussed further in the text, the Jupiter observations after 57090 MJD show large offsets from previous Jupiter measurements and concurrent Moon measurements due to high winds on those dates.  \label{fig:gains_date}%
}
\end{center}
\end{figure}

\begin{figure}
\begin{center}
\includegraphics[width=0.9\columnwidth]{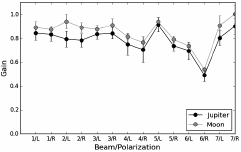}
\caption{Beam gains for the \amm\ (1,1) spectral window averaged over all observing dates as a function of beam and polarization for Jupiter and the Moon. The observations of Jupiter during windy periods have been omitted. Error bars show the $1-\sigma$ variation in the values. Although the gains vary between beams and polarizations, they remained consistent over the observing period. \label{fig:gains_beams}%
}
\end{center}
\end{figure}

\floattable
\begin{deluxetable}{cccccccc}
\tablecolumns{8}
\tablewidth{0pt}
\tablecaption{GAS relative beam calibration \label{tab:gains_15A}}
\tablehead{
\colhead{Pol} & \multicolumn{7}{c}{Beams} \\
\colhead{} & \colhead{1} & \colhead{2} & \colhead{3} & \colhead{4} & \colhead{5} &
\colhead{6} & \colhead{7}
}
\startdata
LL & 0.894 (0.071) & 0.865 (0.101) & 0.847 (0.049) & 0.772 (0.074) & 0.925 (0.062) & 0.677 (0.081) & 0.851 (0.080) \\
RR & 0.845 (0.060) & 0.846 (0.072) & 0.816 (0.074) & 0.745 (0.077) & 0.764 (0.057) & 0.511 (0.046) & 0.970 (0.081) \\
\enddata
\end{deluxetable}
\twocolumngrid

\subsection{Observations}
\label{sec:obs}

For all regions, observations were performed using the seven-beam K-Band Focal Plane Array (KFPA) at the GBT, with the VErsatile GBT Astronomical Spectrometer (VEGAS) backend. GAS uses VEGAS configuration Mode 20, allowing eight separate spectral windows per KFPA beam, each with a bandwidth of 23.44\ MHz and 4096 spectral channels, for a spectral resolution of 5.7\ kHz, or $\sim 0.07$\ \kms\ at 23.7\ GHz. All spectral lines observed are listed in Table \ref{tab:spectral_setup}. Due to limitations on the maximum separation in GHz between spectral lines in a single VEGAS bank, the GAS setup includes six spectral lines observed in each KFPA beam, plus the C$_2$S $2_1 - 1_0$ line in a single central beam. Observations were performed using in-band frequency switching to maximize on-source time, with a frequency throw of 4.11\ MHz. 

GAS maps were generally observed in 10\arcmin\ $\times$ 10\arcmin\ footprints, scanning in Right Ascension (R.A.), with scan rows separated by 13\arcsec\ in Declination (Dec.) to ensure Nyquist sampling. Each 10\arcmin\ square submap was observed once. To best match the observed regions to known sub-millimetre continuum emission, some regions included submaps of 10\arcmin\ $\times$ 5\arcmin, or 5\arcmin\ $\times$ 10\arcmin\ (where scans were in Dec., spaced by 13\arcsec\ in R.A. to minimize overheads). For all maps, the telescope scan rate was 6\farcs2 s$^{-1}$, with data dumped every 1.044\ s. A fast frequency-switching rate of 0.348\ s was used to meet the proposed sensitivity per map in the time allotted, but resulted in a small fraction (approximately a few percent) of the data being blanked due to synthesizer settling. 

\subsection{Flux Calibration}

Flux calibration was performed using observations of the Moon and Jupiter. Because the Moon is large relative to the KFPA footprint, simple on-target and off-target observations allowed relative beam calibrations to be performed. In angular size, Jupiter was comparable to the 32\arcsec\ GBT beam at 23.7\ GHz. For these sources, nod observations were performed to alternate on- and off-source beams, cycling through all beams. Relative beam calibrations were then calculated in GBTIDL\footnote{\url{http://gbtidl.nrao.edu/}}. 
Models for each solar system target, including the effects of solar illumination and apparent size, were used to determine the expected brightness temperature in the gain analysis\footnote{\url{https://safe.nrao.edu/wiki/pub/GB/Knowledge/GBTMemos/GBTMemo273-11Feb25.pdf}}.

The Moon was used as the primary calibrator when available, as the planetary calibrations could vary substantially when the weather was not ideal, particularly when the on-site winds were near the 5\ m\ s$^{-1}$ limits for KFPA operations at 23\ GHz. This can be seen in Figure \ref{fig:gains_date}, where we show the beam- and polarization-averaged gains derived from observations of Jupiter and the Moon over the 15A observing semester. Jupiter and Moon calibrations agree well when the winds are low. The last three dates when Jupiter was observed had relatively high winds ($\gtrsim 5$\ m\ s$^{-1}$), lowering the pointing accuracy of the telescope \added{, which is important to obtain an accurate flux calibration, but was ameliorated for the science observations by more frequent pointing calibration observations.
Moreover, the actual telescope pointing is recorded in the position encoder positions, and these are the values used during the data cube gridding}. 
Jupiter's angular size over the observation dates was $\sim 40$\arcsec, only slightly larger than the 32\arcsec\ (FWHM) GBT beam. Small pointing errors introduced by high winds are thus able to impact greatly the derived gains. In contrast, we show in Figure \ref{fig:gains_date} that the gains determined from the Moon were very stable over the 15A semester. 

Figure \ref{fig:gains_beams} shows the gains as a function of beam and polarization, averaged over the observing semester. The final three Jupiter measurements that were impacted by winds have been omitted from the analysis. Although the gains vary between beams and polarizations, they remained consistent over the observing period with only small variation. The gains determined and used for calibration (next section) are listed in Table \ref{tab:gains_15A}, along with the standard deviation of the measurements for each beam over the semester. 

Two known targets with bright \amm\ (1,1) lines (L1489\_PPC and L43) were also observed frequently in a single beam.
We found that the peak line brightness varied by $< 10$\ \% (1-$\sigma$) for these sources over the semester, in agreement with the calibration uncertainty estimated from our directly-measured beam gain variations.

\subsection{Data reduction and calibration}

All GAS data reduction and imaging codes (next section) are publicly available on GitHub\footnote{\url{https://GitHub.com/GBTAmmoniaSurvey/GAS}} and can be easily adapted for similar observations. 

The data reduction was performed using a Python wrapper that called the GBT KFPA data reduction pipeline \citep{masters11}. The data were calibrated to main beam brightness temperature  ($T_\mathrm{MB}$) units using the relative gain factors for each of the beams and polarizations discussed above and listed in Table \ref{tab:gains_15A}. Based on the 1 $\sigma$ variation in the beam gain calibrations, we estimate our calibration uncertainty to be $\sim 10$\ \%. 

\begin{figure*}
\gridline{\fig{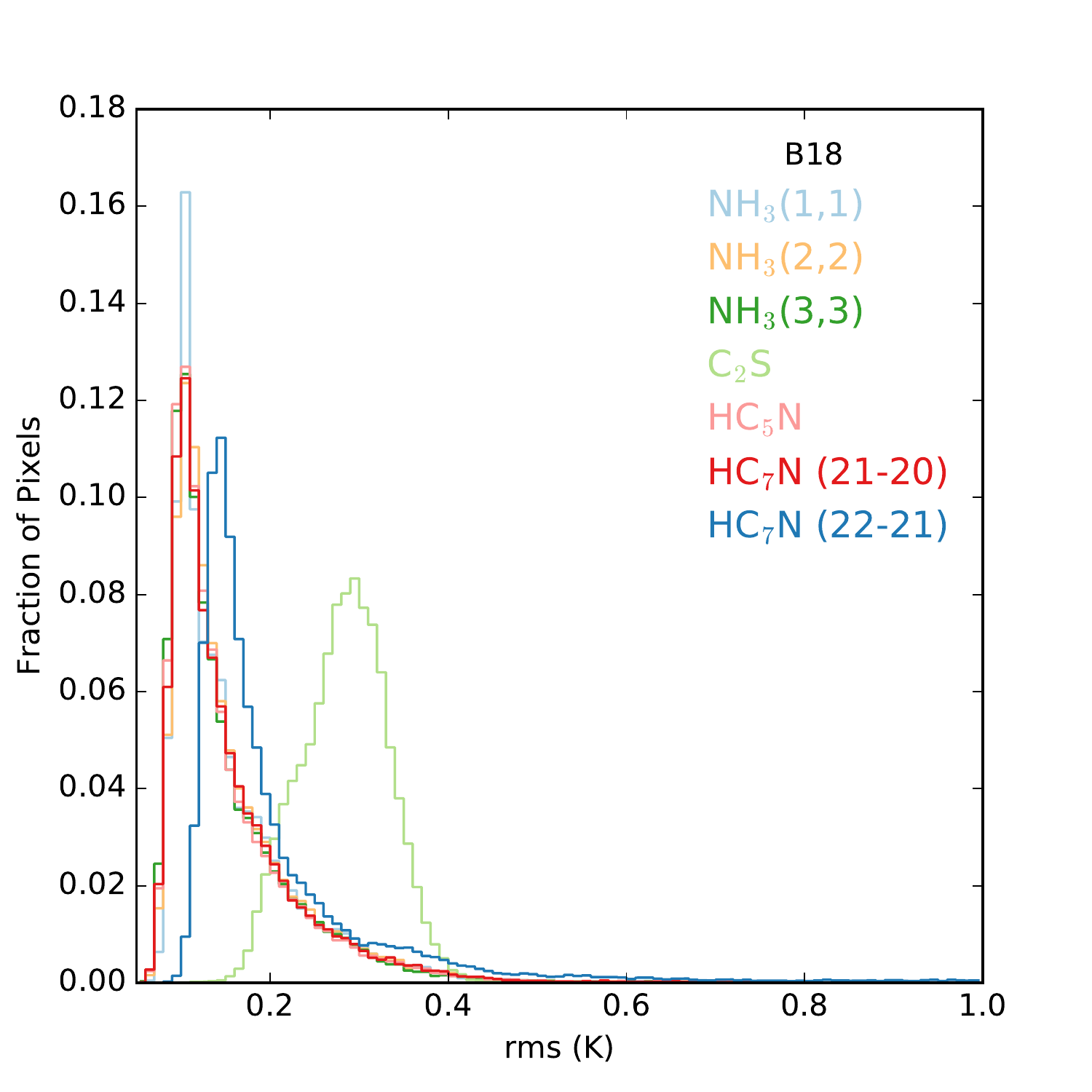}{0.9\columnwidth}{(a)}
          \fig{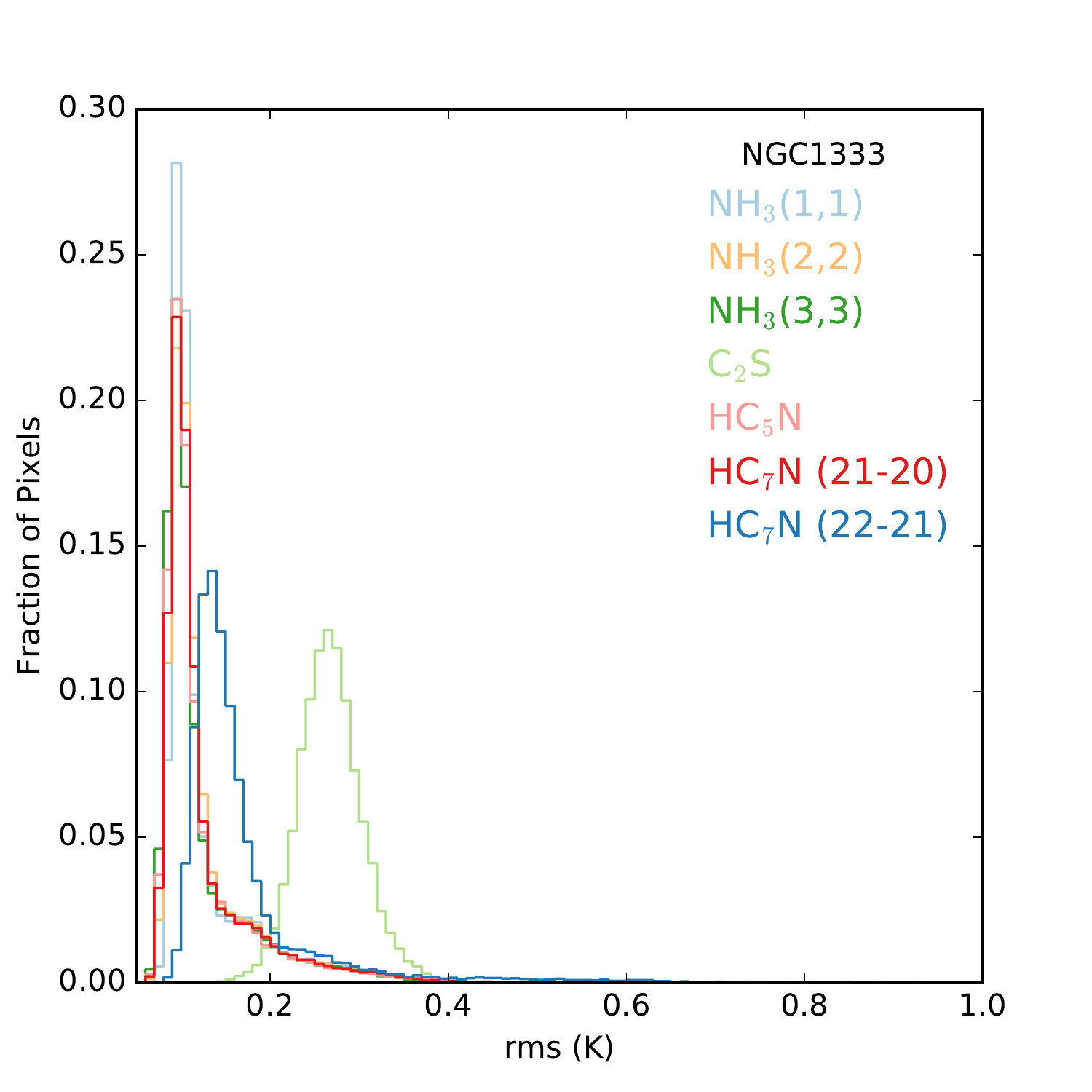}{0.9\columnwidth}{(b)}}
\gridline{\fig{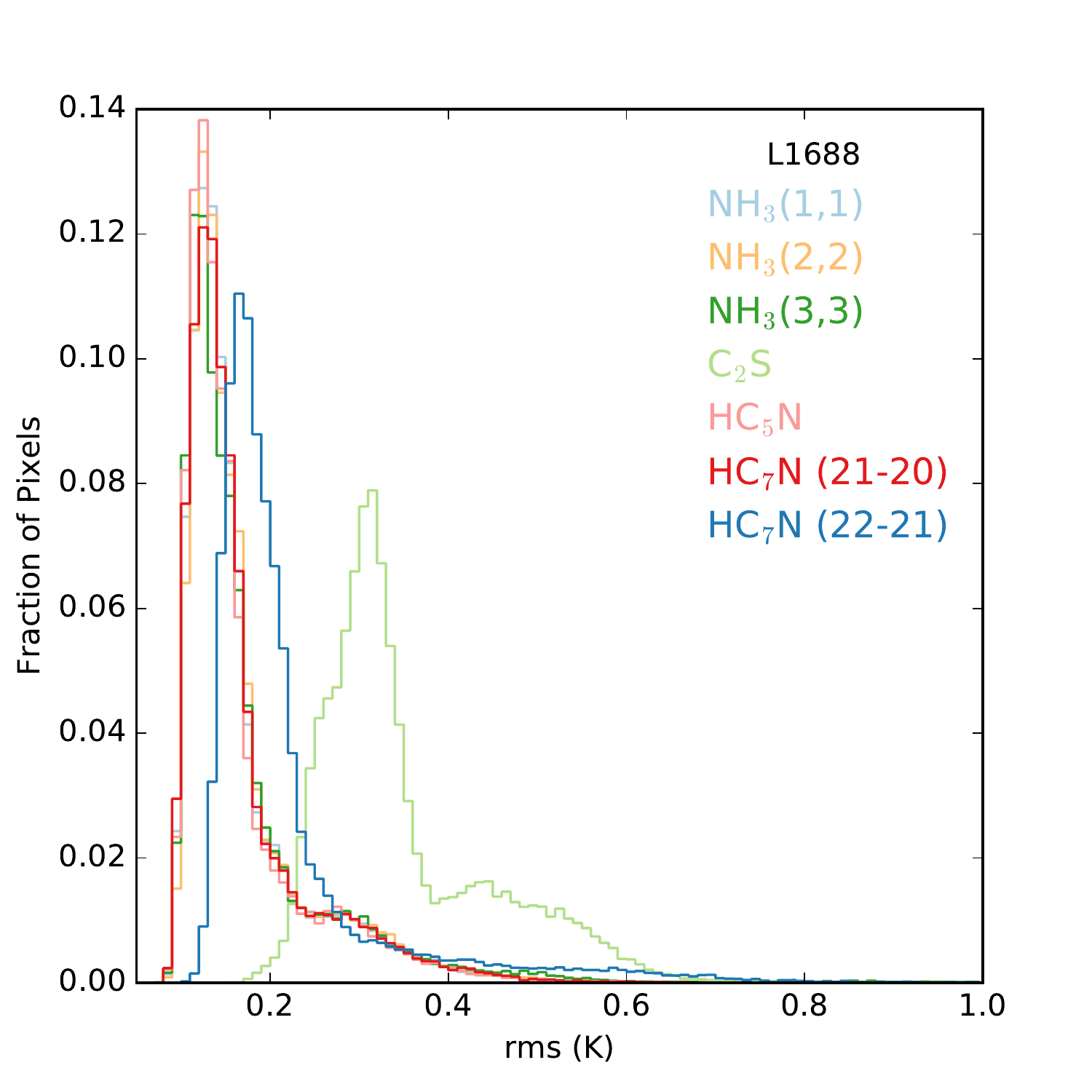}{0.9\columnwidth}{(c)}
          \fig{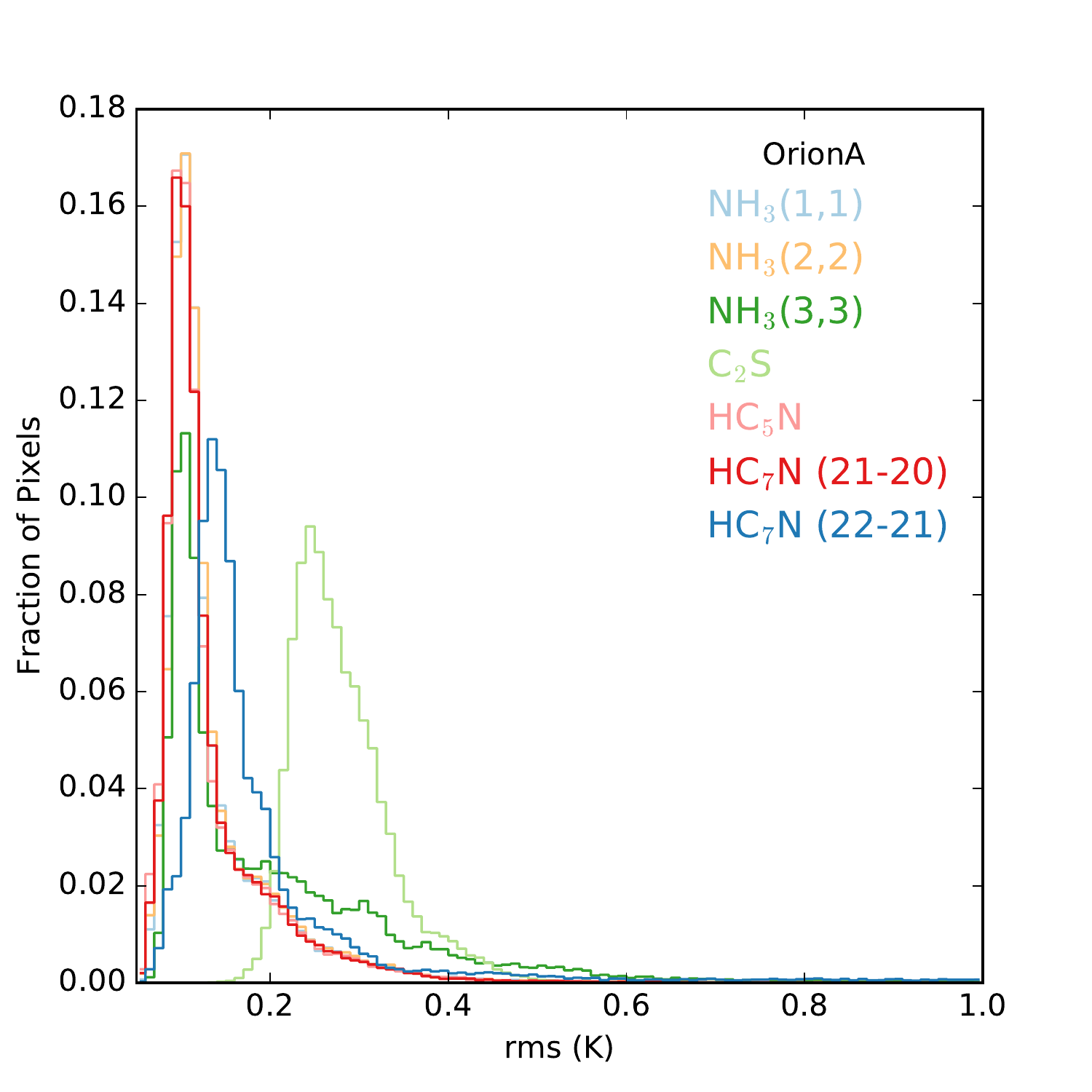}{0.9\columnwidth}{(d)}}
\caption{Histograms of the rms noise maps for all lines observed towards the DR1 regions, where each observed transition is shown in a different color. 
The rms is on the T$_\mathrm{MB}$ scale. \label{fig_allDR1_rms}}
\end{figure*}

\begin{figure*}
\begin{center}
\includegraphics[width=0.9\textwidth]{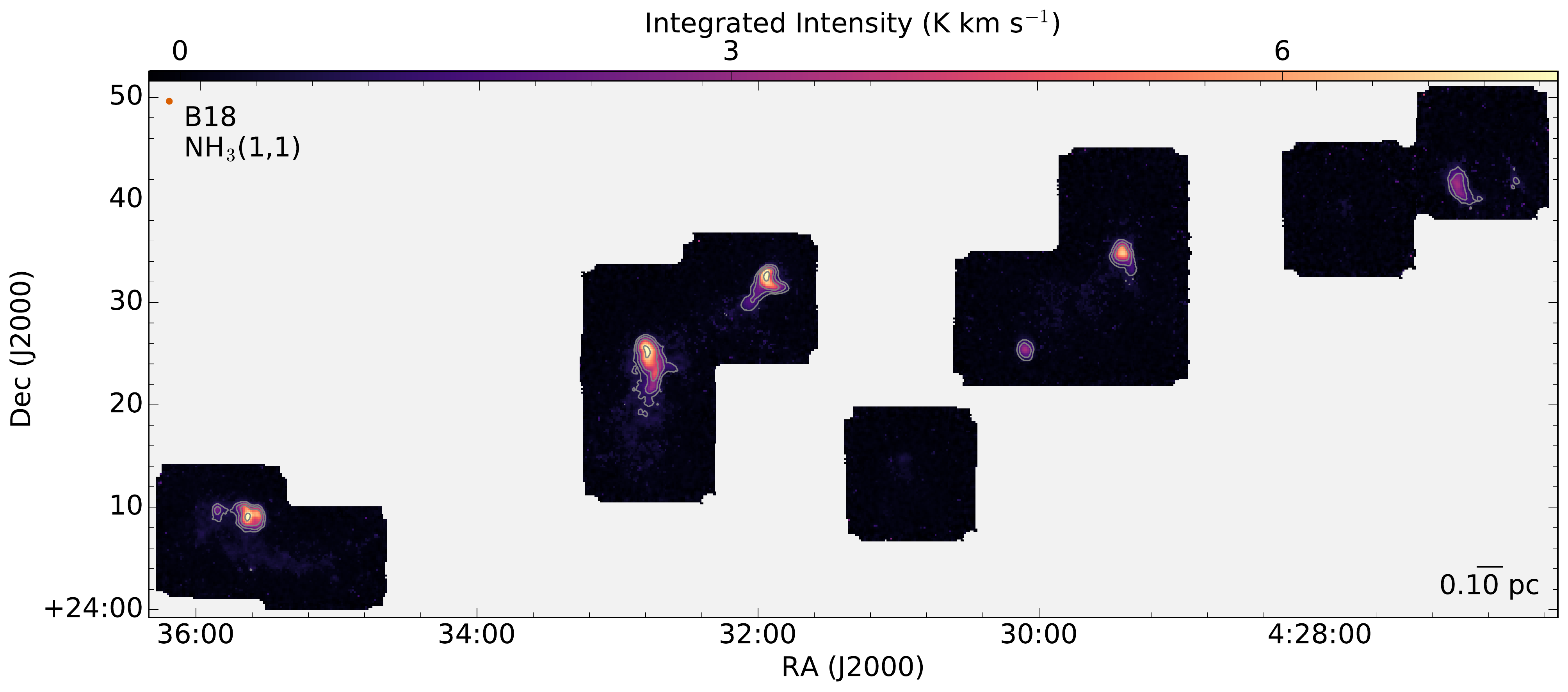}
\includegraphics[width=0.93\textwidth]{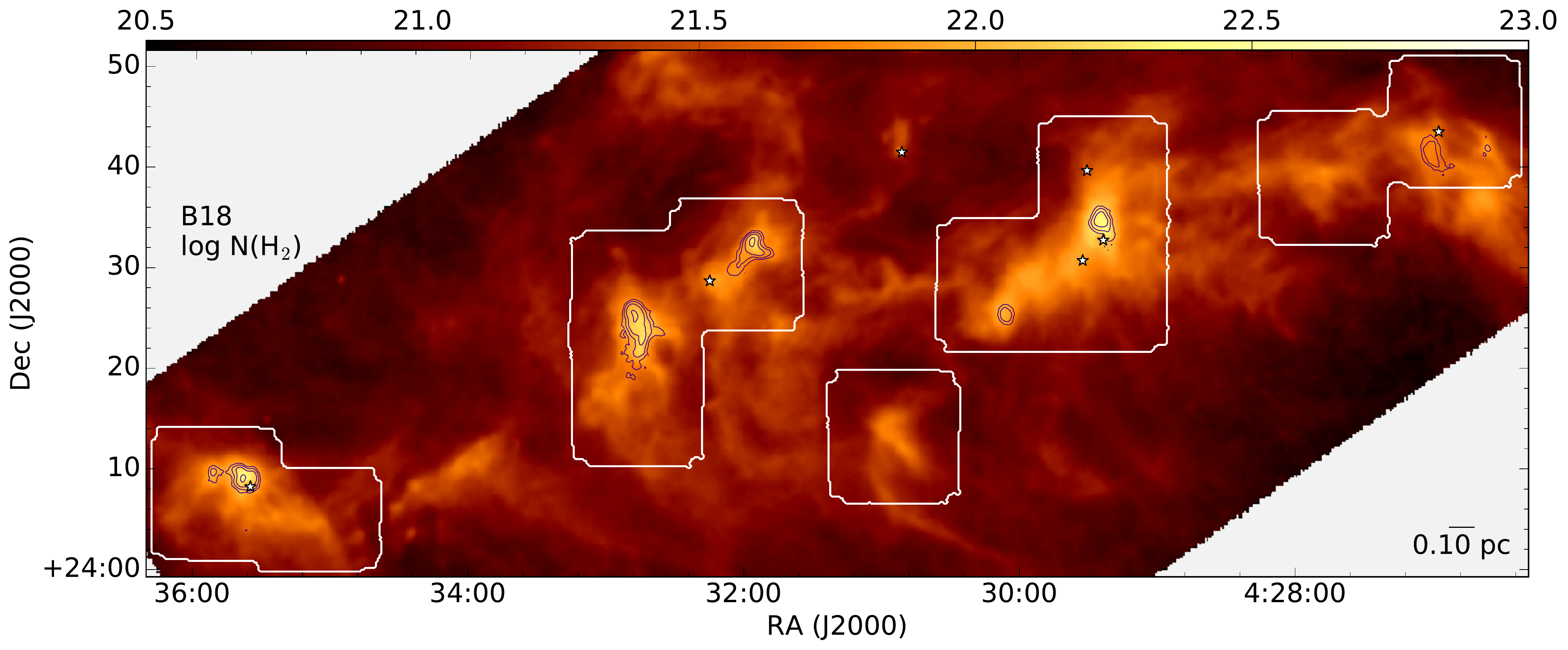}
\caption{\label{fig-b18-NH3-11_TdV} Top: Integrated intensity map of the \amm\ (1,1) line for the B18 region. Contours are drawn at [3,6,12,24,...]-$\sigma$, where $\sigma$ is the rms estimated from emission-free pixels. Beam size and scale bar are shown in the top left and bottom right corners, respectively. %
Bottom: The H$_2$ column density, log$_{10}$ [$N$(H$_2$)/cm$^{-2}$], derived from SED fitting of \textit{Herschel} submillimeter dust continuum data (A. Singh et al., in preparation). Purple contours show \amm (1,1) integrated intensity as at top. The white contour shows the \textit{GAS} map extent. 
Stars show the locations of Class 0/I and flat spectrum protostars \added{with a reliability grade of A- or higher} \citep{Rebull_2010}.
}
\end{center}
\end{figure*}

\subsection{Imaging}
\label{sec:imaging}

We grid the On-The-Fly (OTF) observations following the prescription by \citet{Mangum_2007}.  For each region, our mapping algorithm establishes a world coordinate system and corresponding pixel grid that covers the footprint of the observations, using a pixel size of 1/3 of the beam FWHM ($\theta_\mathrm{beam}$) of the \amm\ (1,1) line.
We then calculate the value of each map pixel as the weighted average of spectra at the center of the map pixel, using a truncated, tapered Bessel function for the weighting scheme.  Each integration is given a weight, $W$, based on the location of the observation relative to the center of the pixel in the resulting map and the observations rms:
\begin{equation}
W (r)  = \frac{J_1(\pi r/a)}{(\pi r/a)} e^{-r^2/b^2} \frac{1}{T_{\mathrm{sys}}^2}
\end{equation}
where $a = 1.55(\theta_{\mathrm{beam}}/3)$, $b=2.52(\theta_{\mathrm{beam}}/3)$, $r$ is the angular distance between the observed datum and the map datum, 
and $T_{\mathrm{sys}}$ is the recorded system temperature.  Integrations with $r>\theta_{\mathrm{beam}}$ are excluded from the average, as suggested by \cite{Mangum_2007}.  For simplicity of comparison, we grid data for all lines to a common world coordinate system.   

Online Doppler tracking is only applied for the \amm\ (1,1) line. For all other lines, we apply a frequency shift in the Fourier domain to each individual integration to align all spectra before convolution (gridding), i.e. Doppler tracking is done in software.  Baseline subtraction is done in both the individual integrations (linear baseline) and in the individual pixels in the final data cubes (Legendre polynomials of order 3).  In the latter case, sections of the spectrum with significant emission or within 2 $\mbox{km s}^{-1}$ of the expected systemic velocity are excluded from the fit.  

The KFPA on the GBT is a sparse array, with seven beams arranged in a hexagonal pattern, and beam centers separated by 94.88\arcsec\ on the sky. The mapping method described in \S\ \ref{sec:obs} ensured Nyquist-spaced sampling over 10\arcmin\ $\times$ 10\arcmin\ footprints on the sky. At the edges of these footprints, coverage is variable, and the rms noise increases substantially as noted previously. For the final moment maps and parameters maps for each region, we mask pixels within 3 pixels of the map edges to remove the noisiest regions. 

In Figure \ref{fig_allDR1_rms}, we show histograms of the rms noise for all regions and lines observed. The rms noise map for the \amm\ (1,1) observations toward each DR1 region are presented in Appendix \ref{sec:noise_maps}. 

In all regions, the noise histograms have strong peaks that agree well for the \amm\ (1,1), (2,2), (3,3), HC$_5$N $9-8$, and HC$_7$N $21-10$ observations, with typical 1 $\sigma$ noise values of $\lesssim 0.1$\ K per velocity channel across the DR1 regions. Note that the typical noise in individual map footprints can vary slightly depending on the conditions in which the region was mapped. Figure \added{Set} \ref{FigSet-A} show that in general this variation is minimal. The C$_2$S observations have greater rms noise values since C$_2$S was observed with a single beam rather than all seven beams in the KFPA. The HC$_7$N $22-21$ data are noisier than the data of other lines observed with all KFPA beams, a systematic difference seen over all observations. 

\begin{figure*}
\begin{center}
\includegraphics[width=0.9\columnwidth]{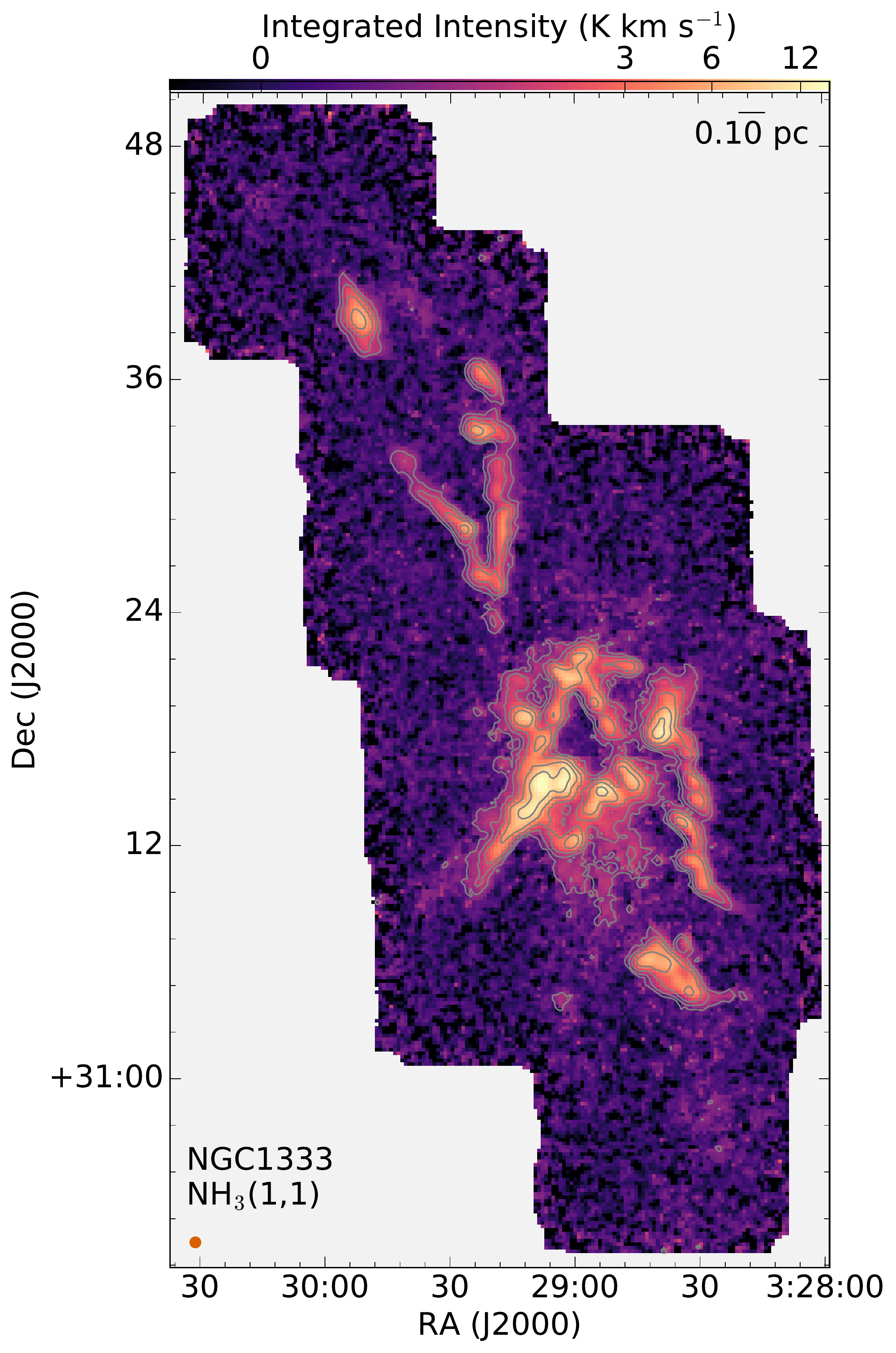}
\includegraphics[width=0.9\columnwidth]{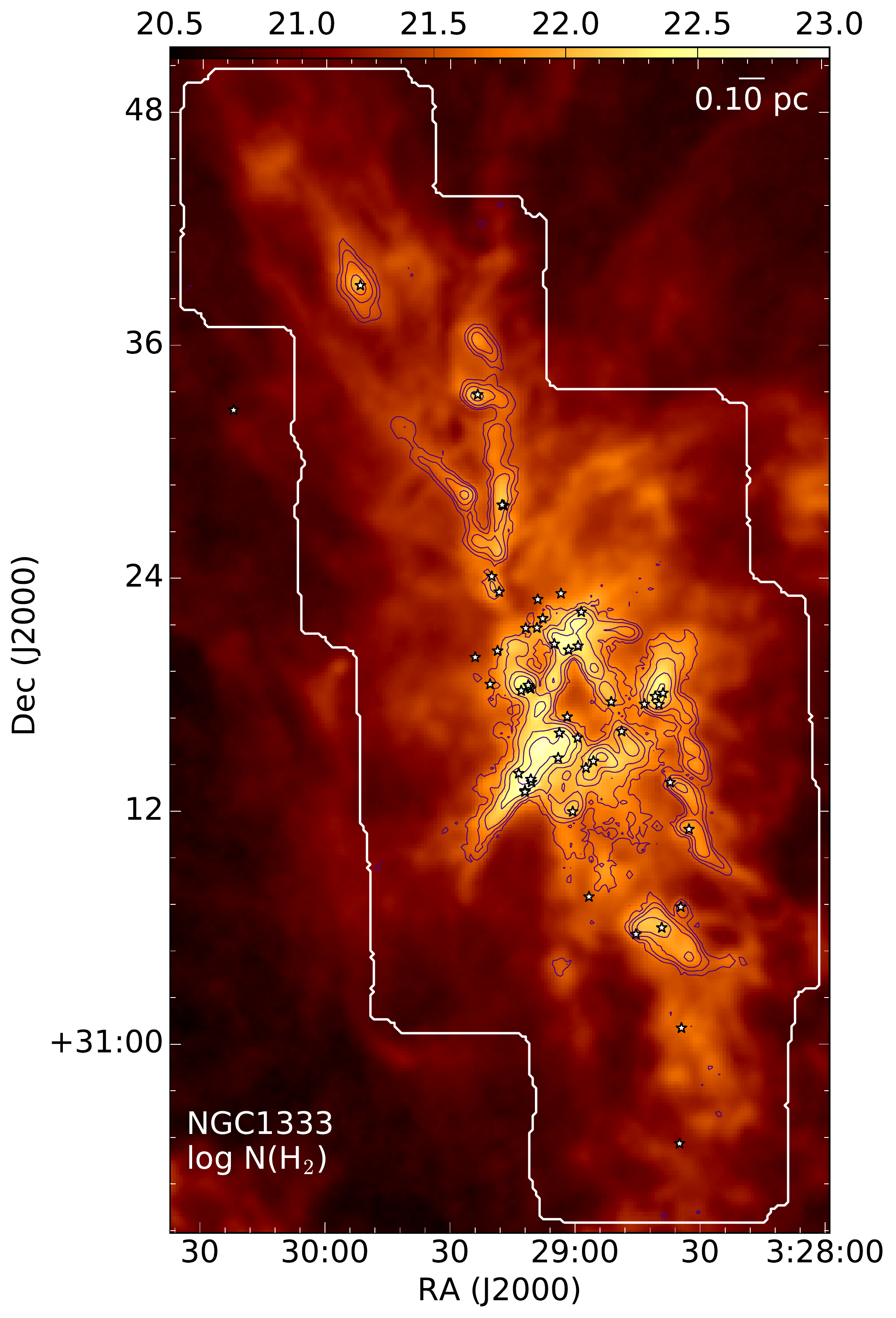}
\caption{\label{fig-ngc1333-NH3-11_TdV}
Like Figure \ref{fig-b18-NH3-11_TdV}, but for NGC 1333. Stars show the locations of Class 0/I and flat spectrum protostars \citep{Dunham_2015}.
}
\end{center}
\end{figure*}

\section{\amm\ line analysis}
\label{sec:linefit}

\subsection{Line fitting}
We calculate the properties of the column of ammonia molecules using a forward-modeling approach in the \verb+pyspeckit+ package \citep{2011ascl.soft09001G}, 
which builds from the works of \citet{Rosolowsky_2008} and \citet{Friesen_2009} relying on the theoretical framework laid out in \citet{Mangum_2015}.  
Under this approach, we consider a beam-filling slab of ammonia gas with a set of physical properties including para-ammonia column density ($N_{\mathrm{p-NH}_3}\equiv N$ for notational compactness in this section), kinetic temperature ($T_{K}$), velocity dispersion $\sigma_v$, and line of sight velocity in the (kinematic) LSR frame ($v_{\mathrm{LSR}}$).  While we observe the $(J,K)=(3,3)$ line of ortho-ammonia, we do not directly incorporate any information from this line into our fit because of unknown ortho-to-para ratios that might differ from LTE \citep[e.g.,][]{Faure_2013}.

Given a total column density for the molecule and a velocity dispersion for the slab, we use standard relations to determine the fraction of molecules found in each of the states, namely $(J,K)$ rotational states of the species and the upper ($u$, antisymmetric inversion symmetry) and lower ($l$, symmetric inversion symmetry) within those states.  
Our model incorporates three distinct temperatures. 
The kinetic temperature $T_K$ describes the velocity distributions of the particles in the system, notably of the collider species H$_2$ and He.  
The rotation temperature, $T_{\mathrm{rot}}$ is the temperature that characterizes the rotational level population.  
Our work usually focuses on the temperature that defines the population ratio between the (1,1) and (2,2) states.  
Finally, $T_{\mathrm{ex}}$ is the radiative excitation temperature of the observed spectral lines.  

\amm\ fitting usually proceeds under the assumption that the (1,1) and the (2,2) line have the same radiative excitation temperature and that the frequencies are approximately equal and $\ll kT_{\mathrm{ex}}/h$.  The former assumption is only approximately true: using RADEX \citep{vandertak_2007} for $n=10^4~\mathrm{cm}^{-3}$ and $T_{K} = 15\ \mathrm{K}$ and $N(\mathrm{p-NH}_3)=10^{14}~\mathrm{cm^{-2}}$  gives $T_{\mathrm{ex},(1,1)} = 8.5~\mathrm{K}$ and $T_{\mathrm{ex},(2,2)} = 6.9~\mathrm{K}$.  Each of the inversion transitions consists of multiple, resolved hyperfine components and we further assume a constant $T_{\mathrm{ex}}$ for all these transitions, which precludes our model representing hyperfine anomalies \citep[e.g.,][]{Stutzki_1984}.  We see, however, no evidence in our survey that these effects are important.

Our observations consist of antenna temperature as a function of frequency, and so we generate a model of the spectral line given the input parameters ($N$, $T_K$, $T_{\mathrm{ex}}$,$\sigma_v$, $v_{\mathrm{LSR}}$).  We assume a single radiative excitation temperature, $T_{ex}$, characterizes both the (1,1) and the (2,2) transitions. We generate a model of optical depth as a function of frequency $\tau(\nu)$ to produce a model spectrum of the main-beam radiation temperature (assuming the beam filling fraction $=1$):
\begin{equation}
T_{\mathrm{MB}}(\nu) = \left[J(T_{\mathrm{ex}})-J(T_{\mathrm{bg}})\right]\left[1-e^{-\tau(\nu)}\right].
\end{equation}

The optical depth, $\tau(\nu)$, is a function of the input parameters.  To derive $\tau(\nu)$ from the input parameters, we first approximate the molecular population as having most of the p-NH$_3$ in the (1,1) and (2,2) states so that we can define a $T_{\mathrm{rot}}$ by the ratio of the column densities in the (2,2) and (1,1) states:
\begin{equation}
 N_{(2,2)} = N_{(1,1)} \frac{5}{3} \exp\left(-\frac{\Delta E}{kT_{\mathrm{rot}}}\right)
\end{equation}
where the $5/3$ is the ratio of the rotational statistical weights of the states ($g_J = 2J+1$), 
and $N(J,K)$ is the sum of the column densities of the two inversion levels. 
From here we have that $T_{\mathrm{rot}}$ can be related to the kinetic temperature by analytically solving the equations of detailed balance \citep{Swift_2005}:
\begin{equation}
T_{\mathrm{rot}} = T_K \left\{1+\frac{T_{K}}{T_{0}} \ln \left[1+\frac{3}{5} \exp\left(-\frac{15.7~\mathrm{K}}{T_K}\right)\right]\right\}.
\end{equation}

Here $T_0=41.5~\mathrm{K}$ is the energy difference between the $(2,2)$ and the $(1,1)$ states.  
We then use this rotational temperature to calculate a partition function for the p-NH$_3$ molecule, though the next metastable level is the $(4,4)$ line at 178 K above the ground state and is negligibly populated ($\tau<1$) for $T_K<65\mbox{ K}$. 
The partition function is given by
\begin{eqnarray}
Z_\mathrm{para} &=& \sum (2 J + 1) \times \\ \nonumber
&\exp&\left(\frac{-h(B_\mathrm{rot} J(J+1) + (C_\mathrm{rot} - B_\mathrm{rot}) J^2)}{k_\mathrm{B}T_\mathrm{rot}}\right),
\end{eqnarray}
where $B_\mathrm{rot}$ and $C_\mathrm{rot}$ are the rotational constants of the ammonia molecule, 298117 and 186726 MHz, respectively \citep{Pickett_1998}.

\begin{figure*}
\begin{center}
\includegraphics[width=0.7\textwidth]{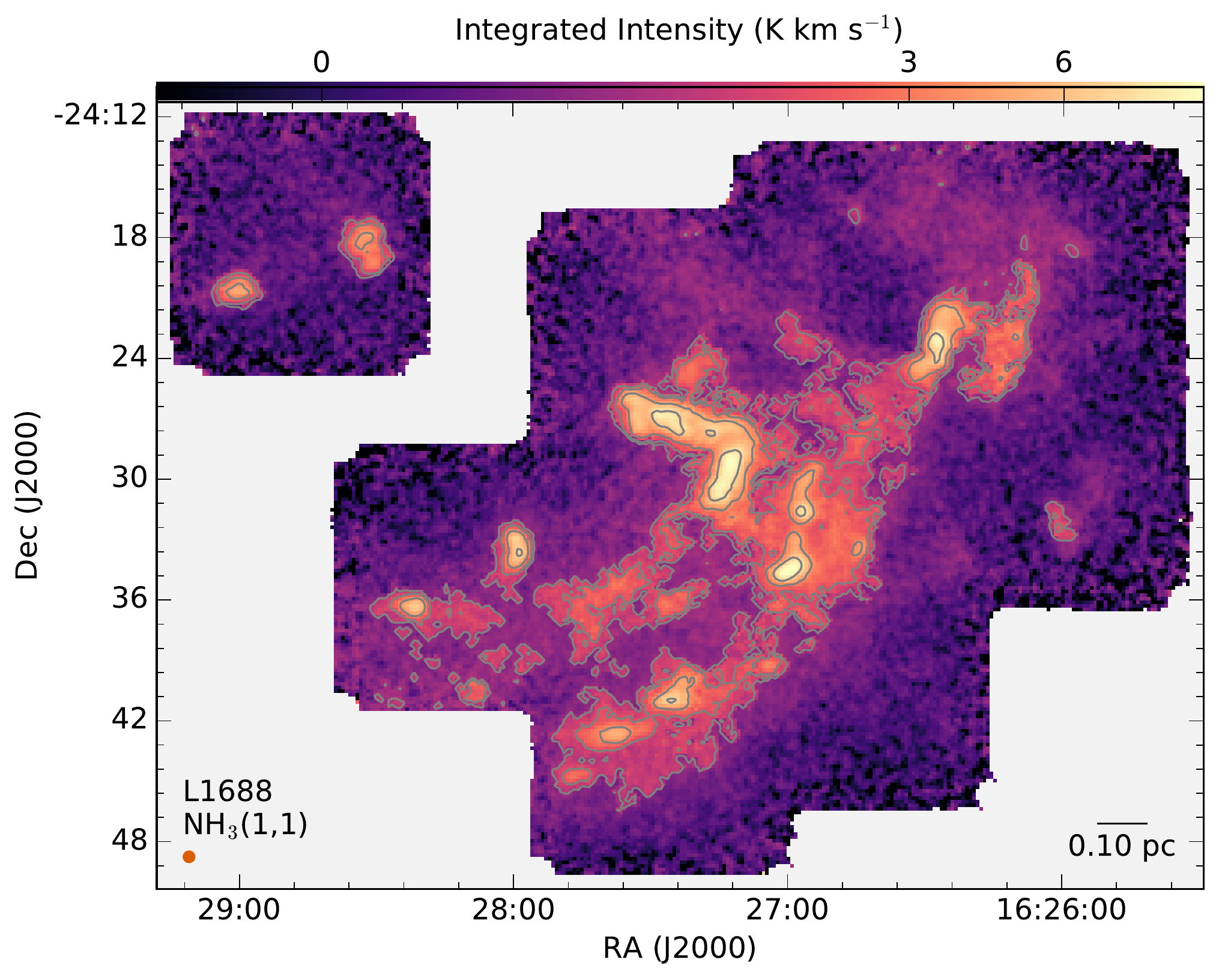}
\includegraphics[width=0.73\textwidth]{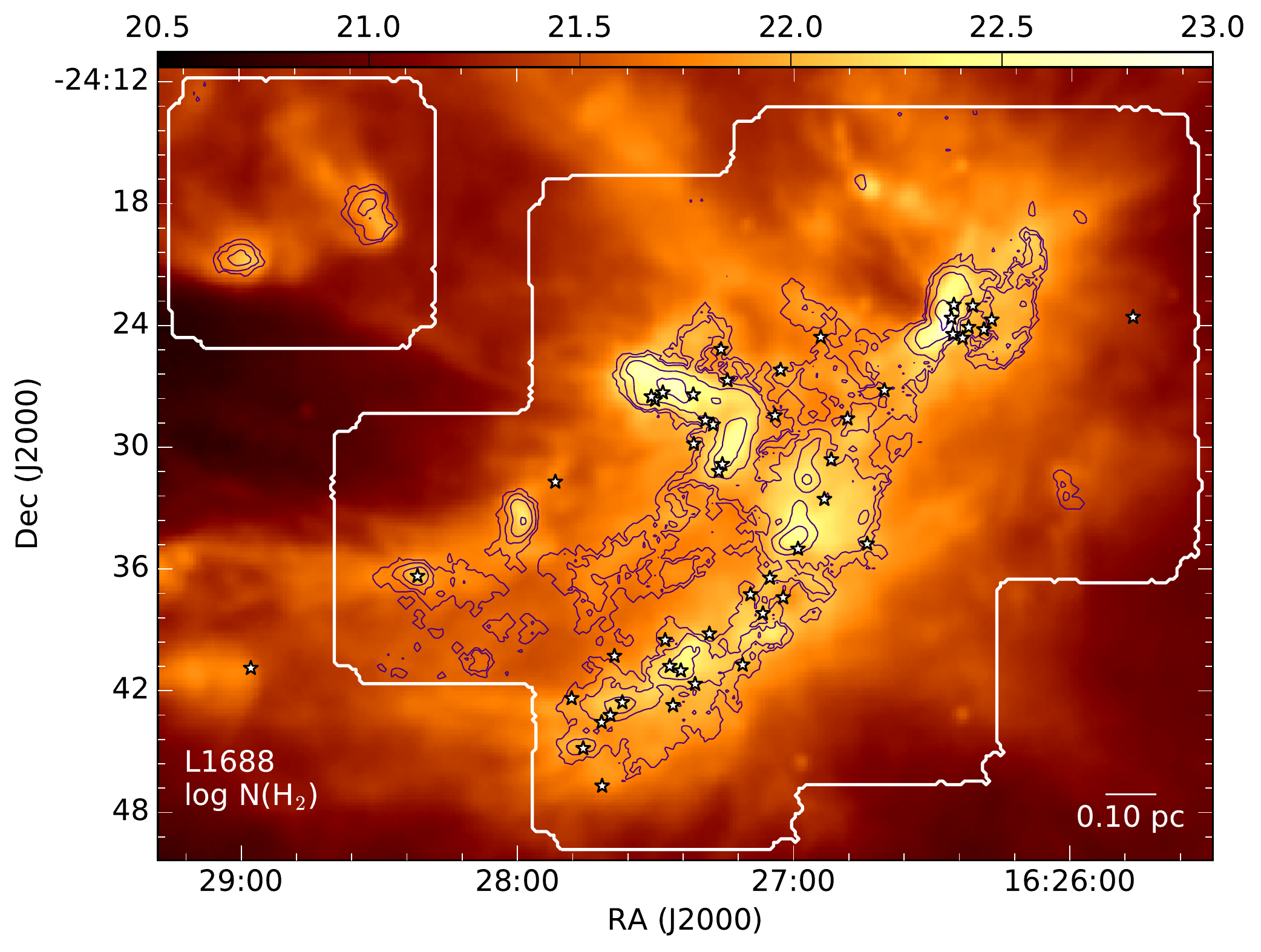}
\caption{\label{fig-l1688-NH3-11_TdV} 
Like Figure \ref{fig-b18-NH3-11_TdV}, but for L1688. 
Stars show the locations of Class 0/I and flat spectrum protostars \citep{Dunham_2015}.
}
\end{center}
\end{figure*}

\begin{figure*}
\begin{center}
\includegraphics[width=0.8\columnwidth]{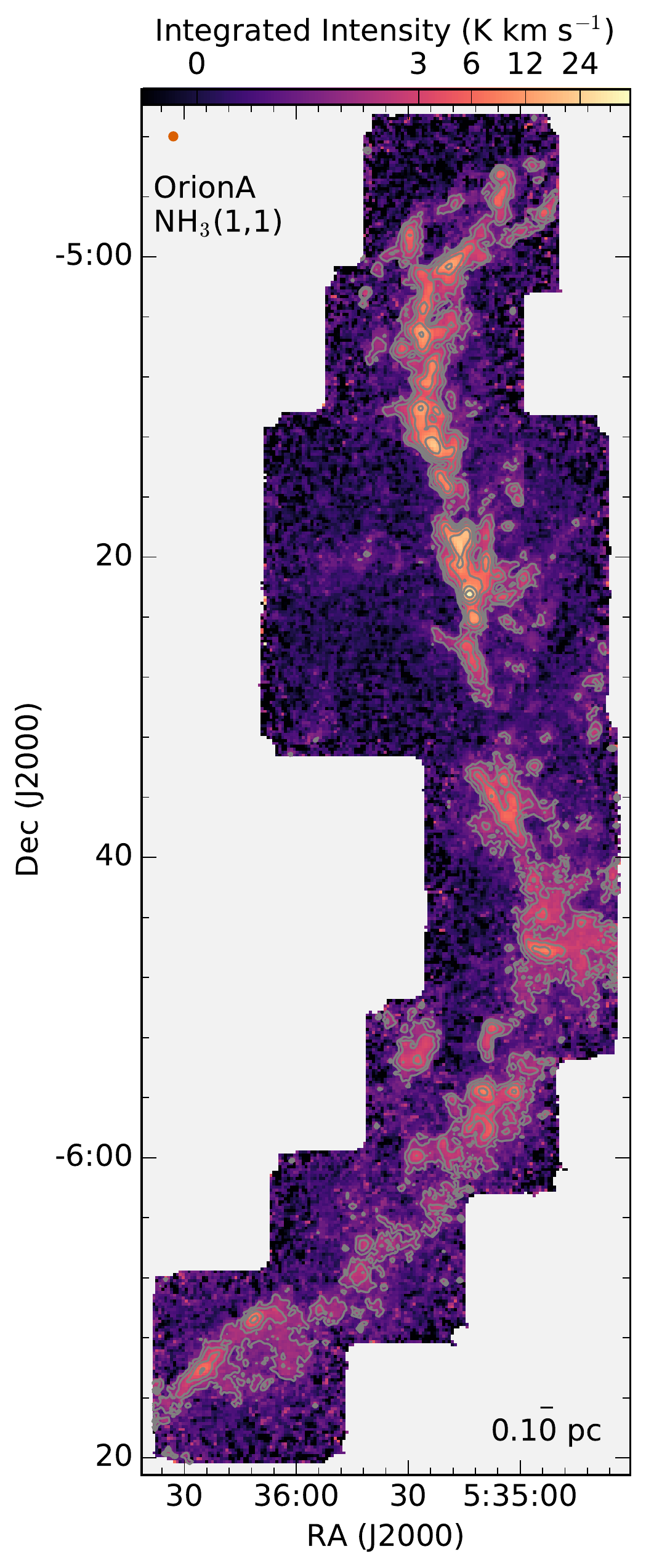}
\includegraphics[width=0.835\columnwidth]{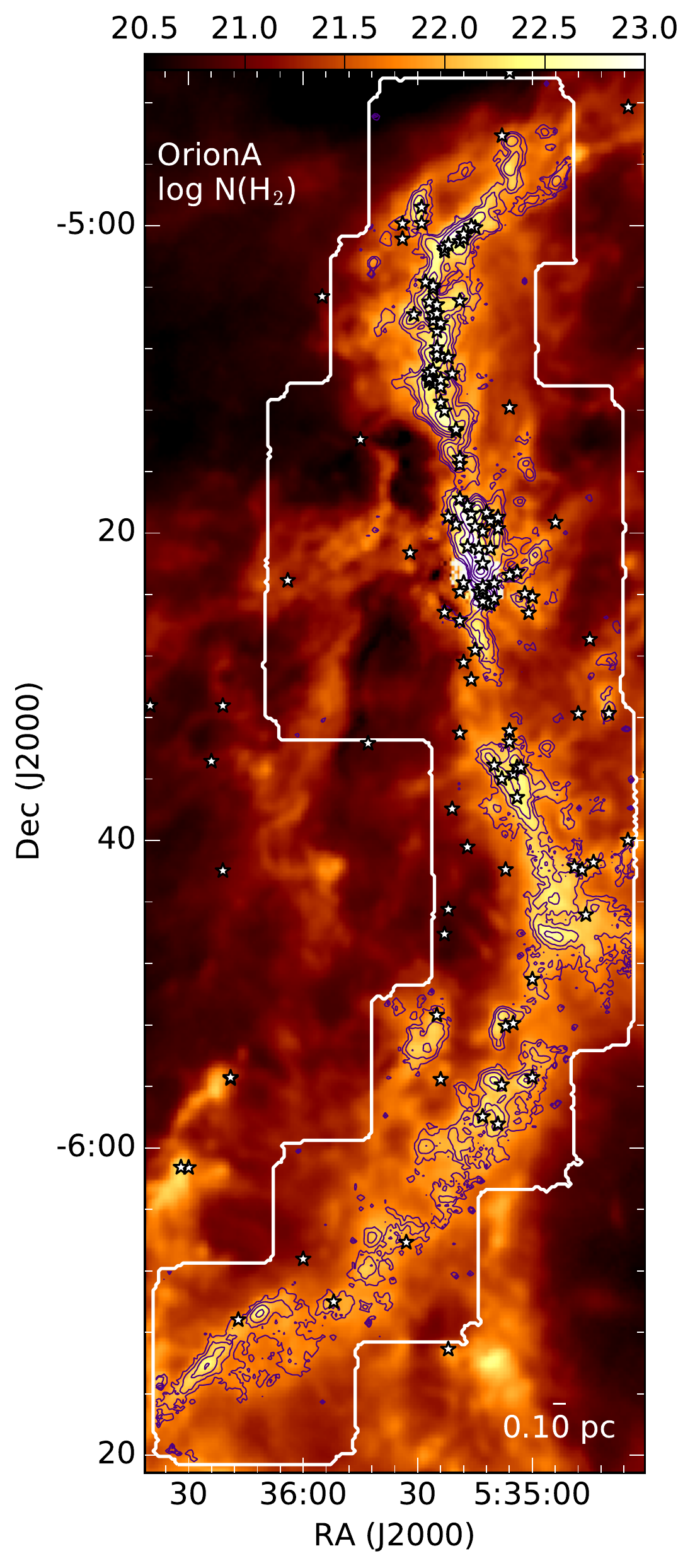}
\caption{\label{fig-OrionA-NH3-11_TdV} 
Like Figure \ref{fig-b18-NH3-11_TdV}, but for Orion A (North). 
Stars show the locations of Class 0/I and flat spectrum protostars \citep{Megeath_2012}. Note that the column density is poorly fit toward the Orion nebula. 
}
\end{center}
\end{figure*}

\begin{figure*}
\includegraphics[trim=10 0 13 0, clip, width=0.5\columnwidth]{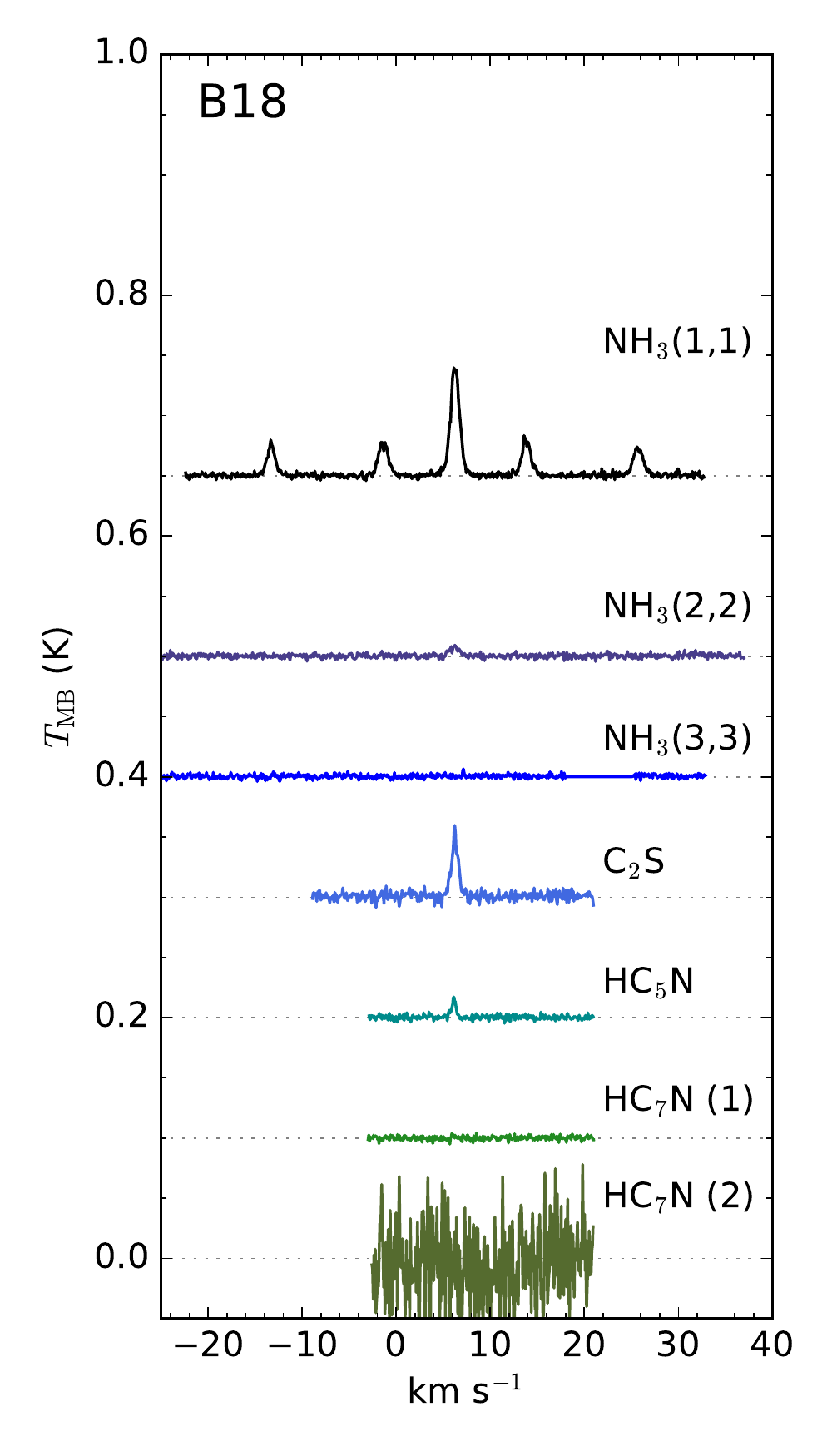}
\includegraphics[trim=10 0 13 0, clip, width=0.5\columnwidth]{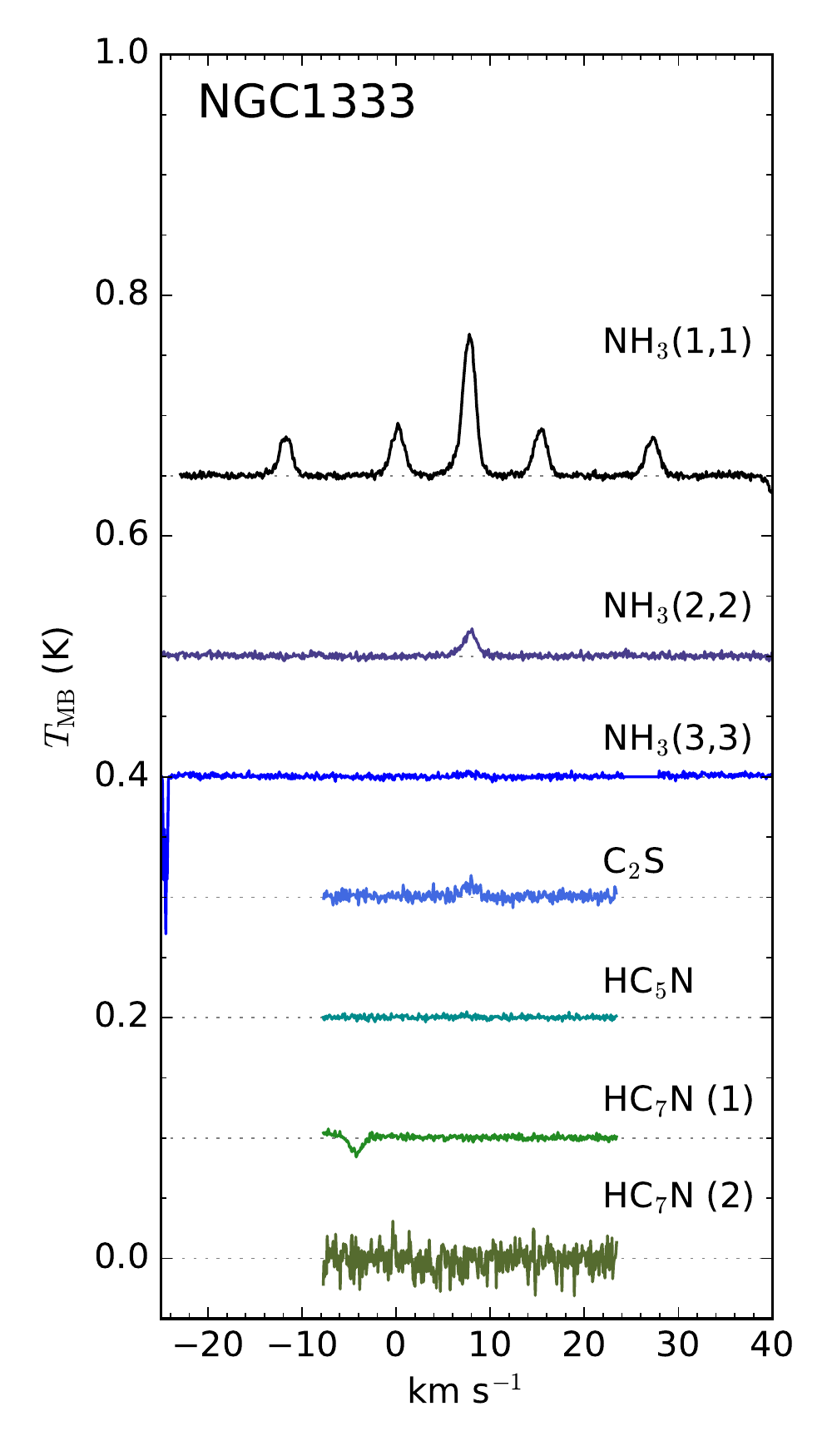}
\includegraphics[trim=10 0 13 0, clip, width=0.5\columnwidth]{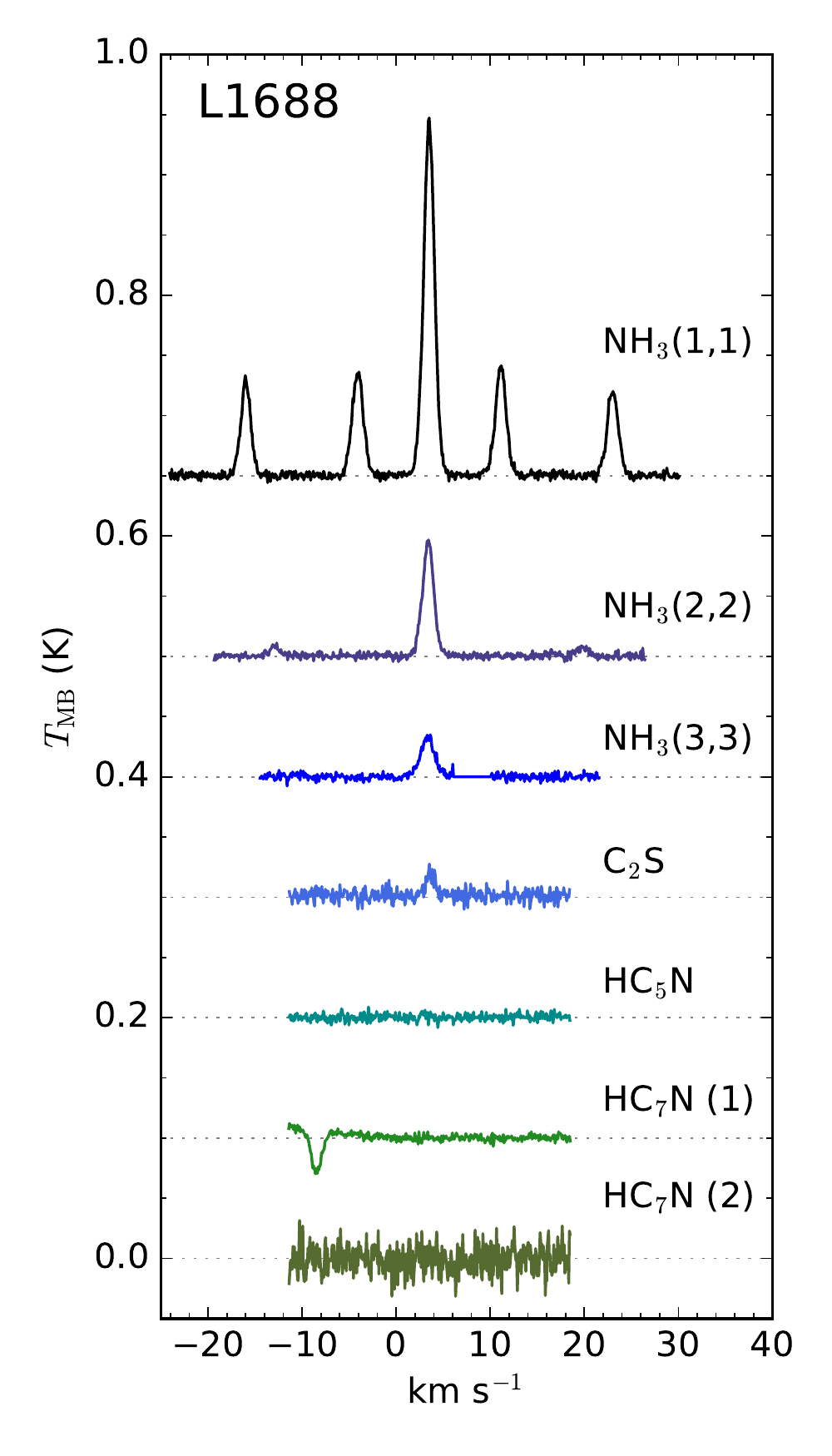}
\includegraphics[trim=10 0 13 0, clip, width=0.5\columnwidth]{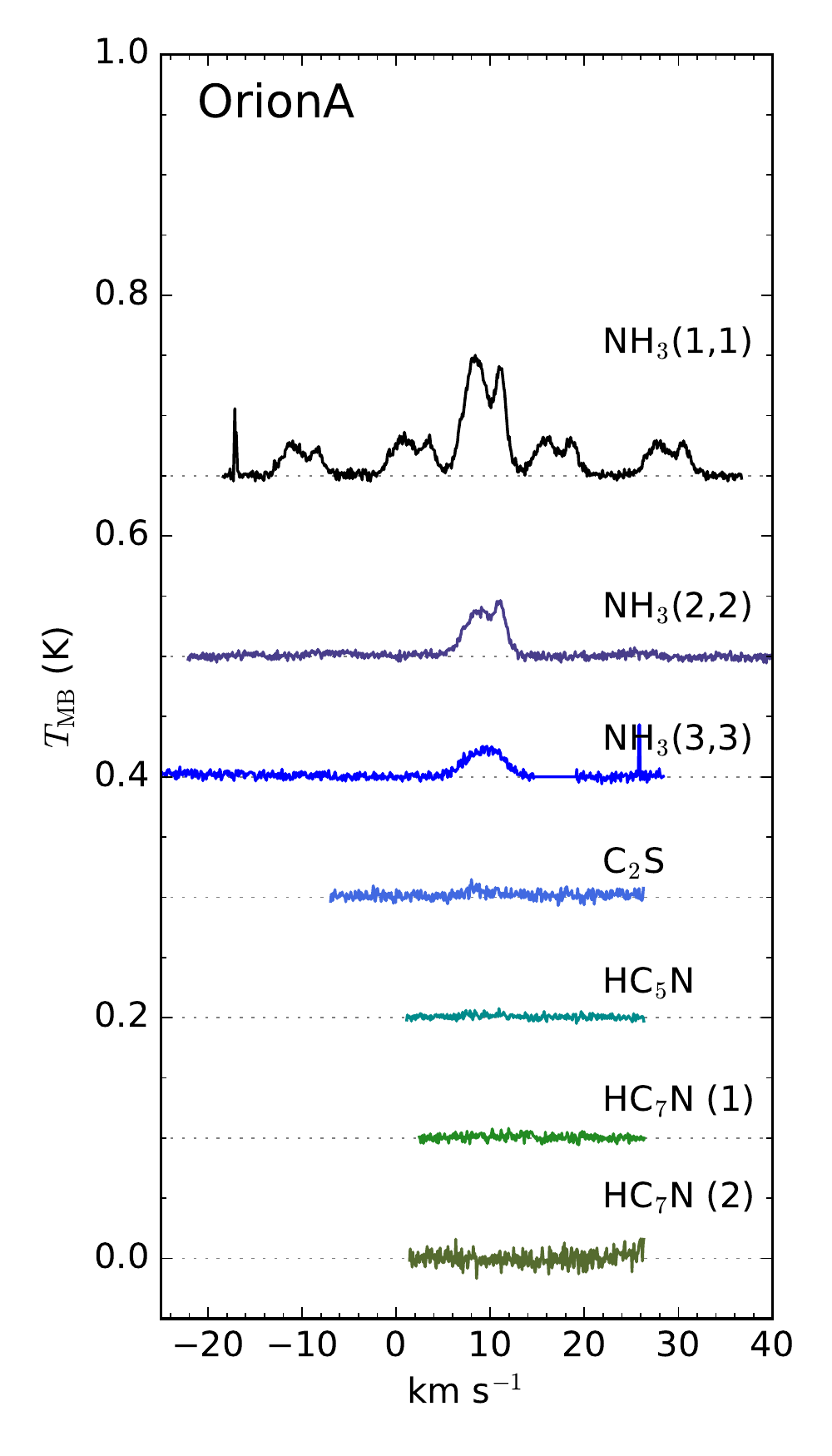}
\caption{Spectra of all lines observed, averaged over each of the DR1 regions. Lines have been shifted in $T_\mathrm{MB}$ for clarity. The spectra labeled HC$_7$N (1) and HC$_7$N (2) are HC$_7$N 21-20 and HC$_7$N 22-21, respectively. The hyperfine structure of the \amm\ (1,1) is clearly detected in the averaged spectra toward all regions. In L1688, the relatively weaker hyperfine structure of the \amm\ (2,2) line is also seen. The HC$_7$N 22-21 spectra have significantly more noise than the other lines, primarily due to greater rms values at the maps' edges, and have been multiplied by a factor of 0.5. In NGC 1333 and L1688, the absorption features in the HC$_7$N 21-20 spectra are artifacts of  frequency-switching while observing, where some of the \amm\ (1,1) emission was switched near the HC$_7$N line. After running the calibration and imaging pipelines, several footprints have greater rms noise values at some velocity channels. For \amm\ (3,3), we have masked the velocity channels where this issue is strongest, but several additional features (visible as sharp, narrow spikes) remain.  \label{fig_all_spectra}}
\end{figure*}

Since the excitation temperature defines the population ratio of the lower to the upper state, we can calculate $N_l$ from the total column density in the $(1,1)$ state, $N_{(1,1)}$.
\begin{equation}
\frac{N_u}{N_l}=\frac{g_u}{g_l}\exp\left(-\frac{h\nu_0}{kT_{\mathrm{ex}}}\right),
\end{equation}
so 
\begin{equation}
N_l=\frac{N_{(1,1)}}{1+\frac{g_u}{g_l}\exp\left(-\frac{h\nu_0}{kT_{\mathrm{ex}}}\right)}~.
\end{equation}

Given the column density in the lower state, we then calculate a total, integrated optical depth in the (1,1) and (2,2) lines using standard radiation relationships.  In particular, the opacity per unit frequency,
\begin{equation}
 \alpha_\nu = \frac{c^2}{8\pi} \frac{1}{\nu_0^2} \frac{g_u}{g_l} n_{l} A_{ul} \left[1-\exp\left(-\frac{h\nu_0}{kT_{\mathrm{ex}}}\right)\right] \phi(\nu),
 \end{equation}
where $\nu_0$ is the line rest frequency, $n_{l}$ is the volume density in the lower state of the transition, $g_u$,$g_l$ are statistical weights of the upper and lower states of a transition and whose ratio is unity for each pair of inversion transitions, $A_{ul}$ is the Einstein $A$ coefficient, and $\phi(\nu)$ is the line profile function defined such that 
\begin{equation}
\int \phi(\nu)\,d\nu = 1.
\end{equation}
We integrate over line of sight and frequency to get the optical depth as a function of the column density in the lower transition of the spectral line:
\begin{eqnarray}
\tau &=& \int \alpha_\nu\, ds\, d\nu  \nonumber \\
&=& \frac{c^2}{8\pi} \frac{1}{\nu_0^2} \frac{g_u}{g_l} N_{l} A_{ul} \left[1-\exp\left(-\frac{h\nu_0}{kT_{\mathrm{ex}}}\right)\right]. 
\end{eqnarray}
or, in terms of $N_{(1,1)}$, which is the total column of molecules in the symmetric and antisymmetric (upper and lower) states of the J=1 K=1 level,
\begin{eqnarray}
\tau_{(1,1)} &=& \int \alpha_\nu\, ds\, d\nu \nonumber\\
&=& \frac{c^2}{8\pi} \frac{1}{\nu_0^2} \frac{g_u}{g_l} N_{(1,1)} A_{ul} \frac{1-\exp\left(-\frac{h\nu_0}{kT_{\mathrm{ex}}}\right)}{1+\frac{g_u}{g_l}\exp\left(-\frac{h\nu_0}{kT_{\mathrm{ex}}}\right)}. 
\end{eqnarray}
The optical depth for the (2,2) transition has the same form. The column densities are related to the total column $N$ through the partition function via $T_{\mathrm{rot}}$.


As per \citet{Mangum_2015}, 
\begin{equation}
A_{ul} = \frac{64 \pi^4 \nu_0^3}{3 h c^3} \frac{1}{g_u} |\mu_{ul}|^2,
\end{equation}
where $\mu_{ul}$ is the dipole moment of a given transition, which is related to the dipole moment of the molecule $\mu^2$ by
\begin{equation}
|\mu_{ul}|^2 = \mu^2 \frac{K^2}{J(J+1)},
\end{equation}
where $\mu = 1.468\times 10^{-18}\mathrm{~esu~cm}$ for ammonia.

Finally, we calculate the line profile function $\phi(\nu)$ from the hyperfine structure of the different transitions.  Since each hyperfine transition is assumed to have the same radiative excitation temperature, $\phi(\nu)$ is given as the weighted sum of Gaussian line profiles:
\begin{equation}
\phi(\nu) = \sum_i \frac{w_i}{\sqrt{2\pi} \sigma_{\nu,i}} \exp\left[-\frac{(\nu-\nu_0-\delta\nu_i)^2}{2\sigma_{\nu,i}}\right].
\end{equation}
The weights, $w_i$, and frequency offsets from the rest frequency, $\delta\nu_i$, are set by quantum mechanics and are tabulated in \citet{Mangum_2015}.  The frequency widths of the lines are calculated in terms of the velocity widths from the Doppler formula: $\sigma_{\nu,i} = (\nu_0+\delta\nu_i) \sigma_v c^{-1}$.  

This model provides a complete spectrum $T_{\mathrm{MB}}$ as a function of input parameters.  We then use a non-linear gradient descent algorithm \citep[\textsc{MPFIT;}][]{Markwardt_2009} implemented in Python to calculate the optimal values of the input parameters and their uncertainties \citep[e.g.,][]{NumericalRecipes}.

\begin{eqnarray}
&\mathrm{argmin}&\, \chi^2 =  \\
&\mathrm{argmin}& \left[\sum_j \frac{[T_{\mathrm{MB},\mathrm{obs}}(\nu_j) - T_{\mathrm{MB}}(\nu_j;N, T_{K}, T_{\mathrm{ex}}, \sigma_v, v_{\mathrm{LSR}})]^2}{\sigma^2(\nu_j)}\right] \nonumber
\end{eqnarray}

Non-linear least squares fitting can be unstable unless the algorithm is provided with a good set of initial conditions.  In particular, the $v_{\mathrm{LSR}}$ value of the fit is critical, so we use guesses based on the velocity centroid of the line (or of the center hyperfine component for ammonia lines).  The velocity dispersion is taken initially to be the intensity weighted second moment of the velocity around the centroid.  We adopt $\log_{10}(N / \mathrm{cm}^{-2})=14.5$, $T_{K}=12$~K, and $T_{\mathrm{ex}}=3$~K as our initial guesses. The fit is performed serially for each position in the map, starting from the pixel with the highest integrated intensity and working down in reverse order of integrated intensity.  If a given position has a neighboring pixel with a good fit, the parameters of the successful fit are used as initial guesses for the next optimization.

In this release, we do not attempt to fit multiple velocity components or non-Gaussian velocity profiles to our spectra.  Such approaches require human supervision under current implementations \citep{Henshaw_2016}.  Our fits produce good quality results for most of the studied regions\added{ ($>95\%$ of the detections in all regions release in DR1)}, but some lines of sight in \added{NGC 1333, L1688, and} Orion A show good evidence for needing improved models for their underlying velocity components.  

Recent work by \citet{Estalella_2016} presented the {\sc HfS} hyperfine fitting tool.  The assumptions embedded in our approach are essentially equivalent to those embedded in that tool, though \citet{Estalella_2016} used a relationship between $T_{\mathrm{rot}}$ and $T_{K}$ based on the collision coefficients in \citet{Maret_2009}, which differs from the \citet{Swift_2005} expression by $>10\%$ at $T>50\mbox{ K}$.  The optimization approach differs significantly, in that we use a gradient descent non-linear regression, whereas the {\sc HfS} code uses a genetic algorithm.  Our approach requires more accurate guesses for the initial conditions, but the strategy we have adopted leads to good convergence.  The uncertainty estimates both follow standard estimates based on $\chi^2$ fitting.

\begin{figure}
\begin{center}
\includegraphics[width=1\columnwidth]{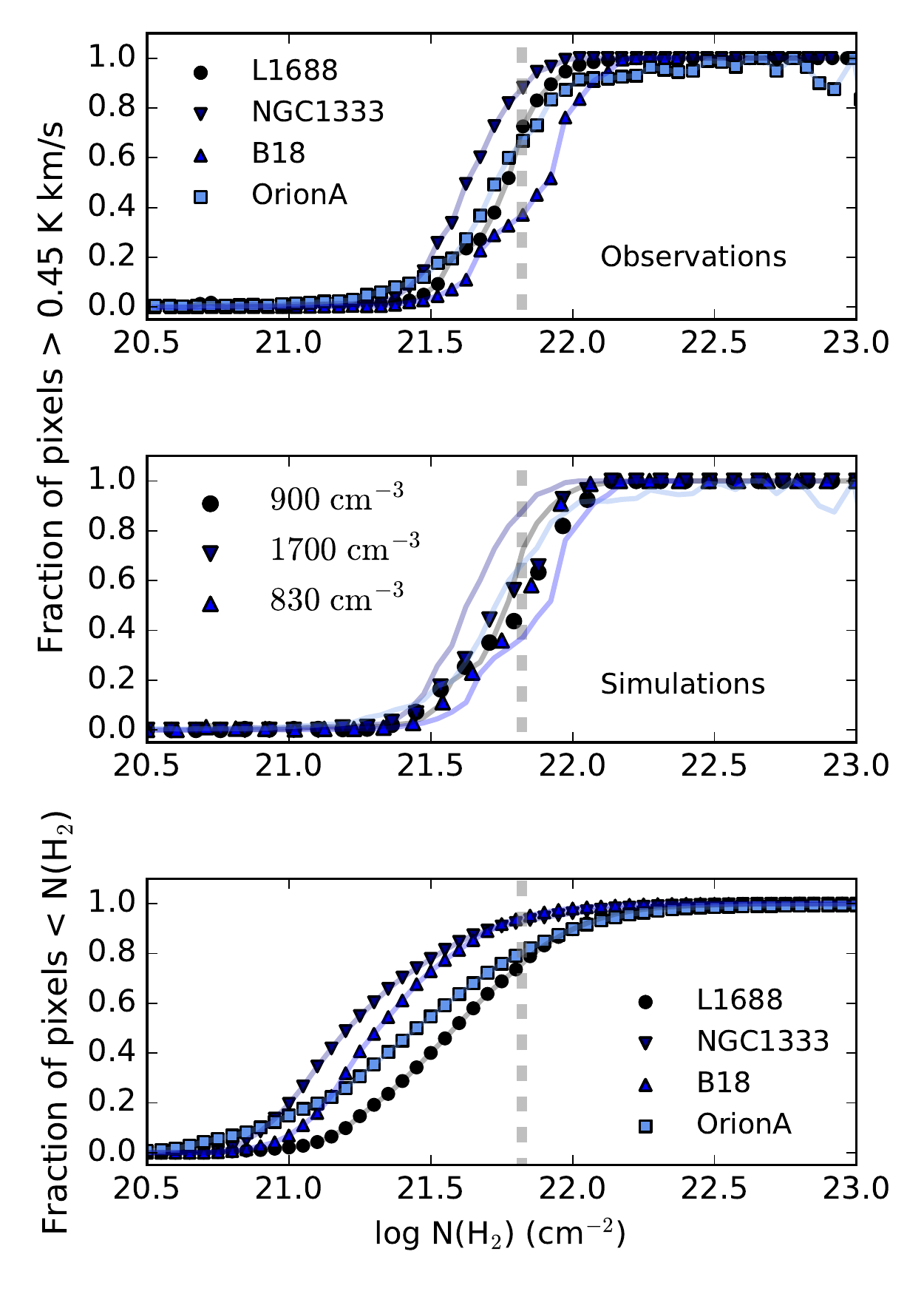}
\caption{\label{fig_nh3_h2_cumulative}
Top: The fraction of pixels with \amm\ (1,1) integrated intensity $\geq 0.45$\ K \ km\ s$^{-1}$ (approximately $3\sigma$) in each DR1 region. We show only the range in \nh\ where significant change in \amm\ detections are seen; the largest $\mathrm{log}$ \nh\ values reach $\sim 26$ in Orion A. In all panels, the vertical dashed grey line shows $A_\mathrm{V} \simeq 7$. Middle: Fraction of \amm\ pixels as a function of \nh\ for the GAS data (lines; colors as at top) and a synthetic observation (markers) based on astrochemical modeling from \citet{Offner_2014}. The simulated values are obtained from maps with average number densities of $n_{\rm H_2}=900$\ \cc, 1700\ \cc, and 830\ cm$^{-3}$. The regions have projected sizes of 2~pc, 0.5pc,  and 1~pc, respectively. The synthetic observation assumes a circular beam of 32\arcsec\ and a distance of 250\ pc with a $\Delta v=0.08$\ \kms\ velocity channel width. Bottom: The cumulative distribution function of $N(\mathrm{H}_2)$ within the observed footprints. 
}
\end{center}
\end{figure}

\begin{figure}
\begin{center}
\includegraphics[width=1\columnwidth]{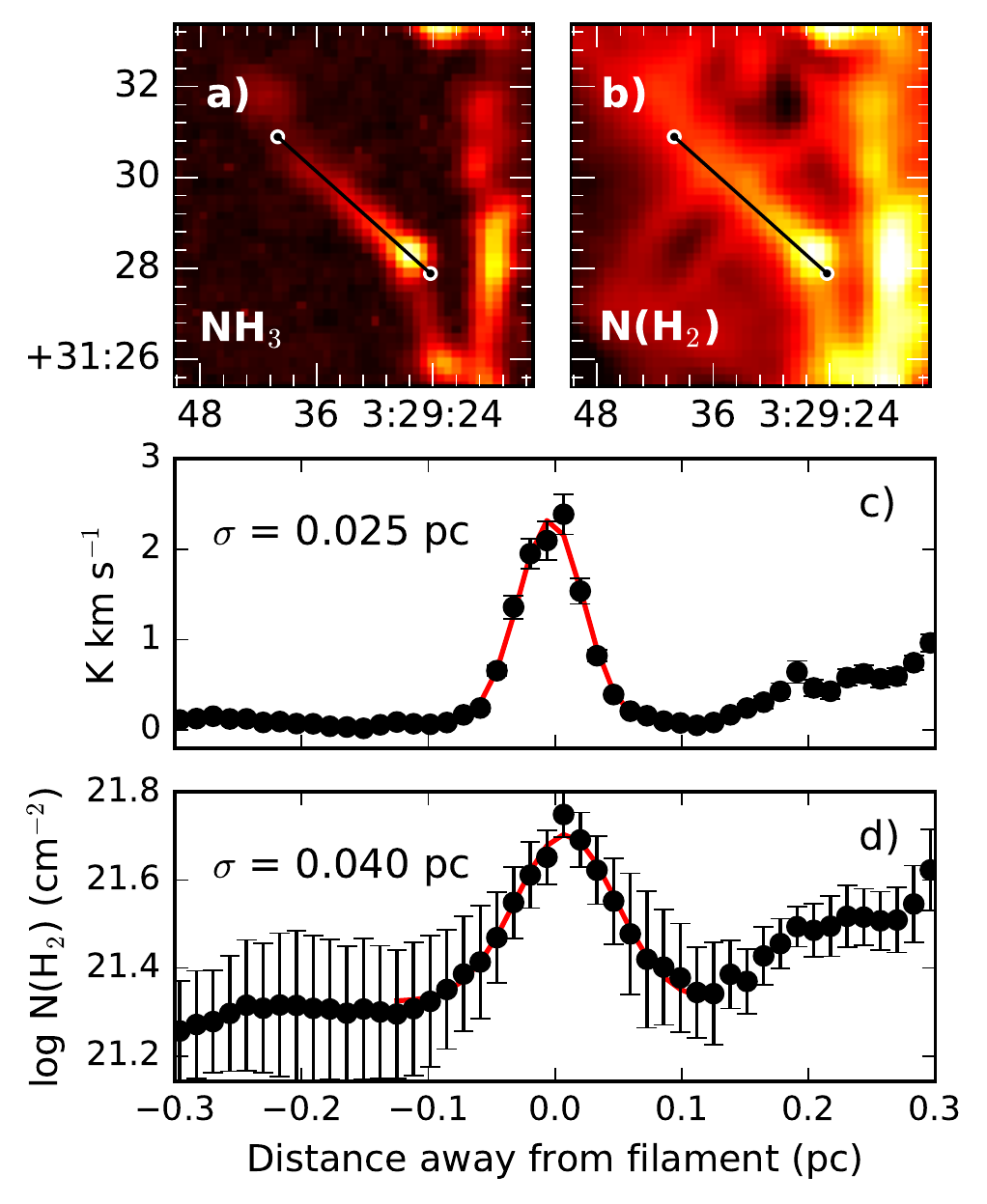}
\caption{\label{fig-ngc1333-filament}
a) \amm\ (1,1) integrated intensity toward a filament in NGC 1333. b) \nh\ over the same region as in a). c) The radial profile of the filament in \amm\ (1,1) integrated intensity, averaged along the straight black line identified in a) and b). d) The radial profile of the filament in \nh\ along the same line. }
\end{center}
\end{figure}

\subsection{\vlsr-tracked integrated intensity maps}
When calculating the integrated intensity map of a specific line, it is common practice to define a fixed velocity range that includes the emission of all hyperfine components for a single region. The maps presented here, however, cover large areas of individual star-forming clouds. In these regions, large velocity gradients make a fixed range sub-optimal.

To create the \amm\ (1,1) and (2,2) integrated intensity maps, we therefore first produce a model emission cube given the best fit models over the entire map. We set a threshold value $> 0.0125$\ K (i.e., slightly greater than machine precision) to identify the spectral channels with model line emission. We then select all the voxels (3D ``pixels") with brightness greater than the threshold to calculate the integrated intensity map. In regions where no line fit was found, we calculate the integrated intensity within a spectral window defined by the mean cloud $v_\mathrm{LSR}$ and $\sigma_v$. We calculate the uncertainty at each pixel based on the channel range thus identified.

This method for calculating the integrated intensity while following the velocity gradients in a star-forming region depends on having a good model, which is usually the case for \amm\ emission in the clouds presented here. There are locations, however, where the emission is not perfectly fitted with a single component. In those cases, this method may not include some of the flux in the final moment maps. A detailed study of the gas kinematics will be done in future papers focusing on individual regions.

Since the spatial and kinematic distribution of \amm\ (3,3) and the carbon-chain molecules do not typically follow that of \amm\ (1,1) and (2,2), and are significantly less extended, where detected we use a single velocity range to produce integrated intensity maps of these transitions.

\subsection{Data masking}
\label{sec:masks}

For all regions, \amm\ (1,1) and (2,2) lines were fit where the \amm\ (1,1) line had a $\mathrm{SNR} > 3$ through the property map analysis described above. Here $\mathrm{SNR} =  T_\mathrm{peak}/\mathrm{rms}$, where $T_\mathrm{peak}$ is the peak brightness temperature in the \amm\ (1,1) cube at a given pixel. At this SNR, however, not all line parameters could be fit well over all pixels. We consequently mask the output parameter maps to remove poor fits to the \amm\ lines. Since \vlsr, \sigv, and \tex\ require good fits to the \amm\ (1,1) line only, we mask pixels with a $\mathrm{SNR} < 3$ in \amm\ (1,1) integrated intensity. The parameters \tkin\ and $N$ require detections of the \amm\ (2,2) line, and for these property maps we additionally mask any pixels with a $\mathrm{SNR} < 3$ in the peak \amm\ (2,2) line intensity. Finally, we apply a mask that removes any remaining isolated pixels that are unconnected with larger-scale features and are generally noisy datapoints. Additional masking based on the returned uncertainty in the fit parameters may be done for detailed analysis in future papers. 

\begin{figure*}
\begin{center}
\includegraphics[width=0.9\textwidth]{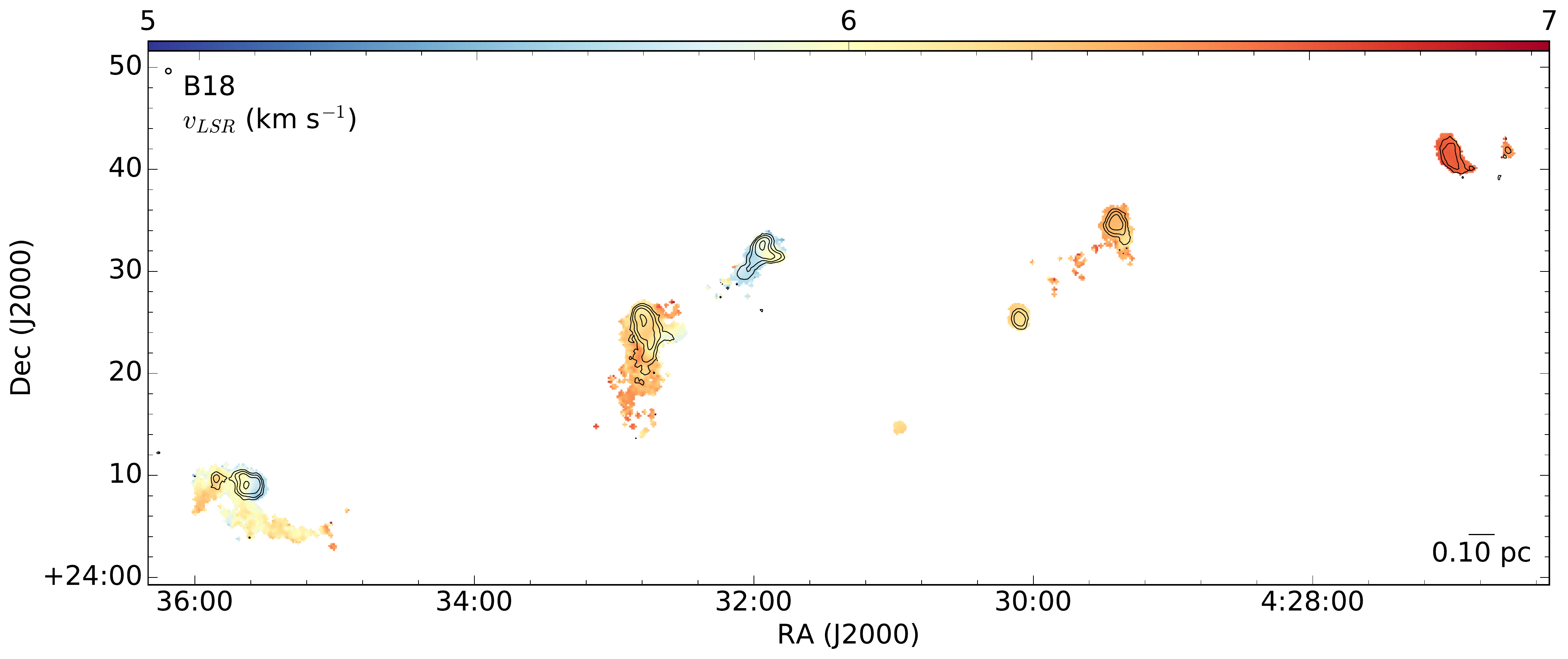}
\includegraphics[width=0.9\textwidth]{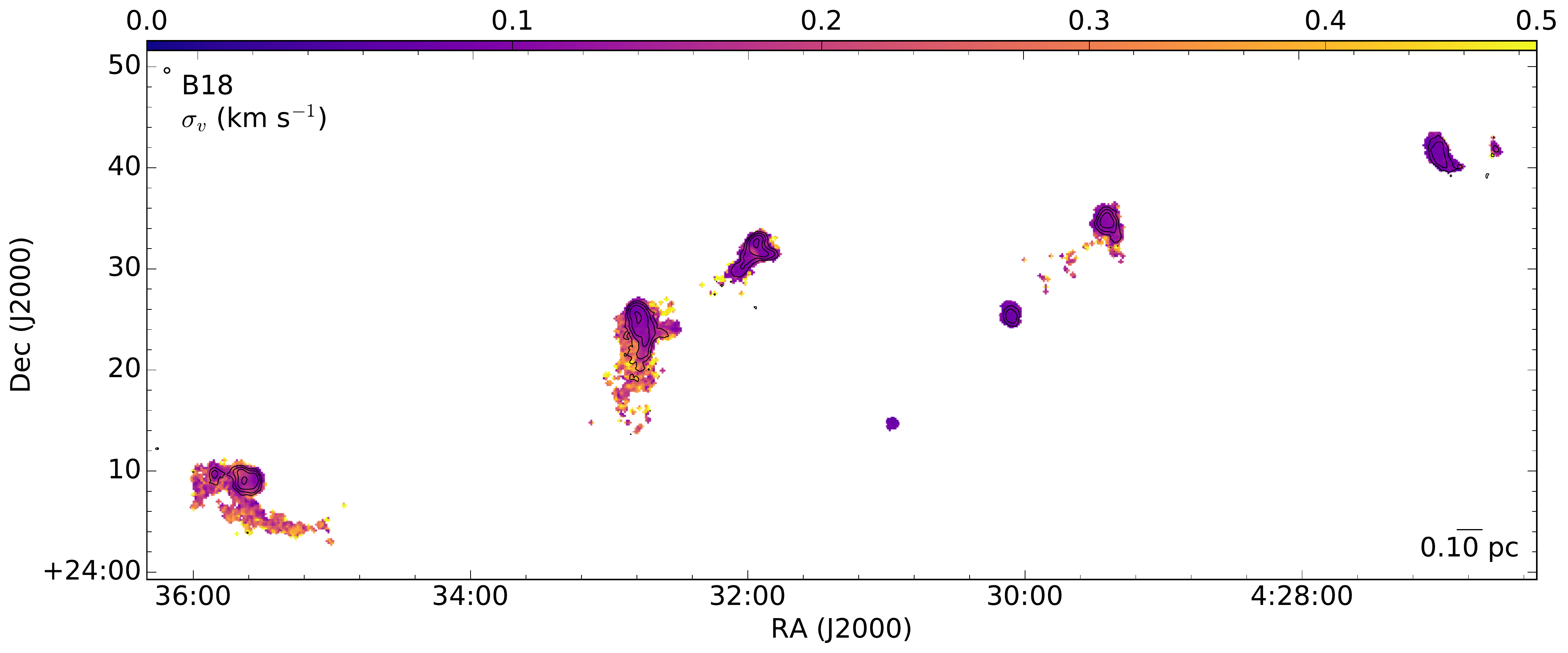}
\caption{\label{fig_b18_vlsr} 
Top: \vlsr\ (\kms) in B18. Contours are \amm\ (1,1) integrated intensity as in Figure \ref{fig-b18-NH3-11_TdV}. %
Bottom: $\sigma_v$ (\kms) in B18, with the color scale chosen to match the spread in $\sigma_v$ across all DR1 regions.  
}
\end{center}
\end{figure*}

\section{Results}
\label{sec:results}

\subsection{\amm\ moment maps}
\label{sec:mom0}

In Figures \ref{fig-b18-NH3-11_TdV}, \ref{fig-ngc1333-NH3-11_TdV}, \ref{fig-l1688-NH3-11_TdV}, and \ref{fig-OrionA-NH3-11_TdV}, we show the \amm\ (1,1) integrated intensity maps of B18, L1688, NGC 1333, and Orion A (North). We further overlay the \amm\ (1,1) integrated intensity contours on maps of the H$_2$ column density, \nh, in each region, derived from spectral energy distribution (SED) modeling of continuum emission from dust observed with the \textit{Herschel Space Observatory} (A. Singh et al., in preparation). The SED modeling is described in more detail in \S\ \ref{sec:threshold}. The locations of Class 0/I and flat spectrum protostars are highlighted, based on infrared analysis of Taurus \citep{Rebull_2010}, Perseus and Ophiuchus \citep{Dunham_2015}, and Orion \citep{Megeath_2012}.  

In general, the \amm\ (1,1) integrated intensity emission follows closely the H$_2$ column density as traced by dust continuum emission. The GAS data highlight individual cores and filamentary structure, as expected, but at the achieved sensitivity level also reveal low-SNR emission that extends between the higher column density structures, and follows, by visual inspection, the lower H$_2$ column density material. As an example, Figure \ref{fig-b18-NH3-11_TdV} shows how \amm\ (1,1) integrated intensity peaks highlight the dense cores in Taurus B18, while extended emission matches the shape of the surrounding material seen in \nh. Toward L1688, the central, high column density cores are surrounded by material at lower \nh\ values that fade away toward the northeast in Figure \ref{fig-l1688-NH3-11_TdV}. Low SNR \amm\ (1,1) emission is similarly more extended in this direction. Extended, low SNR \amm\ emission can also be seen toward the more distant NGC 1333 and Orion A regions. 

We show in Appendix \ref{sec:other_nh3_moment_maps} the integrated intensity maps of \amm\ (2,2) and \amm\ (3,3) (where detected) toward the DR1 regions. 

In appearance, the \amm\ (2,2) integrated intensity maps match closely the \amm\ (1,1) maps, but with lower SNR.

In NGC 1333, L1688, and Orion A, \amm\ (3,3) emission is detected, indicative of warmer dense gas present in these regions. In NGC 1333, the \amm\ (3,3) emission is compact and centered on two well-known sources, the bright Class 0/I object SVS 13 and Herbig-Haro object HH12. In L1688, the \amm\ (3,3) emission is more diffuse, overlapping the \amm\ (1,1) contours in the north-west but primarily visible on the western edge. This region is an interface between the cold, dense gas of the Ophiuchus A core and an edge-on photodissociation region associated with the B2V star HD 147889 \citep{Loren_1986,Liseau_1999}. The \amm\ (3,3) emission in Orion A peaks at the BN/KL region and traces the associated filamentary structure extending north. The \amm\ (3,3) emission is also prominent toward the Orion bar, highlighting this feature significantly better than \amm\ (1,1).

\begin{figure*}
\begin{center}
\includegraphics[width=0.9\columnwidth]{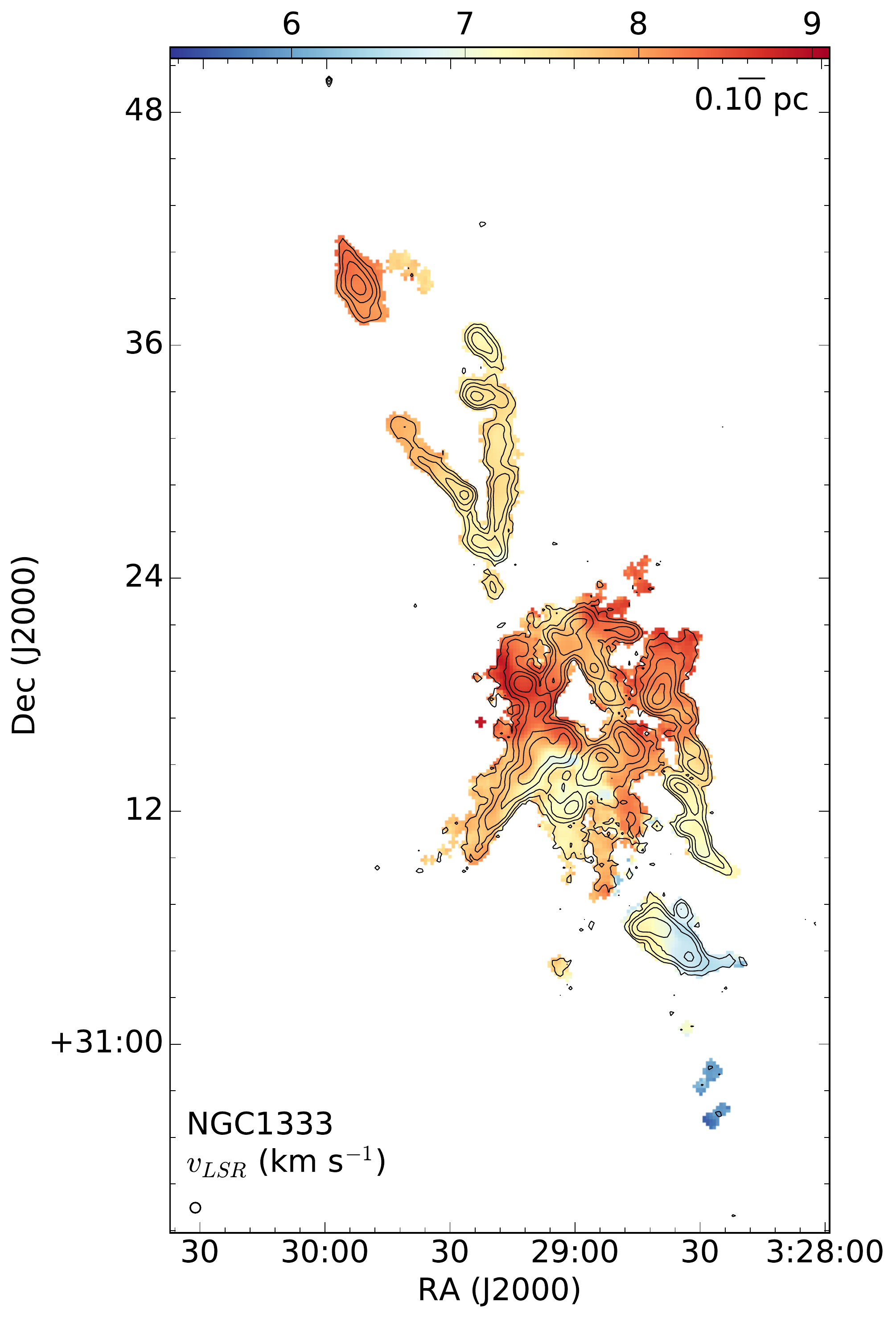}
\includegraphics[width=0.9\columnwidth]{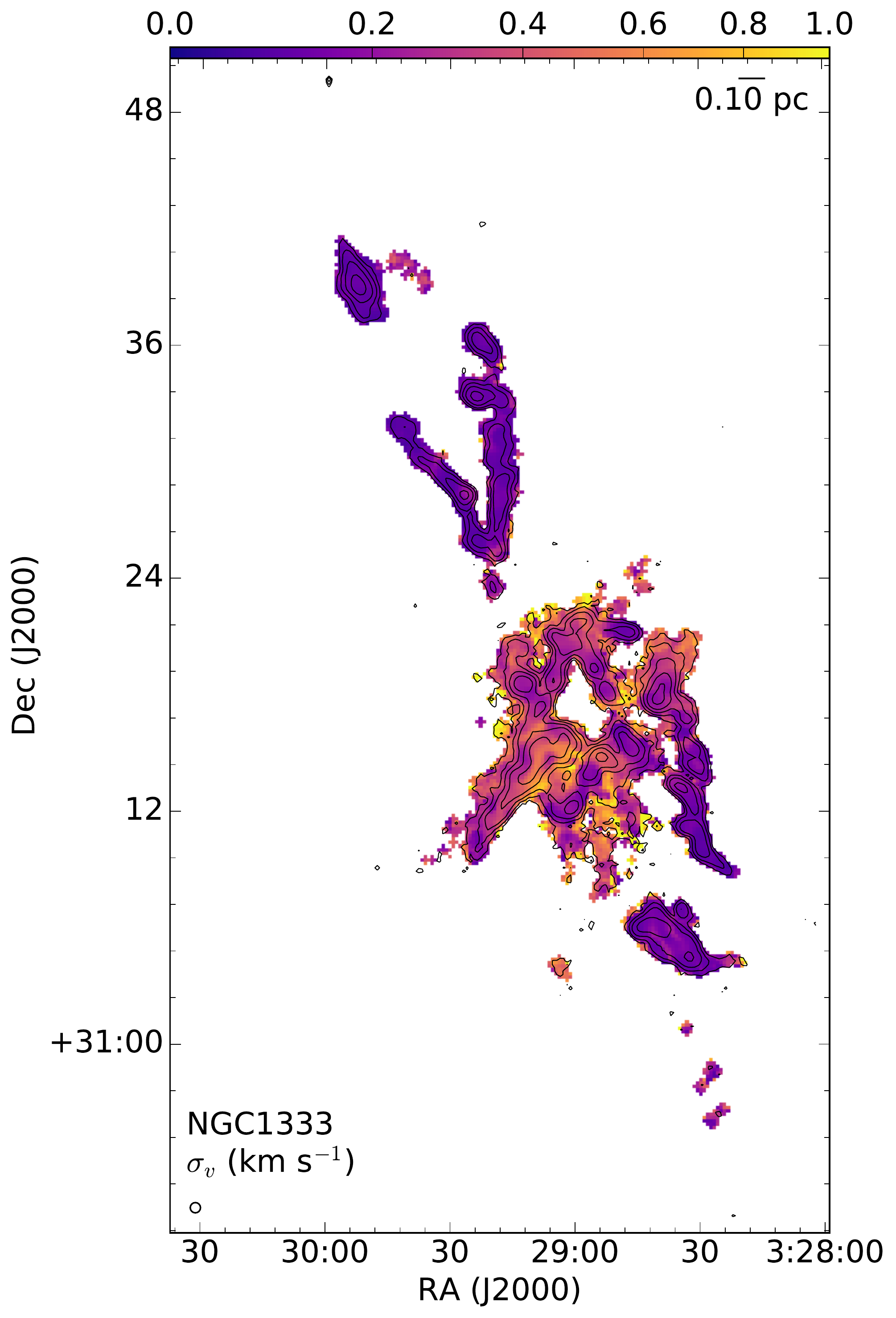}
\caption{\label{fig_ngc1333_vlsr} 
Like Figure \ref{fig_b18_vlsr} but for NGC 1333. 
}
\end{center}
\end{figure*}

\subsection{Detections of other lines}
\label{sec:other_lines}

In the previous sections, we have focused primarily on emission from the \amm\  transitions toward the DR1 regions. We additionally mapped the rotational transitions of several carbon chains, which are listed in Table \ref{tab:spectral_setup}. Not all lines were detected in all regions, and the HC$_7$N transitions were not detected above the rms noise or in an integrated intensity map in any region. We show in Appendix \ref{sec:other_moment_maps} integrated intensity maps of the additional detected lines in each region. 
In Figure \ref{fig_all_spectra}, we show the spectrum of each observed line averaged over each DR1 region after masking of the edge pixels as described in \S\ \ref{sec:imaging}. 

Toward B18, both HC$_5$N 9-8  and C$_2$S $2_1-1_0$  are detected. For both molecules, the distribution of emission is significantly offset from \amm\ (1,1), generally peaking away from the \amm\ emission peaks. While HC$_5$N is seen interior to some \amm\ (1,1) contours, C$_2$S, where seen, appears to surround the \amm\ emission. Similar distributions of \amm\ and C$_2$S have been identified toward the L1495-B218 filaments in Taurus (Seo et al., in preparation). This offset between emission from \amm\ and carbon-chain molecules is common in nearby star-forming regions and is generally explained by the depletion of C-bearing molecules onto dust grains at high densities, while \amm\ remains in the gas phase. Asymmetric offsets like those seen in B18 might be indicative of the contraction of dense gas \citep{Tafalla_2004} or accretion of lower density material onto the dense cores \citep{Friesen_2013}. 

HC$_5$N is also detected toward NGC 1333 in a single compact region (slightly greater in extent than the beam) located to the west of the central structure traced by \amm\ (1,1). While C$_2$S is not visible in the integrated intensity map of NGC 1333 (not shown), the averaged C$_2$S spectrum in Figure \ref{fig_all_spectra} shows a significant detection. Similarly, C$_2$S is clearly detected toward L1688 in the averaged spectrum, while the integrated intensity map shows only a faint, SNR $\sim 3-4$ feature near the map center. In future analysis we will investigate extended, low-level emission of \amm\ and the other observed lines through spectral stacking techniques. 

\subsection{Relationship of \amm\ emission to \nh}
\label{sec:threshold}

\subsubsection{Column density of H$_2$}
\label{subsec:nh2}

The H$_2$ column density, \nh, used here and in the maps above was determined
in two steps (A. Singh et al., in preparation).  First, the optical
depth of dust was obtained by fitting a modified blackbody to the
spectral energy distribution (SED) of the thermal continuum emission
from cold dust using \textit{Herschel} maps at 160\ \micron, 250\ \micron,
350\ \micron, and 500\ \micron.  Second, this is converted to gas column
density by a simple scale factor.  An additional scale factor can be used
to convert to \av.

To implement the first step, we note that for dust temperature $T_d$ the
intensity of thermal dust emission is
\begin{equation}
I_\nu = (1 - e^{-\tau_\nu}) B_\nu (T_d) \,
\end{equation}
and we adopt $\tau_\nu = \tau_{\nu_0} (\nu/\nu_0)^\beta$. The dust opacity
spectral index was taken to be $\beta = 1.62$, the mean value
found by Planck over the entire sky
\citep{planck2013-p06b}.
All maps were first convolved to the 36\farcs3 resolution of the
500\ \micron\ data and put on the same grid. 
Zero offsets were determined for each wavelength by comparison with Planck emission maps.  
Minor color corrections were applied during the SED fitting.
The SED parameters $\tau_{\nu_0}$ and $T_d$ were determined at each
pixel using $\chi$-squared minimization.

The required scale factor to convert to \nh\ can be seen via the relation
\begin{equation}
\tau_{\nu_0} = \kappa_{\nu_0} \mu m_\mathrm{H} \, N(\mathrm{H}_2) \,
\end{equation}
where $\mu = 2.8$, $m_\mathrm{H}$ is the mass of a Hydrogen atom, and
$\kappa_{\nu_0}$ is the opacity (incorporating a gas-to-dust mass ratio
of 100), which is taken to be 0.1~cm$^{2}$\ g$^{-1}$ at $\nu_0 =
1000$\ GHz \citep{Hildebrand_1983}.  The considerable systematic
uncertainty in the opacity propagates uniformly throughout the map into
a systematic uncertainty in the scale of \nh.

Ideally, one could scale the dust-based $\tau_{\nu_0}$ directly into
\av, which is also proportional to dust column density.  This scaling
too is uncertain \citep{planck2013-p06b}.  Here we simply use
$N(\mathrm{H}_2) = 9.4 \times 10^{20}$\ cm$^{-2}$ (\av/mag)), which is
based on data for the diffuse interstellar medium, not molecular clouds,
and so the two-figure precision given belies the actual uncertainty.

Regional variations in dust spectral properties could impact the values
determined through SED modeling (and thus the threshold for
\amm\ emission discussed below), although this is likely a small effect
compared to the uncertainty in the scaling: for example, a change in
$\beta$ of 0.05 produces a change in \nh\ of only $\sim 5$\ \%.


\subsubsection{\nh\ threshold for \amm\ emission}
\label{subsec:threshold}

Figure \ref{fig_nh3_h2_cumulative} (top) shows a smooth increase in the
fraction of pixels with \amm\ emission as \nh\ increases, with some
variation between regions in the rate of rise in the correlation, and
in the H$_2$ column density where \amm\ is found universally. The survey
initially targeted regions with extinctions $A_\mathrm{V} > 7$, based on previous
observations of \amm\ in nearby star-forming regions. That this was a
good choice is confirmed by the DR1 data, where above this extinction
value (highlighted by the gray dashed line in the Figure, assuming $N(\mathrm{H}_2) 
= 9.4 \times 10^{20}$\ cm$^{-2}$ (\av/mag)), $\gtrsim 60$\ \% of pixels have 
detectable \amm\ emission above the survey noise
levels in all DR1 regions, with the exception of B18, where the threshold 
appears to be slightly higher. Furthermore, the
survey footprints were rectangular, not following the high column
density contours precisely, and so lines of sight corresponding to lower
column density were surveyed too.

\begin{figure*}
\begin{center}
\includegraphics[width=0.7\textwidth]{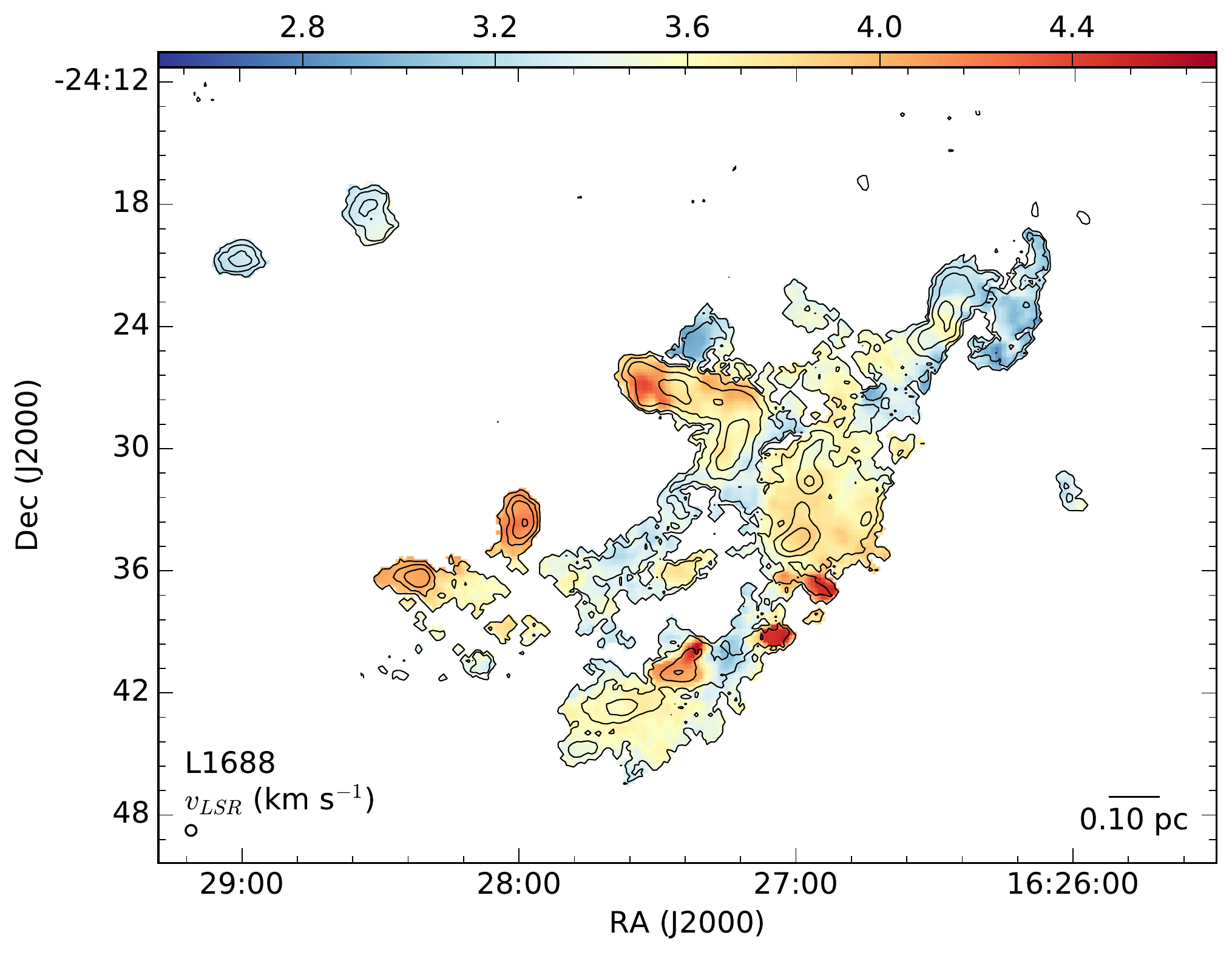}
\includegraphics[width=0.7\textwidth]{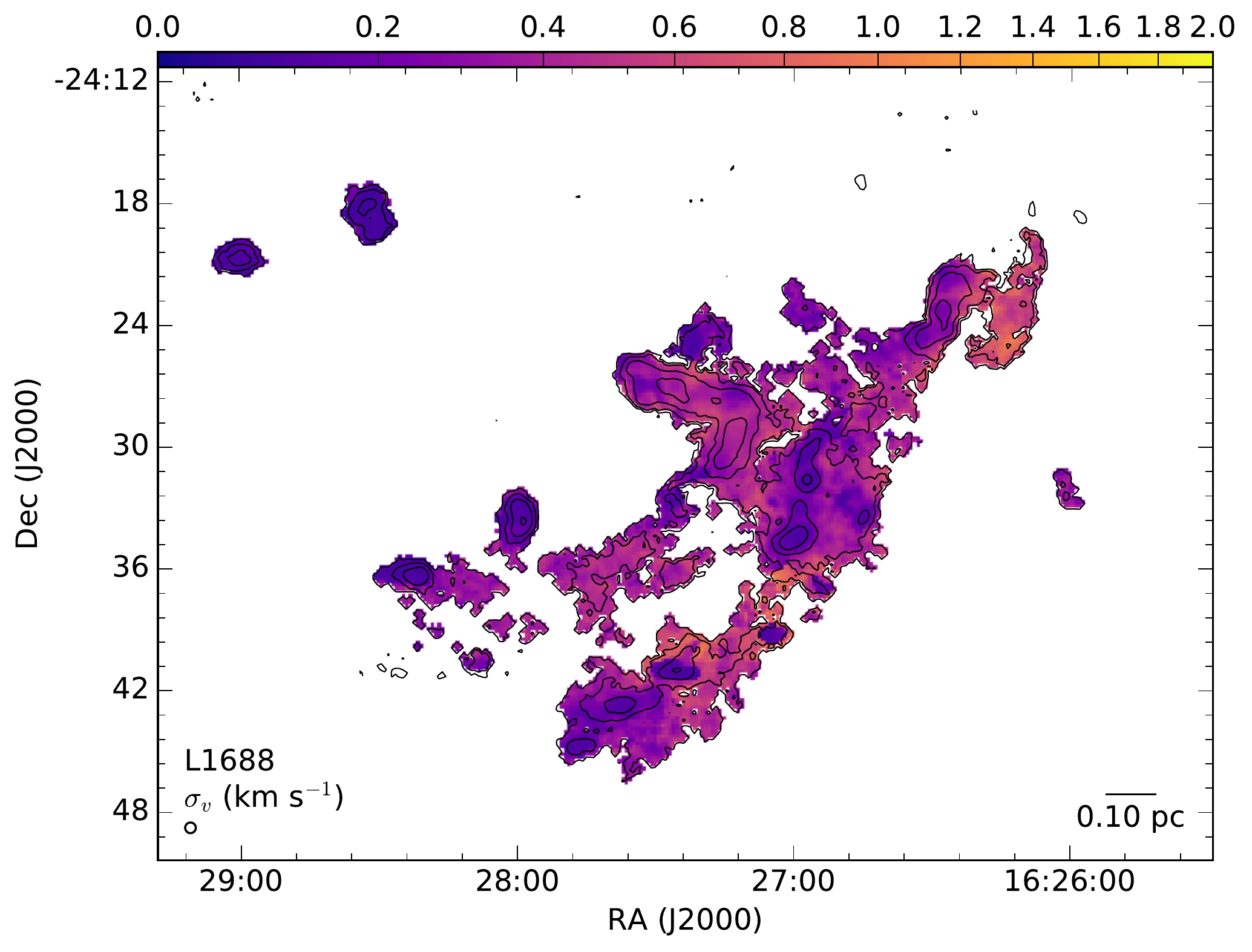}
\caption{\label{fig_l1688_vlsr} 
Like Figure \ref{fig_b18_vlsr} but for L1688. 
}
\end{center}
\end{figure*}

\begin{figure*}
\begin{center}
\includegraphics[width=0.8\columnwidth]{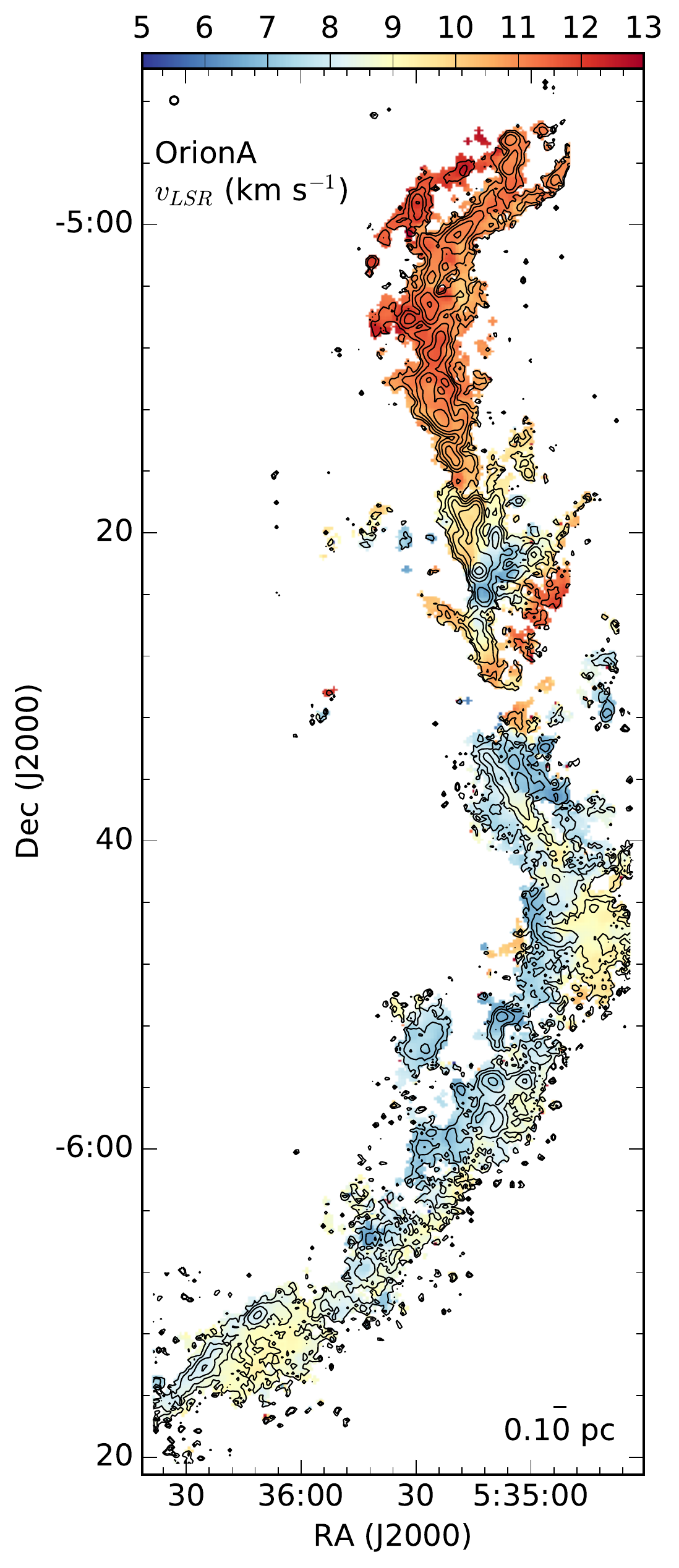}
\includegraphics[width=0.8\columnwidth]{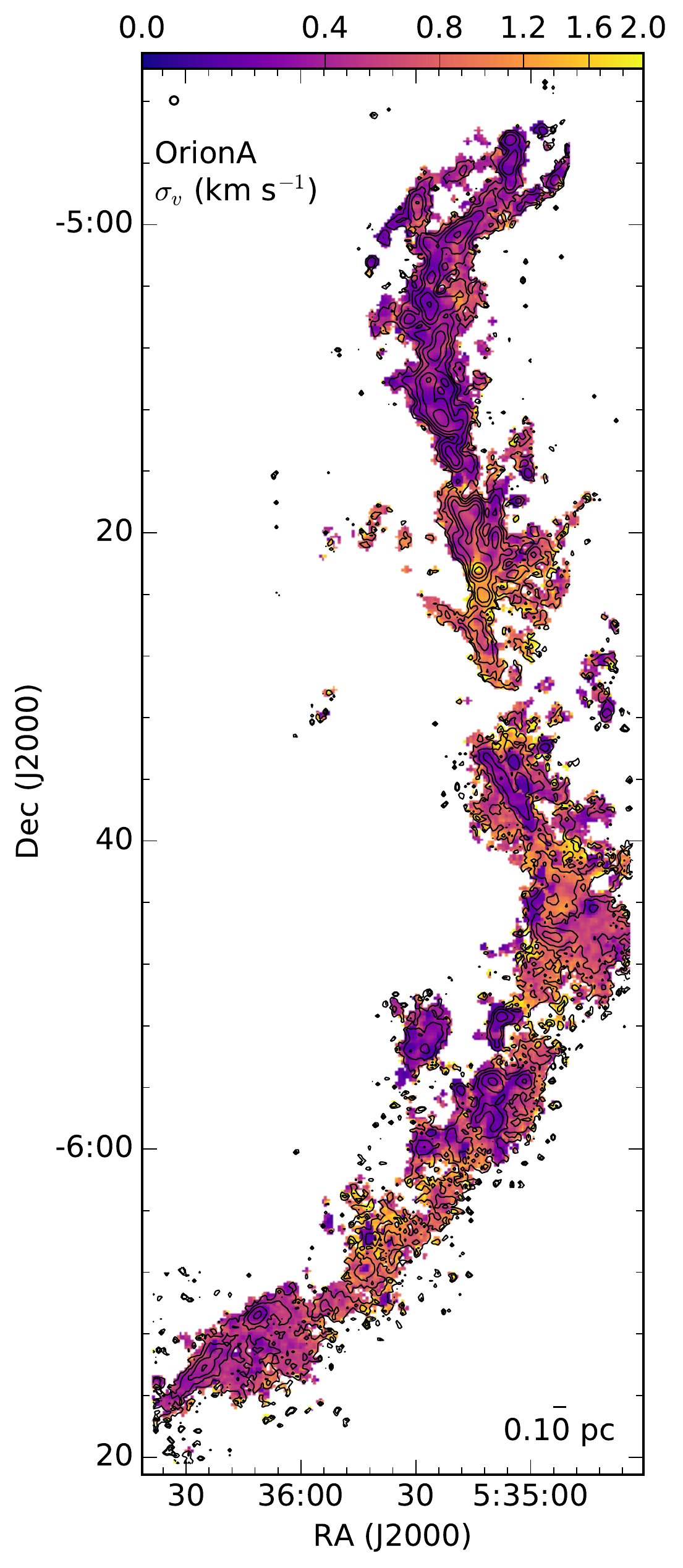}
\caption{\label{fig_OrionA_vlsr} 
Like Figure \ref{fig_b18_vlsr} but for Orion A (North). 
}
\end{center}
\end{figure*}


\subsubsection{Comparison to simulations}
\label{subsec:simulations}


We compared the GAS data to results calculated from a synthetic
observation of NH$_3$ emission produced by a 3D hydrodynamic simulation.
The simulated emission map is based on astrochemical modeling performed
by \citet{Offner_2014}. The abundances were computed by post-processing
the simulation with {\sc 3d-pdr}, a three-dimensional photodissociation
region code \citep{Bisbas_2012}, assuming the simulation is irradiated
with a uniform 1 Draine field. We next computed the emission using {\sc
  radmc-3d}\footnote{\url{http://www.ita.uni-heidelberg.de/~dullemond/software/radmc-3d}}.
Finally, to model instrumental affects we convolved the emission map with
a 32\arcsec\ beam for a distance of 250 pc and added Gaussian noise with
$\sigma_{\rm rms}=$ 0.15\ K\ \kms, a typical value from the GAS
\amm\ (1,1) integrated intensity maps. 

Figure \ref{fig_nh3_h2_cumulative} (middle)
shows the emission fraction for the total
simulation domain ($n_{\rm H_2}=900$ cm$^{-3}$) and for two sub-regions
($n_{\rm H_2}=1,700$ cm$^{-3}$ and $n_{\rm H_2}=830$ cm$^{-3}$). These
distributions agree well with the GAS distributions and demonstrate that
variation in the underlying gas densities can account for some of the
spread between regions. Differences between regions may also result from
variations in the local UV radiation field and mean column density,
which we will explore in future work. 

Lastly, Figure \ref{fig_nh3_h2_cumulative} (bottom) shows the cumulative
distribution of \nh\ in the observed footprints of the GAS DR1
regions. A more gradual increase is seen relative to the top panel,
where the presence of \amm\ rises steeply above $\mathrm{log}_{10}
N(\mathrm{H}_2) / [\mathrm{cm}^{-2}] \sim 21.5$. 

Figure \ref{fig_nh3_h2_cumulative} shows that observations of \amm\ inversion transitions become an
excellent indicator of where there is gas at higher column densities,
and thus probably higher volume volume densities, but miss material at
lower column density that can be detected via the thermal dust continuum
emission.

\begin{figure*}
\begin{center}
\includegraphics[width=0.9\textwidth]{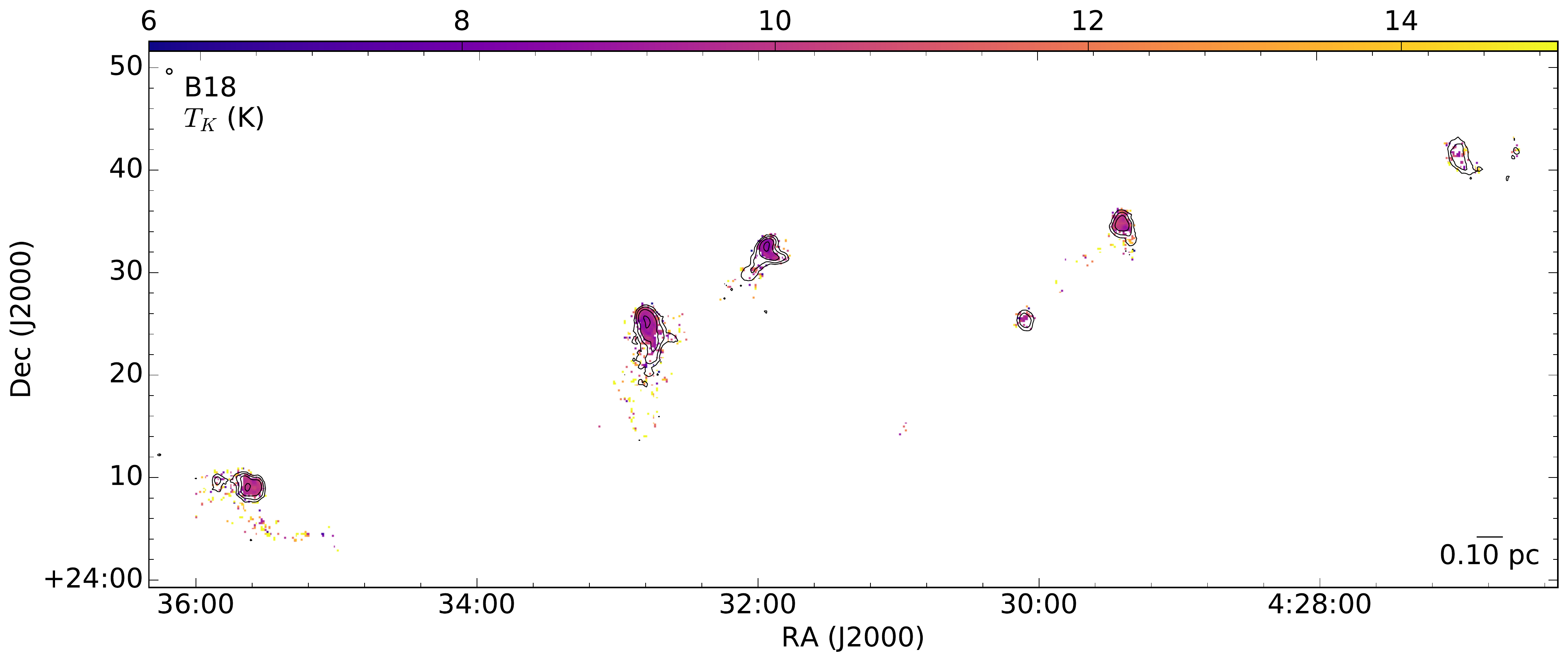}
\caption{\label{fig_b18_tkin} \tkin\ (K) in B18, with the color scale chosen to match the spread in \tkin\ across all DR1 regions.   Contours are \amm\ (1,1) integrated intensity as in Figure \ref{fig-b18-NH3-11_TdV}.%
}
\end{center}
\end{figure*}


\subsubsection{The appearance of filaments}
\label{subsec:filaments}

The ability of \amm\ to highlight higher density structures enables
analyses of structure sizes, masses and concentrations that are
complementary to similar analyses using dust continuum emission.  For
example, we noted previously that filaments are prevalent within
star-forming regions, with a typical filament width of $\sim 0.1$\ pc
suggested by profile-fitting of the mid- and far-infrared dust emission
\citep{Arzoumanian_2011}. The authors define the filament `width' as the 
FWHM of a Gaussian fit to the innermost section 
of the filament's radial profile. 
In Figure \ref{fig-ngc1333-filament}, we show
a filament in NGC 1333 in a) \amm\ (1,1) integrated intensity, and b)
\nh. In panels c) and d), we show the filament radial profile in \amm\ and \nh, respectively, 
averaged along the long axis shown by the straight line in panels a) and b). 
We then fit a Gaussian, plus constant offset, to the innermost 0.1\ pc of each radial profile, matching the
analysis of \citeauthor{Arzoumanian_2011}, with the results shown in red.
Similar to the authors' result, we find a filament FWHM of 0.086\ pc in \nh, 
but show that the filament is narrower in \amm\ emission by a factor $\sim 1.4$. 

Locating the range of \nh\ values from panel d) in the top panel
of Figure \ref{fig_nh3_h2_cumulative} shows that the \amm\ emission
does not measure the full column density, only the denser core.
Furthermore, in \nh\ the filament is embedded
within lower column density material and crossed by additional, narrow
features, while the \amm\ emission highlights the filament only. 

Future work using the GAS dataset will identify filamentary structures
systematically using algorithms such as \verb-filfinder-
\citep{Koch_2015} and \verb-DisPerSE- \citep{Sousbie_2011}. This will
probe the properties of filamentary structures at different spatial
scales and determine what dominates the kinematics within the
filamentary structures at each scale, thus providing insight into their
origin.


\begin{figure}
\begin{center}
\includegraphics[width=0.9\columnwidth]{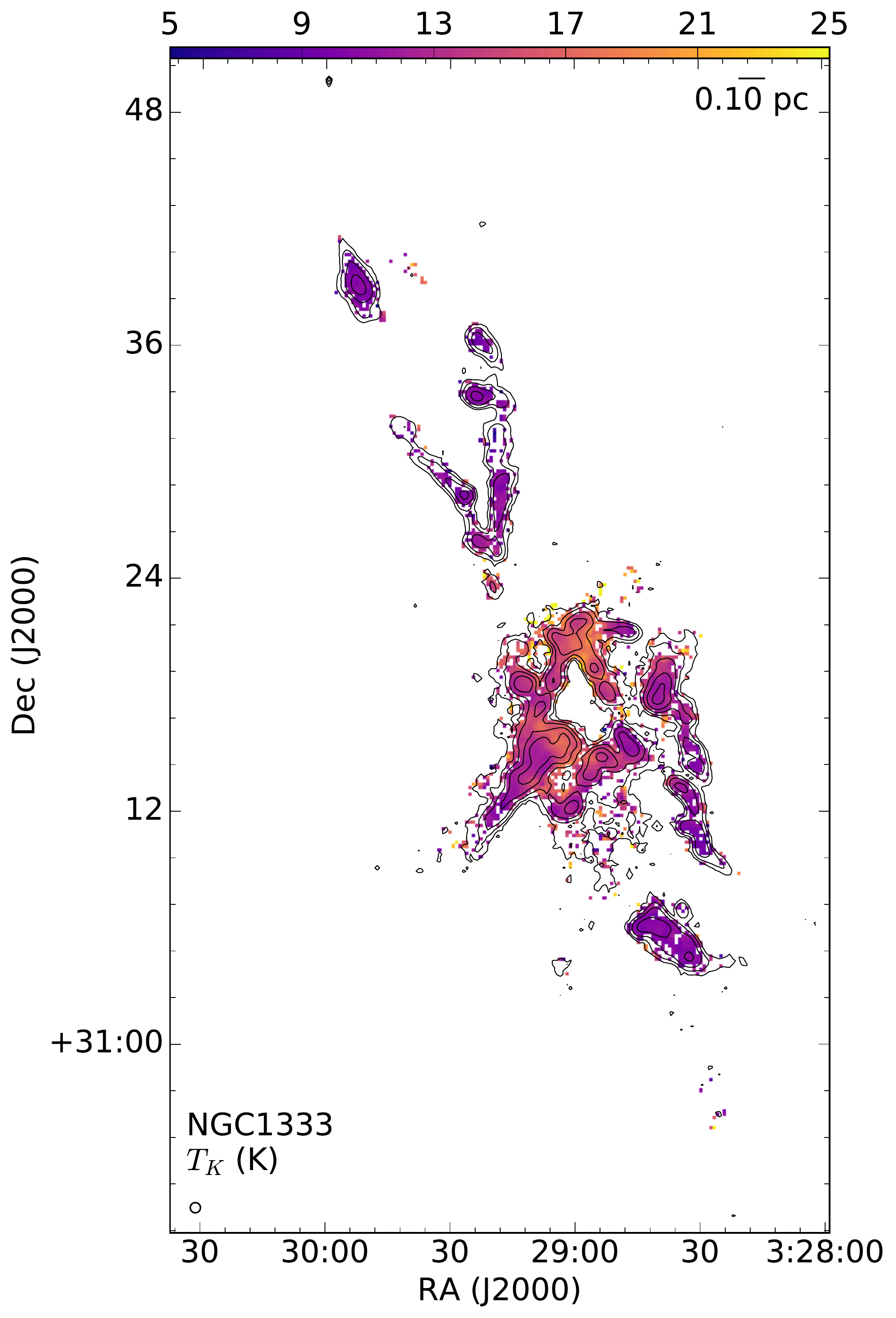}
\caption{\label{fig_ngc1333_tkin} 
Like Figure \ref{fig_b18_tkin} but for NGC 1333. 
}
\end{center}
\end{figure}

\subsection{Property Maps}

\begin{figure*}
\begin{center}
\includegraphics[width=0.7\textwidth]{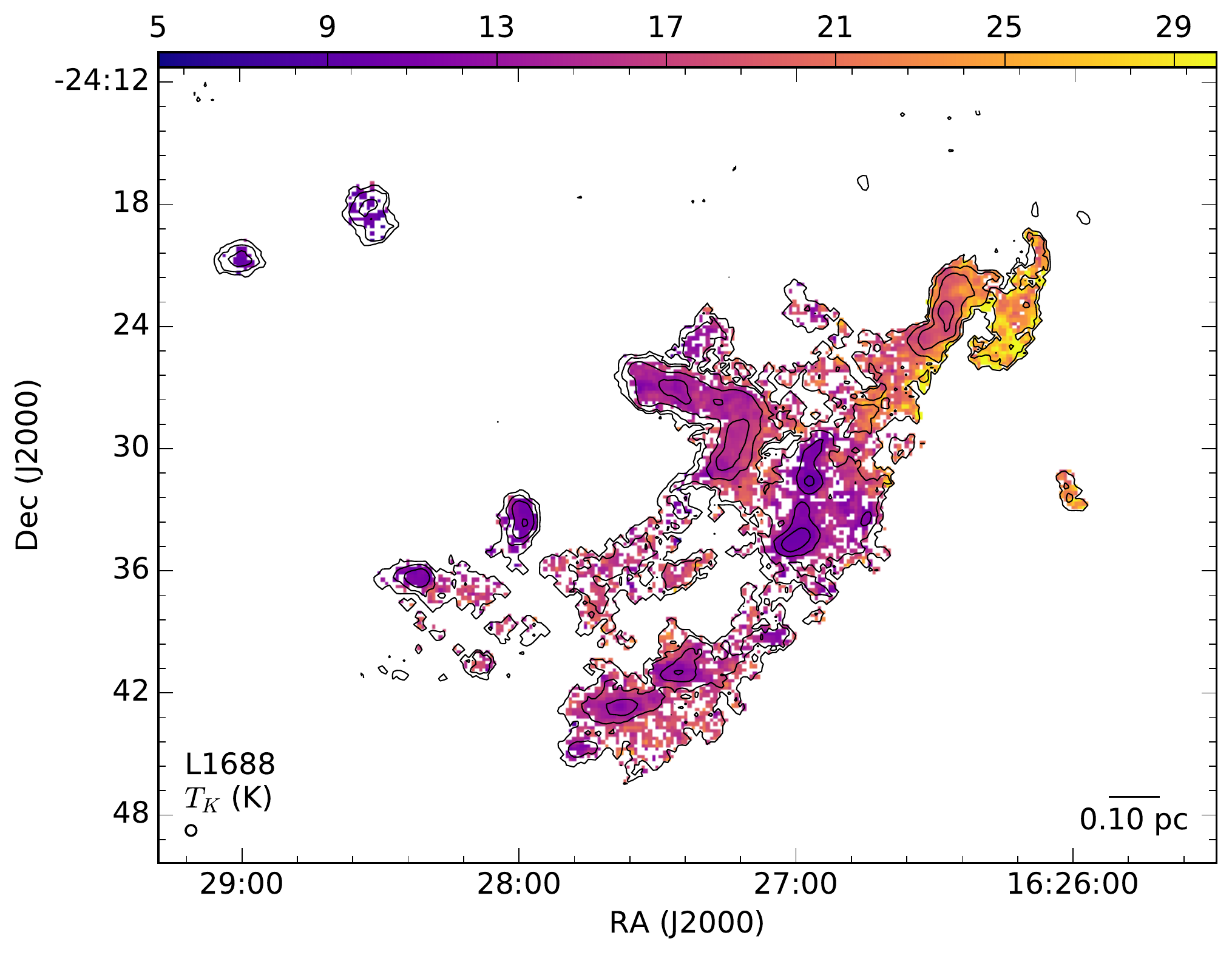}
\caption{\label{fig_l1688_tkin} 
Like Figure \ref{fig_b18_tkin} but for L1688. 
}
\end{center}
\end{figure*}

We present the property maps resulting from the \amm\ line fitting described in \S\  \ref{sec:linefit}. In Figures \ref{fig_b18_vlsr}, \ref{fig_ngc1333_vlsr}, \ref{fig_l1688_vlsr}, and \ref{fig_OrionA_vlsr}, we show maps of \vlsr\ and \sigv\ for each of the DR1 regions. The gas temperature \tkin\ is presented for all regions in Figures \ref{fig_b18_tkin}, \ref{fig_ngc1333_tkin}, \ref{fig_l1688_tkin}, and \ref{fig_OrionA_tkin}. Maps of the \amm\ column density, \namm, are shown for all regions in Appendix \ref{sec:n_nh3_maps}.  

While detailed analyses of the dense gas properties traced by \amm\ in the Gould Belt will be presented in future papers, we discuss here some general results and trends in the data. 

Figures \ref{fig_b18_vlsr} - \ref{fig_OrionA_vlsr} show significant variations in the kinematic properties traced by \amm\ between the DR1 regions. In both B18 and L1688, gas velocities are concentrated around the mean cloud value with only a small spread. In B18, at the survey sensitivity, we detect \amm\ toward only a small fraction of the extended gas surrounding the \amm\ peaks, and the \vlsr\ values shown represent mainly the \vlsr\ of the densest \amm\ structures in the region. In the other DR1 regions, we detect more of the extended \amm\ emission. The \vlsr\ map of L1688 in Figure \ref{fig_l1688_vlsr} shows that distinct variations in \vlsr\ are seen between some of the \amm\ structures and the surrounding dense gas, but the velocity changes are small ($\lesssim 1$\ \kms). Toward NGC 1333, a broader distribution in \vlsr\ is  visible in Figure \ref{fig_ngc1333_vlsr}. Unlike the variation in \vlsr\ between compact and extended emission in L1688, gradients in \vlsr\ are seen along the filamentary structures in NGC 1333. A large gradient in \vlsr\ is also clearly visible extending along the Orion A filament from north to south, with a smaller velocity gradient perpendicular to the filament in the south. 

\begin{figure*}
\gridline{\fig{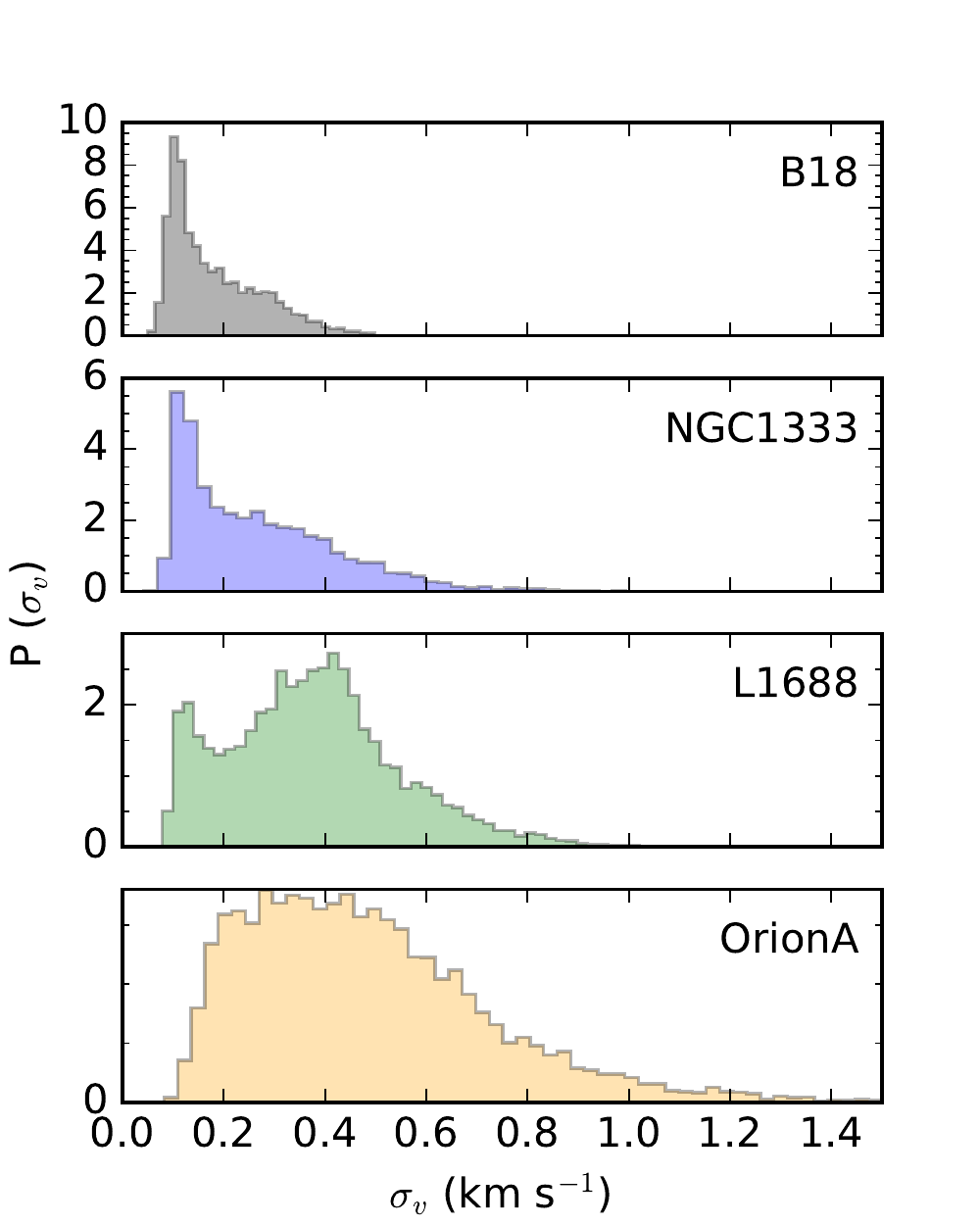}{0.9\columnwidth}{(a)}
          \fig{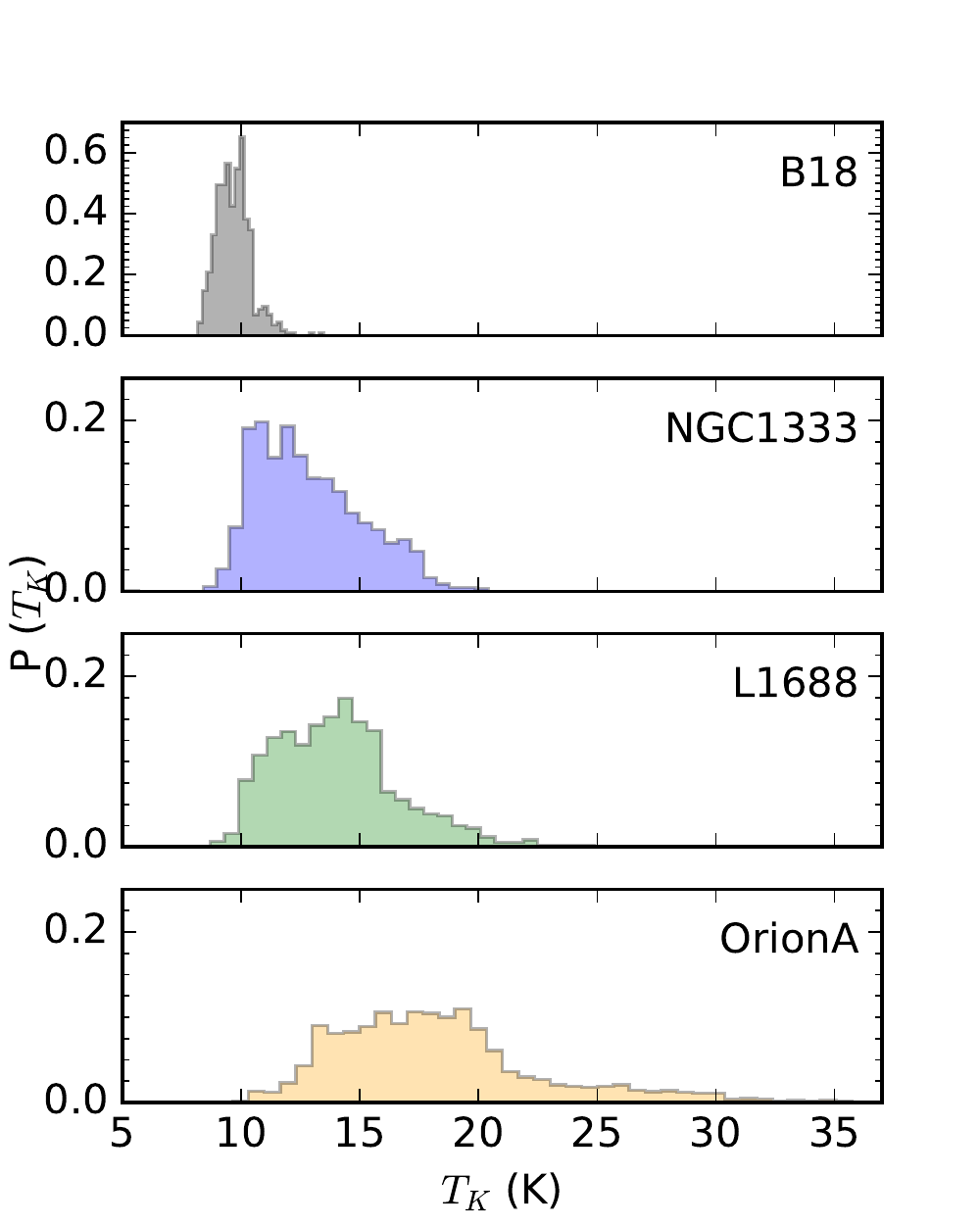}{0.9\columnwidth}{(b)}}
\gridline{\fig{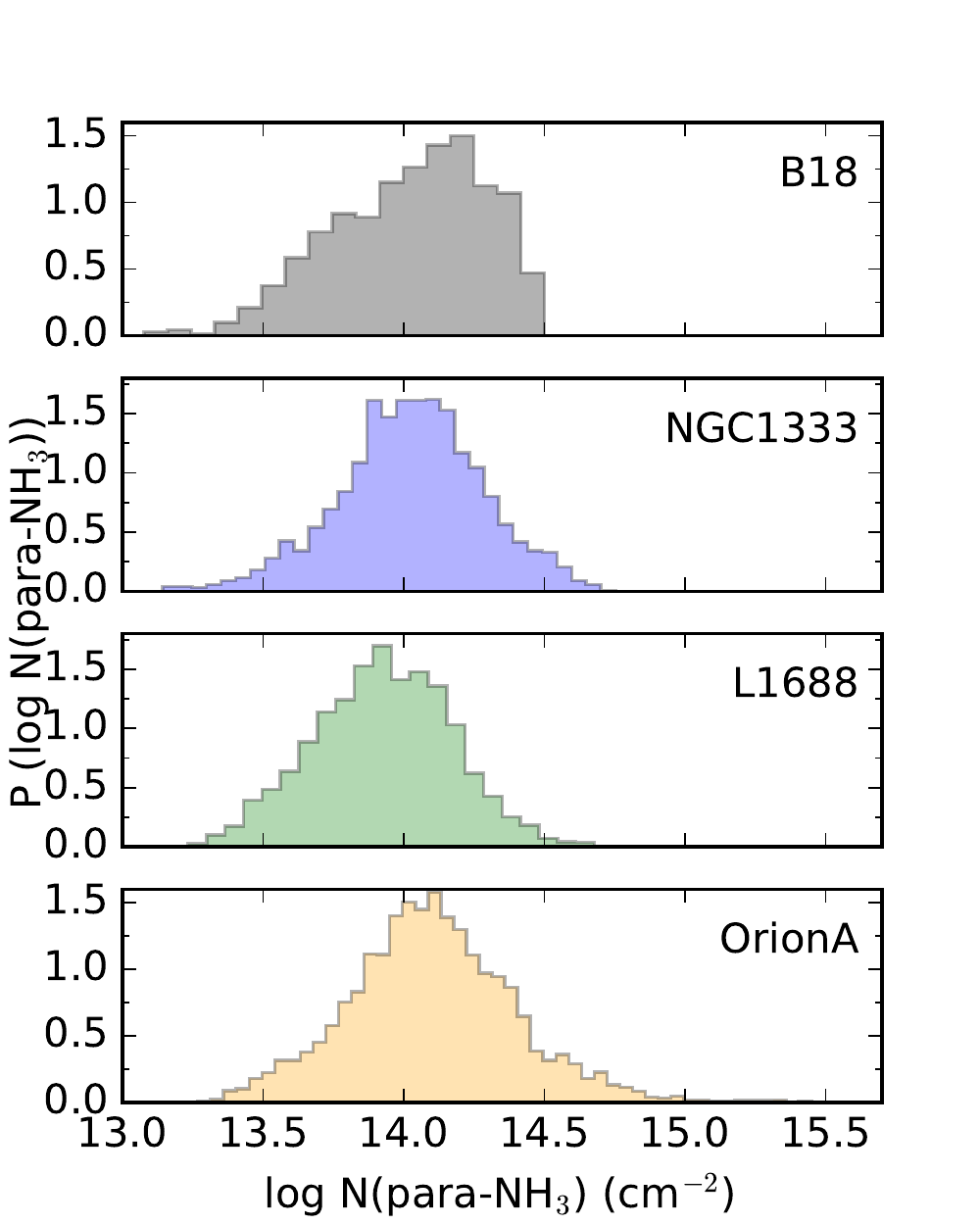}{0.9\columnwidth}{(c)}
          \fig{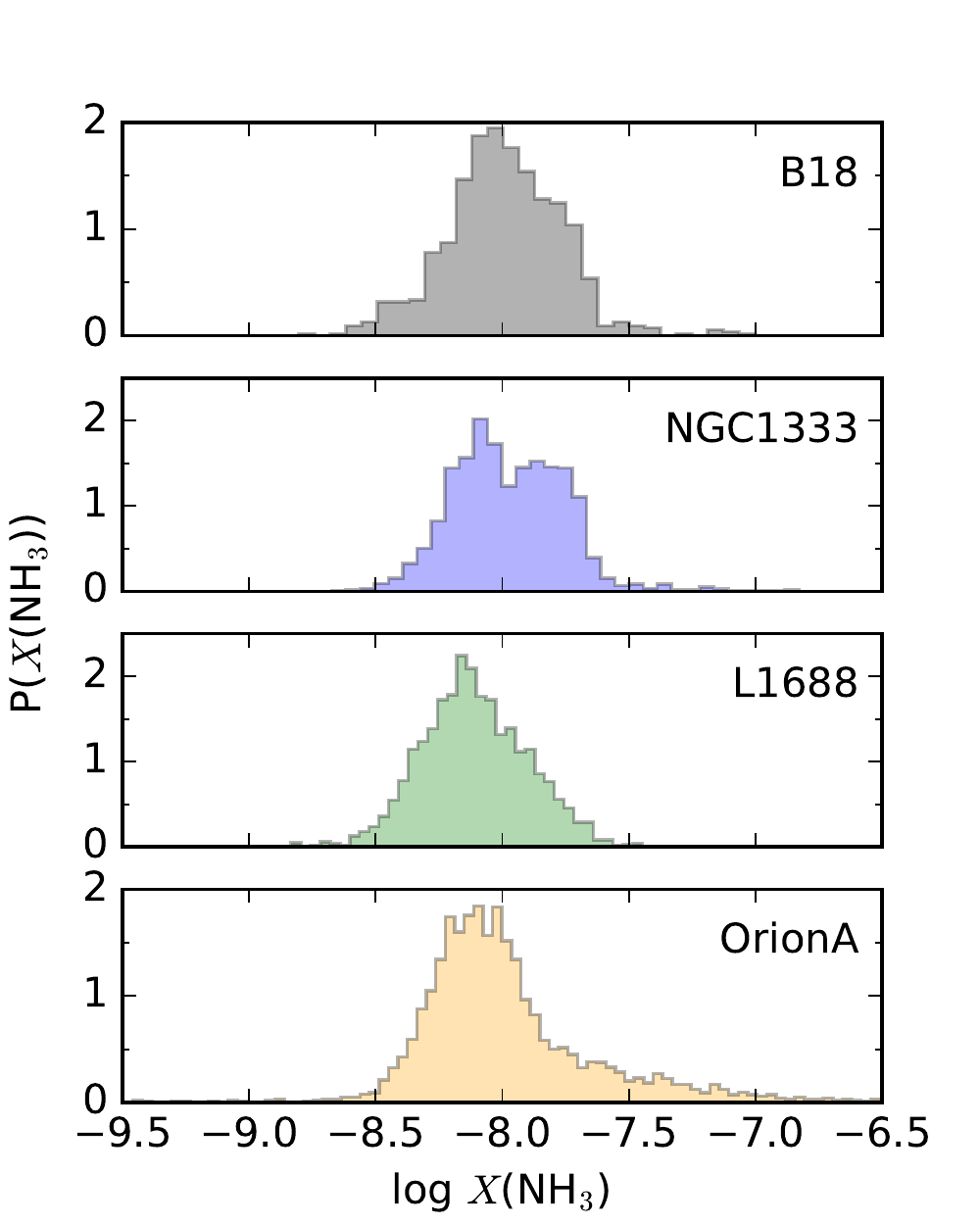}{0.9\columnwidth}{(d)}}
\caption{(a) Histograms of \sigv\ for each DR1 region. Pixels where the \amm\ \sigv\ is fit with an uncertainty $\leq 0.1$\ \kms\ are included. In all panels we show the value of the probability density function, normalized such that the integral over the range is 1 for each region. (b) Histograms of \tkin\ for each DR1 region. Pixels where \tkin\ is fit with an uncertainty $\leq 1$\ K are included. (c) Histograms of \namm\ for each DR1 region. Pixels where $\mathrm{log}\  N (\mathrm{NH}_3)$ is fit with an uncertainty $\leq 0.25$ are included. (d) Histograms of $X(\mathrm{NH}_3)$ for each DR1 region. Pixels where $\mathrm{log}\  N (\mathrm{NH}_3)$ is fit with an uncertainty $\leq 0.25$ are included. \label{fig_hist_all}}
\end{figure*}

We show in Figure \ref{fig_hist_all} the distributions of \sigv, \tkin, \namm, and $X(\mathrm{NH}_3)$ over the DR1 regions. For each parameter we show the probability density function normalized such that the integral over the range is one for each region. We include only those pixels with small uncertainties in the fitted values, as described in the Figure captions. 

Figure \ref{fig_hist_all} (a) reveals a systematic increase in the spread of the \amm\ velocity dispersion in the DR1 regions that reflects the increase in the level of star-formation activity from the least active region, B18, to the most active region, Orion A. In B18, NGC 1333, and L1688, the \sigv\ distribution shows a narrow peak at small velocity dispersions, and a `tail' of greater velocity dispersions that increases in both the distribution amplitude relative to the narrow \sigv\ peak, as well as the maximum \sigv\ present in the region. In Orion A, the narrow peak might still exist, but the population of larger \sigv\ values is substantially greater. The lower limit of measured \sigv\ in all regions is greater than the velocity resolution of the \amm\ data, and will be explored further in a future paper (Pineda et al., in preparation).

\begin{figure}
\begin{center}
\includegraphics[width=0.8\columnwidth]{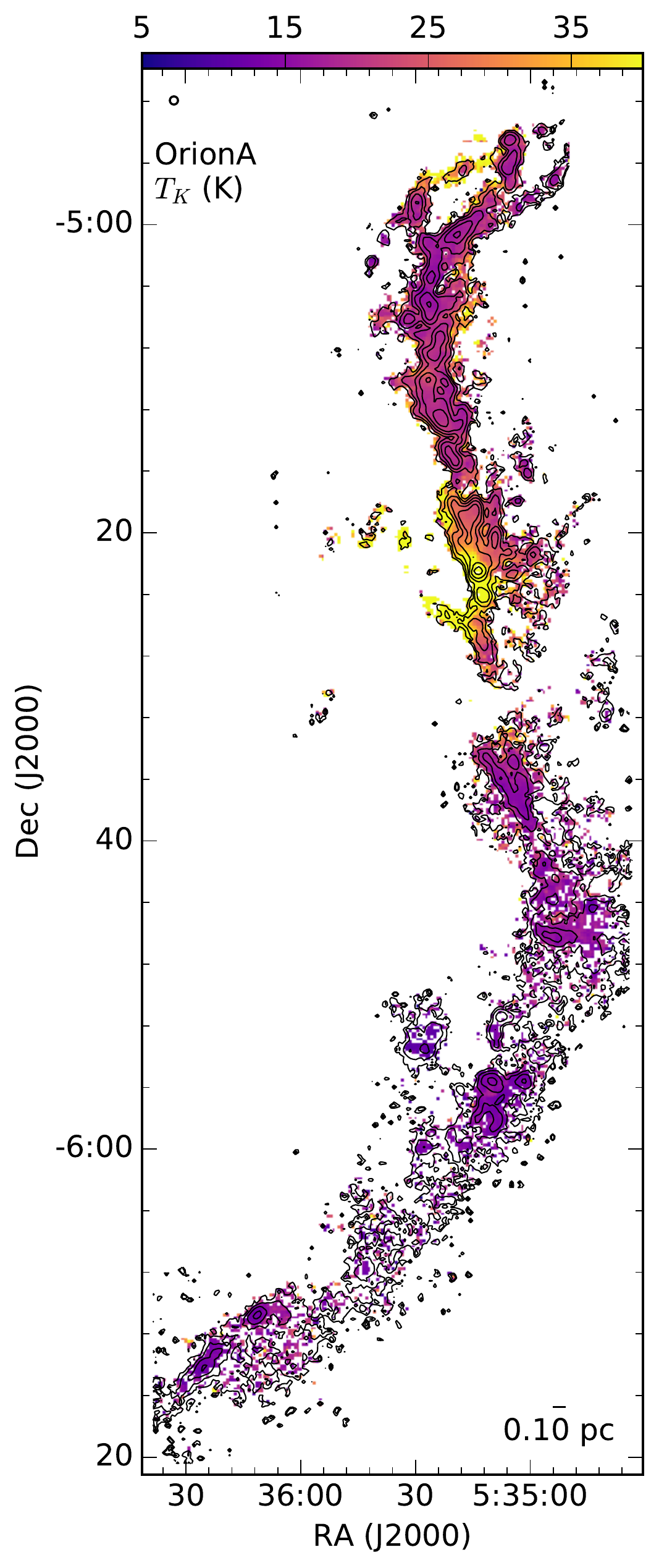}
\caption{\label{fig_OrionA_tkin} 
Like Figure \ref{fig_b18_tkin} but for Orion A (North).
}
\end{center}
\end{figure}

Similarly to the distributions of \sigv, both the mean gas kinetic temperature and the spread of temperatures measured varies systematically between the DR1 regions with increasing star-formation activity. Figure \ref{fig_hist_all} (b) shows that B18 contains the coldest dense gas of the DR1 regions, with a mean \tkin $\lesssim 10$\ K and only a small variation about that value. In both NGC 1333 and L1688, a moderate fraction of the dense gas is similarly cold ($T_\mathrm{K} \sim 10$\ K), but with increasing spread to higher temperatures. Previous pointed \amm\ observations of dense cores in Perseus found a typical temperature of 11\ K for these objects, in agreement with our data \citep{Rosolowsky_2008}, whereas in NGC 1333 warm gas is concentrated around known young stellar objects. In L1688, dense gas is more likely to be $\sim 14-15$\ K, as seen previously in smaller \amm\ maps of some of the cores in this region \citep{Friesen_2009}. Toward Orion A, only a small fraction of the dense gas is as cold as 10\ K, with typical temperatures $T_\mathrm{K} \sim 15-20$\ K, up to very large values near the Orion KL region (seen in Figure \ref{fig_OrionA_tkin}). 
At these temperatures ($>$30\,K), we expect significant populations of higher-order \amm\ (J,K) 
levels, and our temperature estimates from the (1,1) and (2,2) lines become inaccurate. 

Figure \ref{fig_hist_all} (c) shows the distribution of \amm\ column densities over the DR1 regions. 
For all regions, we find a similar increase in $N(p-\mathrm{NH}_3)$\footnote{In most works the total column density of \amm\  
(including ortho- and para-states) is reported, $N(\mathrm{NH}_3)$. 
A simple way to estimate total column density of \amm\ is: $N(\mathrm{NH}_3) \approx 2 \times N(p-\mathrm{NH}_3)$, if the ortho-to-para ratio is the LTE value of 1.} 
at low column densities, and a similar peak at $\mathrm{log}\ N(p-\mathrm{NH}_3) \sim 14$. 
B18 shows a relatively sharp cutoff in the maximum column density present. $N(p-\mathrm{NH}_3)$ extends to higher values in the other regions, however, with the largest column densities found in Orion A. 

The resulting distribution of \amm\ abundances, $X(\mathrm{NH}_3) = N(\mathrm{NH}_3)/N(\mathrm{H}_2)$, shown in Figure \ref{fig_hist_all} (d), have very similar peak values of $\mathrm{log} X(\mathrm{NH}_3) \sim -8.5 \ \mathrm{to} \ -8.0$. A tail to very high \amm\ abundances is seen in Orion A, but might result from difficulties in accurately fitting the submillimetre continuum SED in this active region. In NGC 1333, the abundance distribution appears bimodal. We will present more detailed analysis of the dust properties and \amm\ abundance distributions in an upcoming paper. 

\section{Summary}
\label{sec:summary}

We have presented data from the first release (DR1) for the Green Bank Ammonia Survey, including cubes, moment maps, and property maps, toward B18 in Taurus, L1688 in Ophiuchus, NGC 1333 in Perseus, and Orion A North in the Orion molecular cloud. We furthermore describe in detail the DR1 calibration, imaging, and line fitting pipelines. The four Gould Belt clouds observed span a range of H$_2$ column density and star formation activity. The extensive and sensitive \amm, HC$_5$N, HC$_7$N, and C$_2$S observations of these regions greatly increase the total contiguous areal coverage of dense molecular gas tracers in these well-studied star-forming regions. 

All cubes and maps are publicly available \added{through \url{https://dataverse.harvard.edu/dataverse/GAS_DR1}}, as are the imaging and analysis pipelines.

Some highlights from the sensitive \textit{GAS} data include:

\begin{enumerate}

\item
Extended \amm\ (1,1) emission is detected toward all DR1 regions, revealing the physical properties of both the dense gas associated with prestellar and star-forming cores and filaments, as well as their moderately dense molecular envelopes. 

\item
\amm\ (1,1) emission is a particularly good tracer of material at higher $N(\mathrm{H}_2)$, tracking well the H$_2$ column density as traced by dust continuum emission. In three of four DR1 regions, \amm\ is present above our map rms noise levels in $\gtrsim 60$\ \% of pixels at $N(\mathrm{H}_2) \gtrsim 6 \times 10^{21}$\ cm$^{-2}$, or $A_\mathrm{V} \gtrsim 7$.  

\item 
We observe some variation between the regions in the fraction of pixels with \amm\ as H$_2$ column density increases. We compare the GAS data to results calculated from a synthetic observation of \amm\ emission produced by a 3D hydrodynamic simulation. The simulated distributions agree well with the GAS distributions and demonstrate that variation in the underlying gas densities can account for some of the spread between regions. Differences between regions may also result from variations in the local UV radiation field and mean column density, which we will explore in future work.

\item
Previously unmapped \amm\ (3,3) emission is detected toward NGC 1333, L1688, and Orion A. Highlighting warmer, dense gas, \amm\ (3,3) emission is coincident with YSOs and PDR edges. 

\item
The carbon-chain molecules C$_2$S and HC$_5$N are strongly detected toward Taurus B18, but are faint or undetected in the integrated intensity maps of the other DR1 regions. In Taurus, the spatial distributions of C$_2$S and HC$_5$N are offset from the \amm\ emission. We detect C$_2$S toward NGC 1333 and L1688 only in spectra that were averaged over the entire observation extent. The observed HC$_7$N lines are not detected toward any DR1 target. 

\item
The distributions of gas temperature and velocity dispersion from the \amm\ line fits vary systematically between DR1 regions, in step with increasing star formation activity. Lower \tkin\ is found overall in B18, increasing through NGC 1333 and L1688, and is greatest in Orion A. Similarly, \sigv\ reaches greater values in L1688 and Orion A relative to B18 and NGC 1333.  

\end{enumerate}

\acknowledgments
RKF is a Dunlap Fellow at the Dunlap Institute for Astronomy \& Astrophysics. The Dunlap Institute is funded through an endowment established by the David Dunlap family and the University of Toronto. 
JEP, AP, ACT, and PC acknowledge the financial support of the European Research Council (ERC; project PALs 320620). 
EWR, PGM and CDM are supported by Discovery Grants from NSERC of Canada. 
SSRO acknowledges support from NSF grant AST-1510021. 
The National Radio Astronomy Observatory is a facility of the National Science Foundation operated under cooperative agreement by Associated Universities, Inc. 

\facility{Green Bank Telescope} 

\software{Astropy \citep{Robitaille_2013}, Matplotlib \citep{Hunter_2007}, pyspeckit \citep{2011ascl.soft09001G}\added{, 
GBTIDL (\url{http://gbtidl.nrao.edu}), 
GBT KFPA data reduction pipeline \citep{masters11}, 
RADMC-3D (\url{http://www.ita.uni-heidelberg.de/~dullemond/software/radmc-3d})}
}

\bibliographystyle{aasjournal}
\bibliography{AAS03939_submit_v2.bib}

\clearpage
\appendix

\section{Maps of the rms noise in the \amm\ (1,1) observations for all DR1 regions}
\label{sec:noise_maps}

%
%
\figsetstart
\figsetnum{A}
\figsettitle{\amm\ (1,1) rms maps\label{FigSet-A}}

\figsetgrpstart
\figsetgrpnum{A.1}
\figsetgrptitle{B18 rms \amm\ (1,1)}
\figsetplot{B18_NH3_11_DR1_rebase3_rms_map}
\figsetgrpnote{The rms noise (K) over the \amm\ (1,1) map toward B18. Scale bar is at lower right.\label{fig-b18-rms}}
\figsetgrpend

\figsetgrpstart
\figsetgrpnum{A.2}
\figsetgrptitle{NGC1333 rms \amm\ (1,1)}
\figsetplot{NGC1333_NH3_11_DR1_rebase3_rms_map}
\figsetgrpnote{Like Figure \ref{fig-b18-rms} but for NGC 1333.\label{fig-ngc1333-rms}}
\figsetgrpend

\figsetgrpstart
\figsetgrpnum{A.3}
\figsetgrptitle{L1688 rms \amm\ (1,1)}
\figsetplot{L1688_NH3_11_DR1_rebase3_rms_map}
\figsetgrpnote{Like Figure \ref{fig-b18-rms} but for L1688.\label{fig-l1688-rms}}
\figsetgrpend

\figsetgrpstart
\figsetgrpnum{A.4}
\figsetgrptitle{Orion A rms \amm\ (1,1)}
\figsetplot{OrionA_NH3_11_DR1_rebase3_rms_map}
\figsetgrpnote{Like Figure \ref{fig-b18-rms} but for Orion A (North).\label{fig-oriona-rms}}
\figsetgrpend

\figsetend

\begin{figure*}
\figurenum{A}
\includegraphics[width=0.9\textwidth]{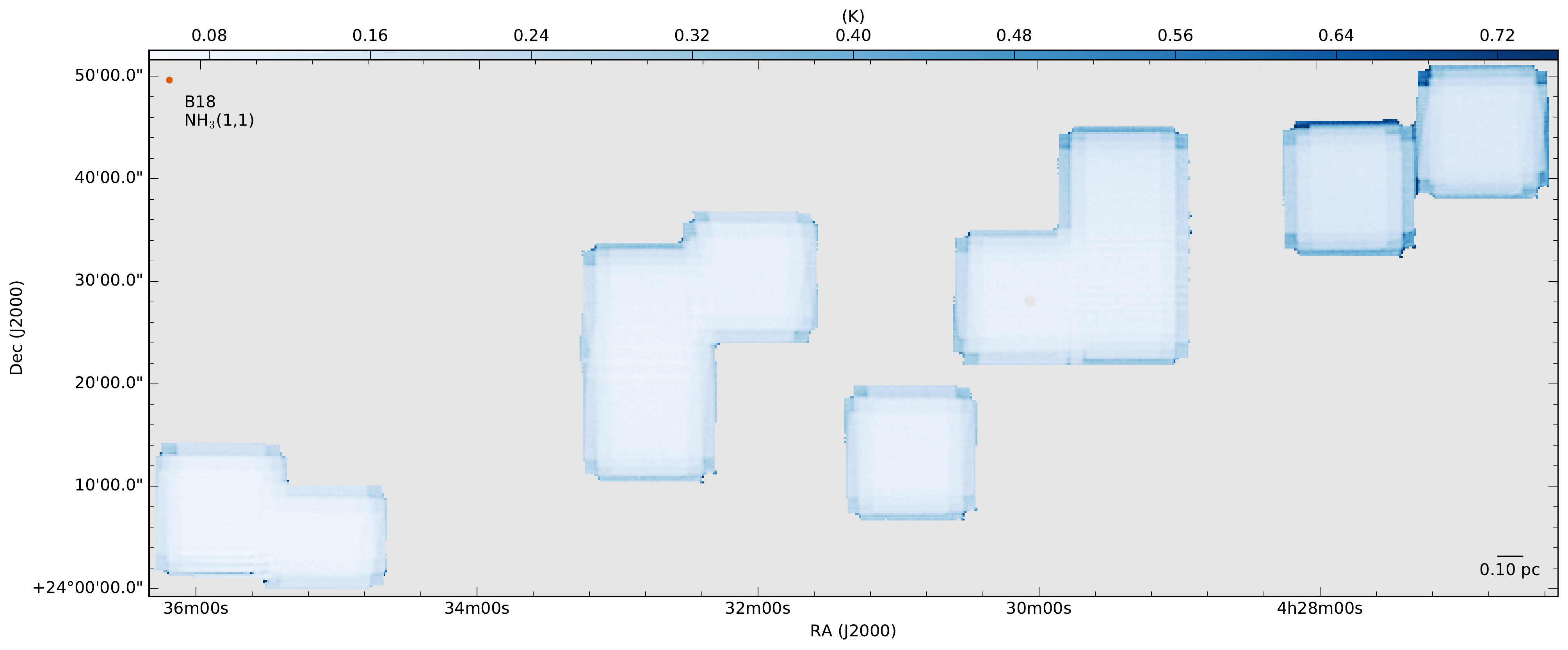}
\caption{The rms noise (K) over the \amm\ (1,1) map toward B18. Scale bar is at lower right.}
\end{figure*}

\section{\amm\ (2,2) and (3,3) integrated intensity maps, where detected}
\label{sec:other_nh3_moment_maps}

%
%
\figsetstart
\figsetnum{B}
\figsettitle{\amm\ (2,2) Integrated Intensity maps\label{FigSet-B}}

\figsetgrpstart
\figsetgrpnum{B.1}
\figsetgrptitle{B18 \amm\ (2,2) Integrated Intensity}
\figsetplot{B18_NH3_22_DR1_rebase3_mom0_map}
\figsetgrpnote{The integrated intensity of \amm\ (2,2) toward B18. Contours are \amm\ (1,1) integrated intensity as in Figure \ref{fig-b18-NH3-11_TdV}. Beam size and scale bar are at upper left and lower right, respectively.
\label{fig-b18-NH3-22_TdV}}
\figsetgrpend

\figsetgrpstart
\figsetgrpnum{B.2}
\figsetgrptitle{NGC1333 \amm\ (2,2) Integrated Intensity}
\figsetplot{NGC1333_NH3_22_DR1_rebase3_mom0_map}
\figsetgrpnote{Like Figure \ref{fig-b18-NH3-22_TdV} but for NGC 1333. \label{fig-ngc1333-NH3-22_TdV}}
\figsetgrpend

\figsetgrpstart
\figsetgrpnum{B.3}
\figsetgrptitle{L1688 \amm\ (2,2) Integrated Intensity}
\figsetplot{L1688_NH3_22_DR1_rebase3_mom0_map}
\figsetgrpnote{Like Figure \ref{fig-b18-NH3-22_TdV} but for L1688. \label{fig-l1688-NH3-22_TdV}}
\figsetgrpend

\figsetgrpstart
\figsetgrpnum{B.4}
\figsetgrptitle{Orion A \amm\ (2,2) Integrated Intensity}
\figsetplot{OrionA_NH3_22_DR1_rebase3_mom0_map}
\figsetgrpnote{Like Figure \ref{fig-b18-NH3-22_TdV} but for Orion A (North). \label{fig-OrionA-NH3-22_TdV}}
\figsetgrpend

\figsetend

\begin{figure*}[h!]
\figurenum{B}
\includegraphics[width=0.9\textwidth]{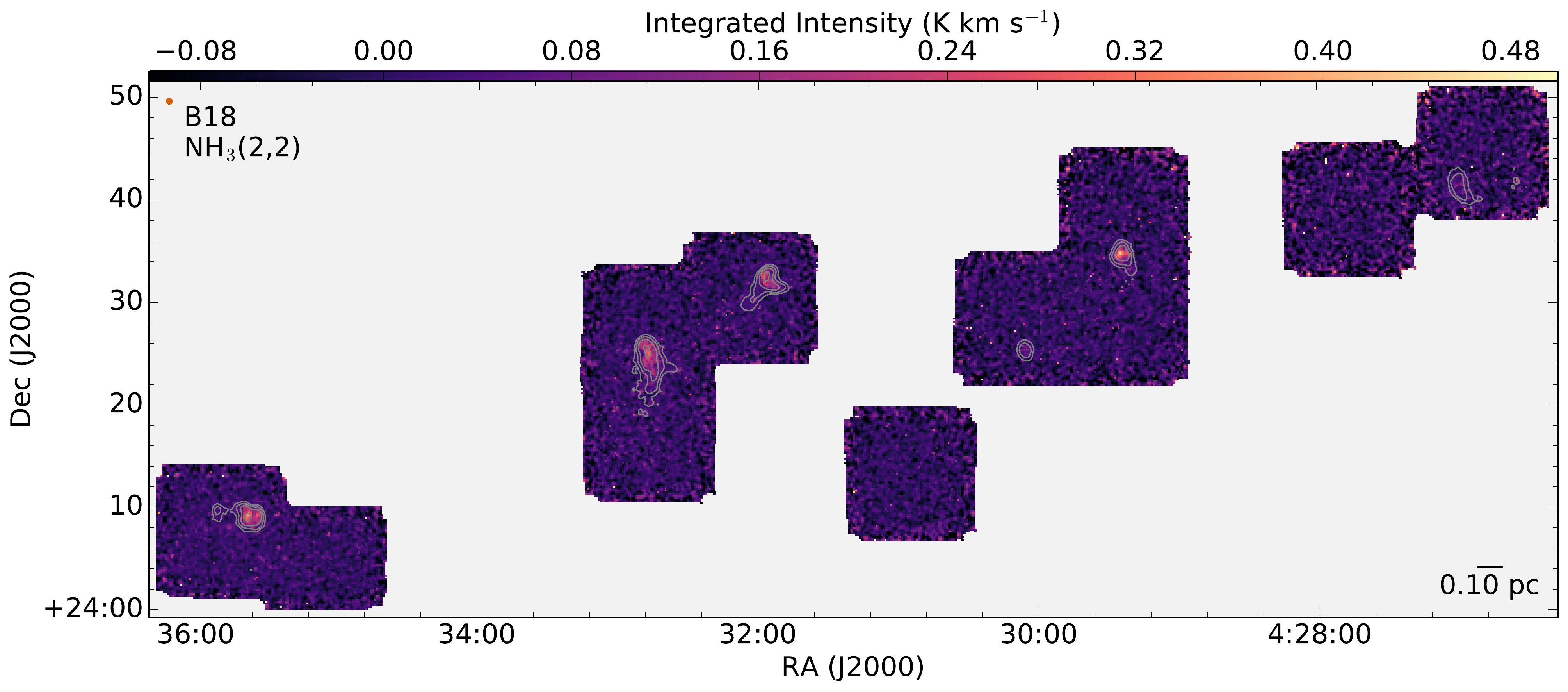}
\caption{The integrated intensity of \amm\ (2,2) toward B18. Contours are \amm\ (1,1) integrated intensity as in Figure \ref{fig-b18-NH3-11_TdV}. Beam size and scale bar are at upper left and lower right, respectively.}
\end{figure*}

%
%
\figsetstart
\figsetnum{C}
\figsettitle{\amm\ (3,3) Integrated Intensity rms maps for L1688, and Orion A\label{FigSet-C}}

\figsetgrpstart
\figsetgrpnum{C.1}
\figsetgrptitle{NGC1333 \amm\ (3,3) Integrated Intensity}
\figsetplot{NGC1333_NH3_22_DR1_rebase3_mom0_map}
\figsetgrpnote{\amm\ (3,3) integrated intensity toward NGC 1333. Contours are \amm\ (1,1) integrated intensity as in Figure \ref{fig-ngc1333-NH3-11_TdV}. The locations  of the Class 0/I object SVS 13 and Herbig-Haro object HH12 are identified by the upward and downward facing triangles, respectively. \label{fig-ngc1333-NH3-33_TdV}}
\figsetgrpend

\figsetgrpstart
\figsetgrpnum{C.2}
\figsetgrptitle{L1688 \amm\ (3,3) Integrated Intensity}
\figsetplot{L1688_NH3_33_DR1_rebase3_mom0_map}
\figsetgrpnote{Like Figure \ref{fig-ngc1333-NH3-33_TdV} but for L1688. 
The white arrow points toward the B2V star HD 147889, just off-map (R.A. 16:25:24.32, Dec. -24:27:56.6, J2000).}
\figsetgrpend

\figsetgrpstart
\figsetgrpnum{C.3}
\figsetgrptitle{Orion A \amm\ (3,3) Integrated Intensity}
\figsetplot{OrionA_NH3_33_DR1_rebase3_mom0_map}
\figsetgrpnote{Like Figure \ref{fig-ngc1333-NH3-33_TdV} but for Orion A (North). 
The Orion BN/KL and Orion bar regions are identified.}
\figsetgrpend

\figsetend

\begin{figure}
\figurenum{C}
\includegraphics[width=0.45\textwidth]{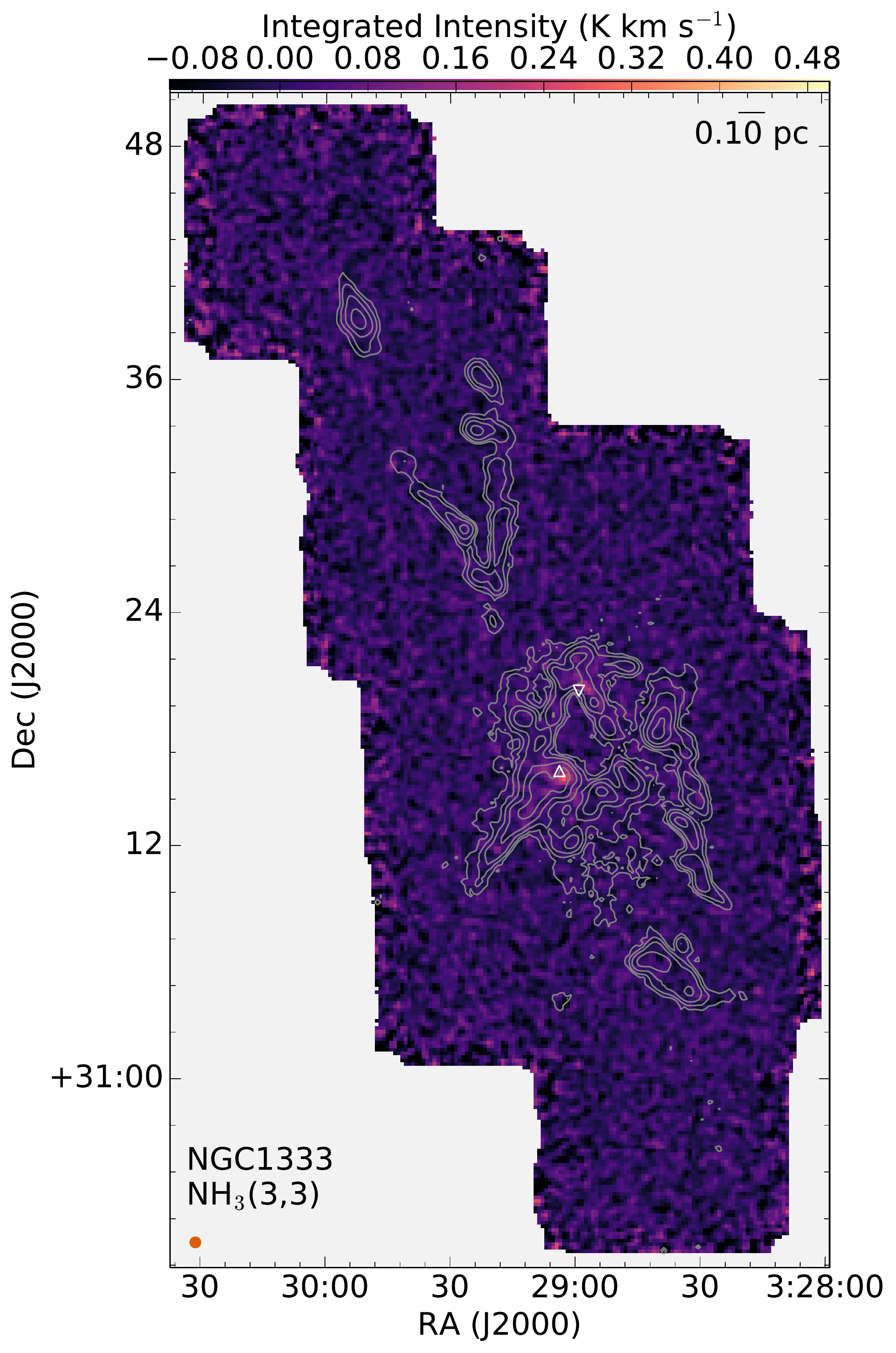}
\caption{\amm\ (3,3) integrated intensity toward NGC 1333. Contours are \amm\ (1,1) integrated intensity as in Figure \ref{fig-ngc1333-NH3-11_TdV}. The locations  of the Class 0/I object SVS 13 and Herbig-Haro object HH12 are identified by the upward and downward facing triangles, respectively.}
\end{figure}


\section{HC$_5$N $9-8$ {and C$_2$S $2_1-1_0$} integrated intensity maps, where detected}
\label{sec:other_moment_maps}

\begin{figure*}[h!]
\includegraphics[width=0.9\textwidth]{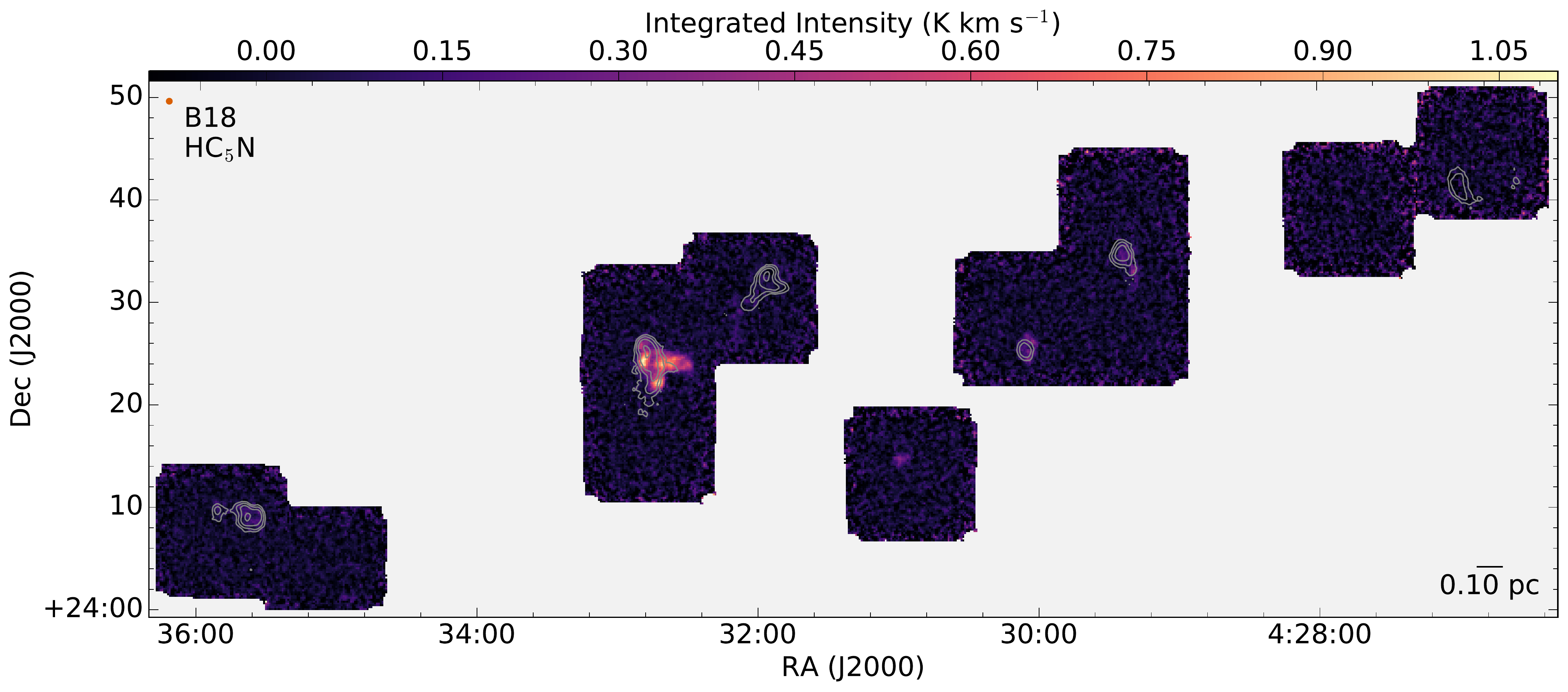}
\caption{HC$_5$N $9-8$ integrated intensity in B18. Contours are \amm\ (1,1) integrated intensity as in Figure \ref{fig-b18-NH3-11_TdV}. \label{fig-b18-hc5n}}
\end{figure*}

\begin{figure*}
\includegraphics[width=0.9\textwidth]{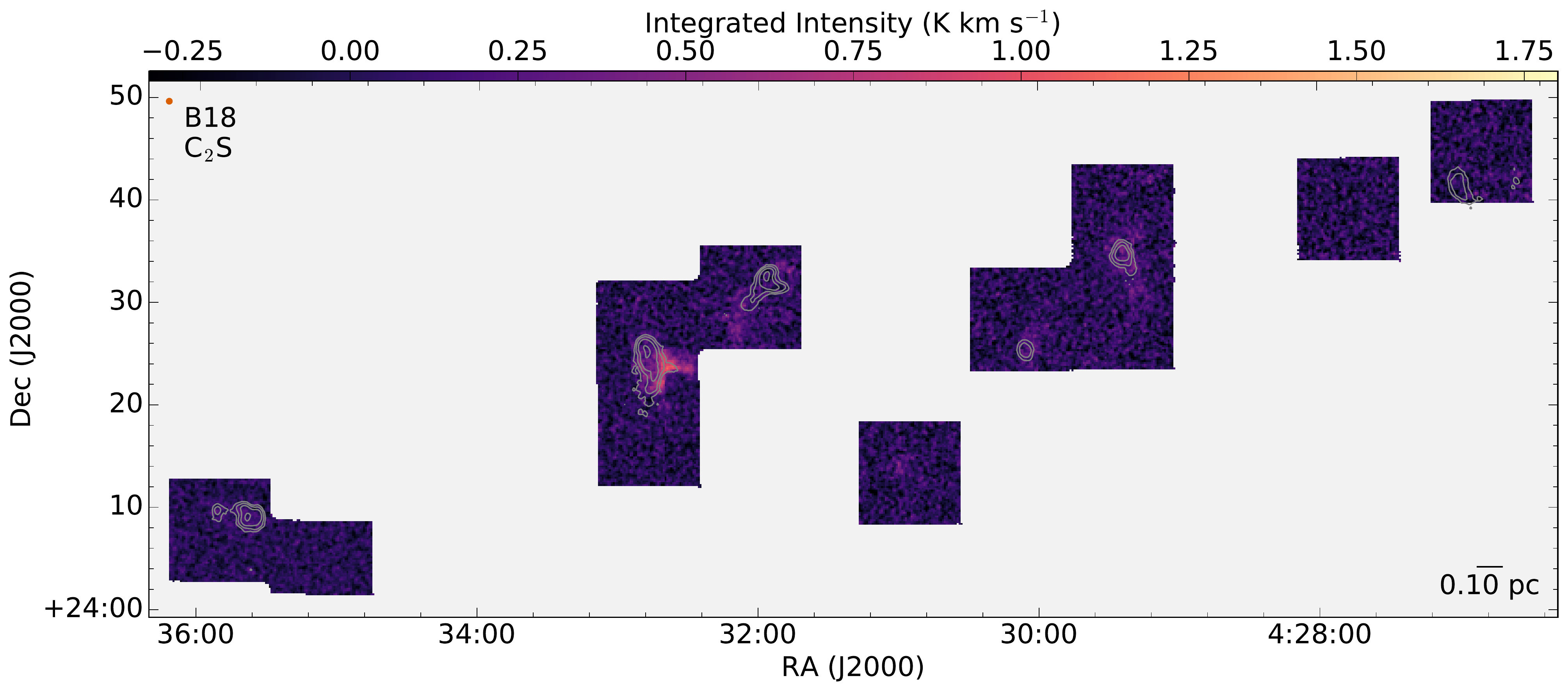}
\caption{C$_2$S $2_1-1_0$ integrated intensity toward B18. Contours are \amm\ (1,1) integrated intensity as in Figure \ref{fig-b18-NH3-11_TdV}. Note that C$_2$S was observed in a single beam only, and hence has a smaller footprint and higher noise values than the other mapped lines, but remains Nyquist-sampled.}
\end{figure*}

\begin{figure}
\includegraphics[width=0.45\textwidth]{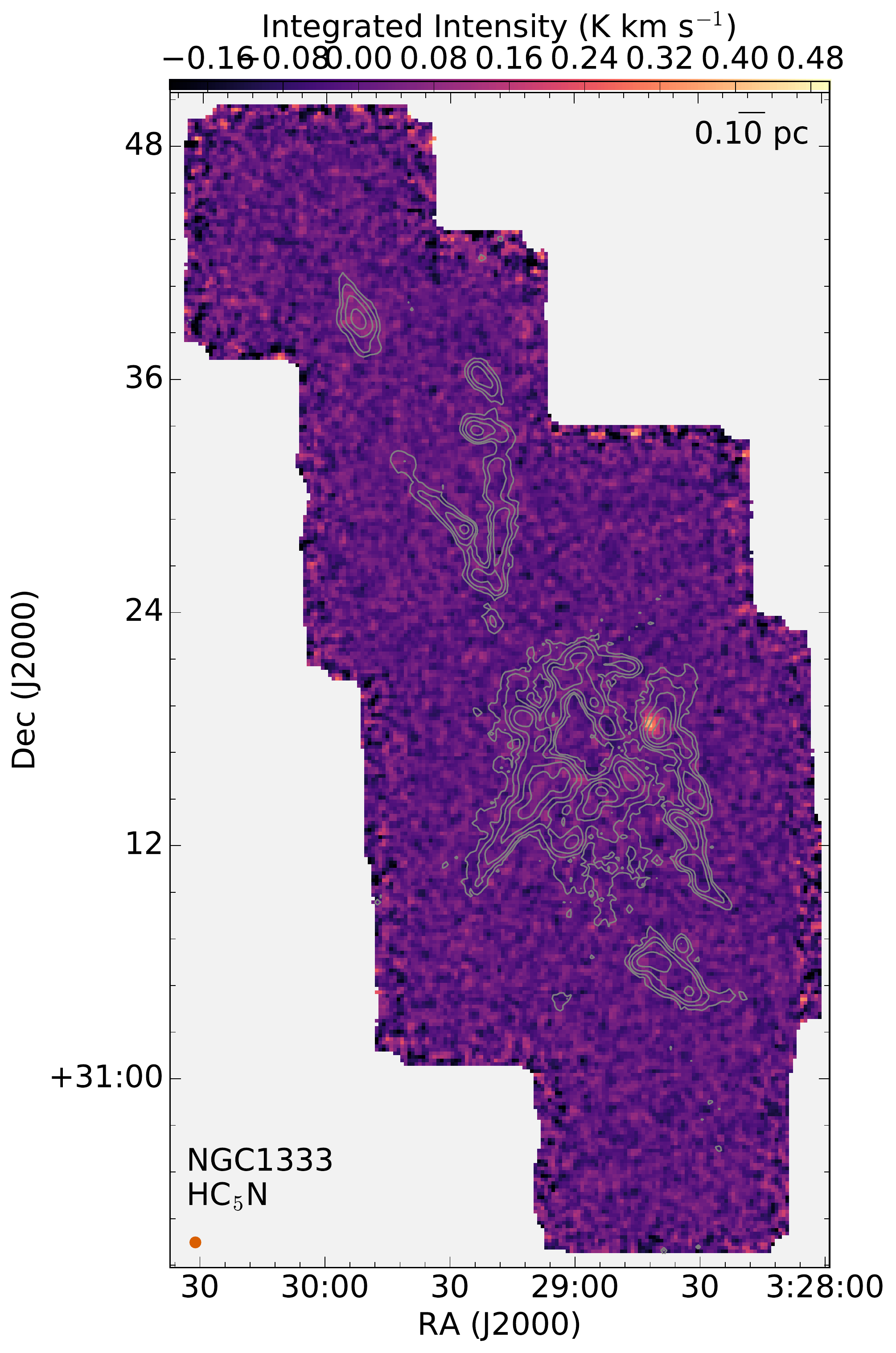}
\caption{Like Figure \ref{fig-b18-hc5n} but for NGC 1333. \label{fig-ngc1333-hc5n}}
\end{figure}

\begin{figure*}
\includegraphics[width=0.7\textwidth]{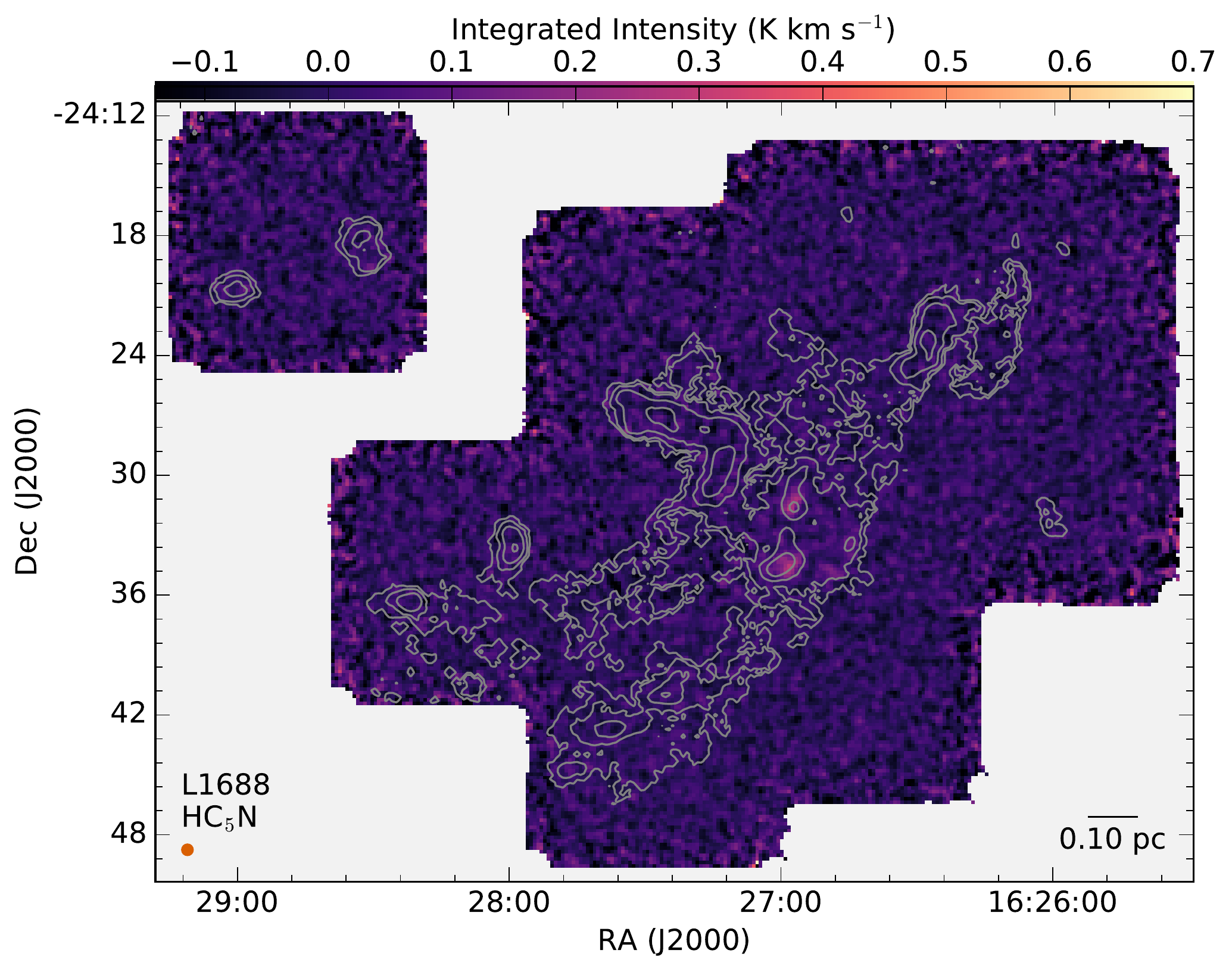}
\caption{Like Figure \ref{fig-b18-hc5n} but for L1688. \label{fig-l1688-hc5n}}
\end{figure*}

\clearpage
\section{\amm\ column density maps for all DR1 regions}
\label{sec:n_nh3_maps}

\begin{figure*}[h!]
\includegraphics[width=0.9\textwidth]{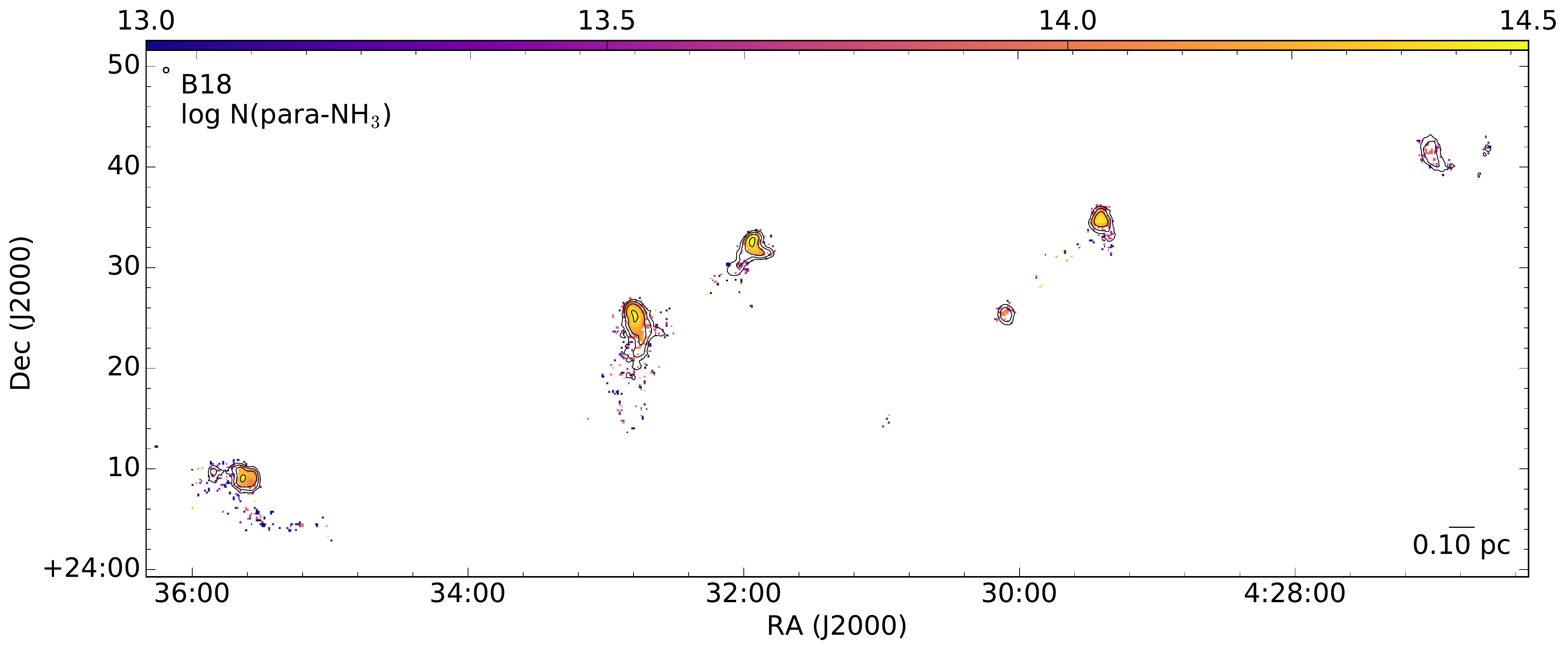}
\caption{\namm\ (cm$^{-2}$) in B18. Contours are \amm\ (1,1) integrated intensity as in Figure \ref{fig-b18-NH3-11_TdV}. \label{fig-b18-N_NH3}}
\end{figure*}
 
\begin{figure}
\includegraphics[width=0.45\textwidth]{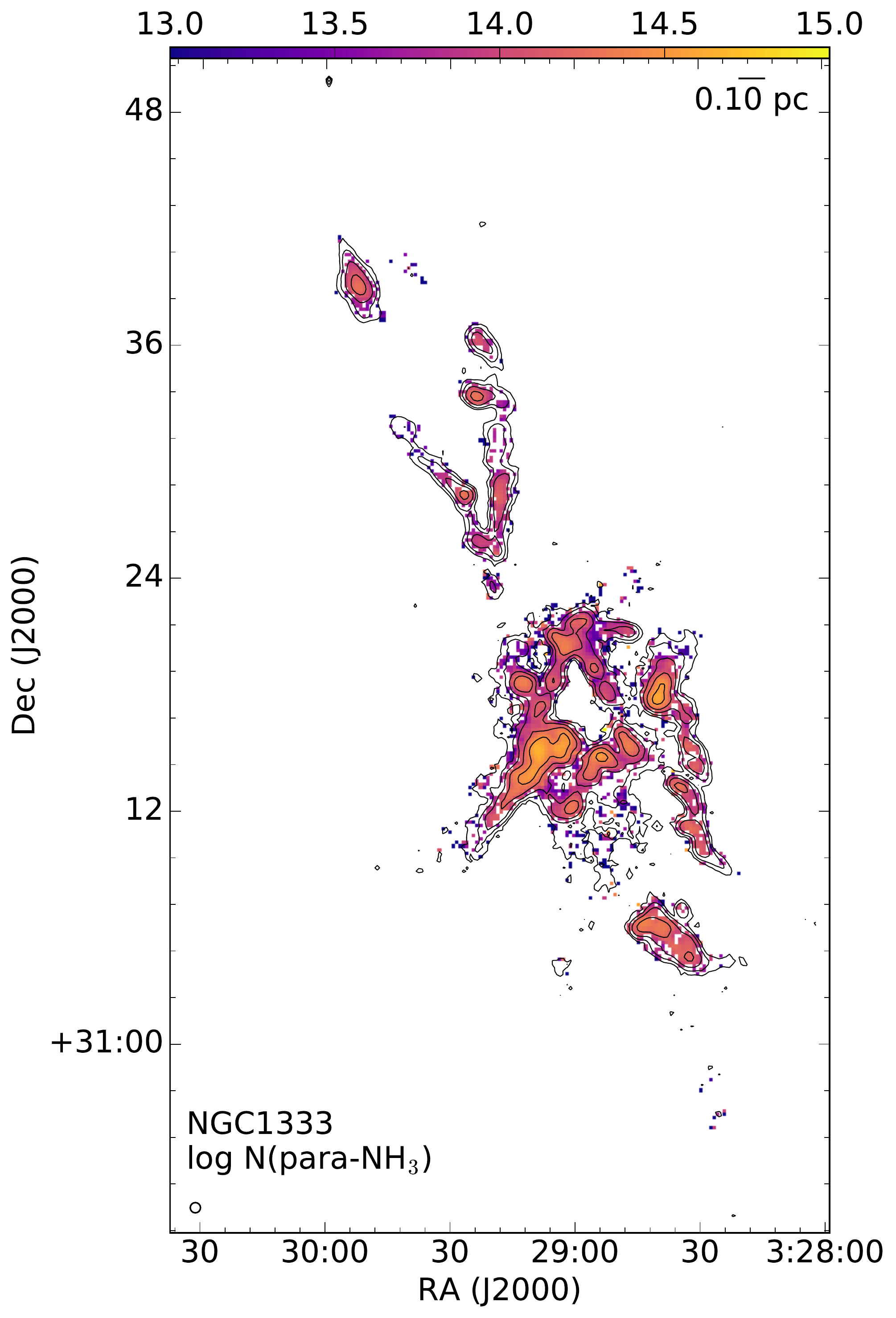}
\caption{Like Figure \ref{fig-b18-N_NH3} but for NGC 1333. \label{fig_ngc1333_nnh3}
}
\end{figure}

\begin{figure*}
\includegraphics[width=0.7\textwidth]{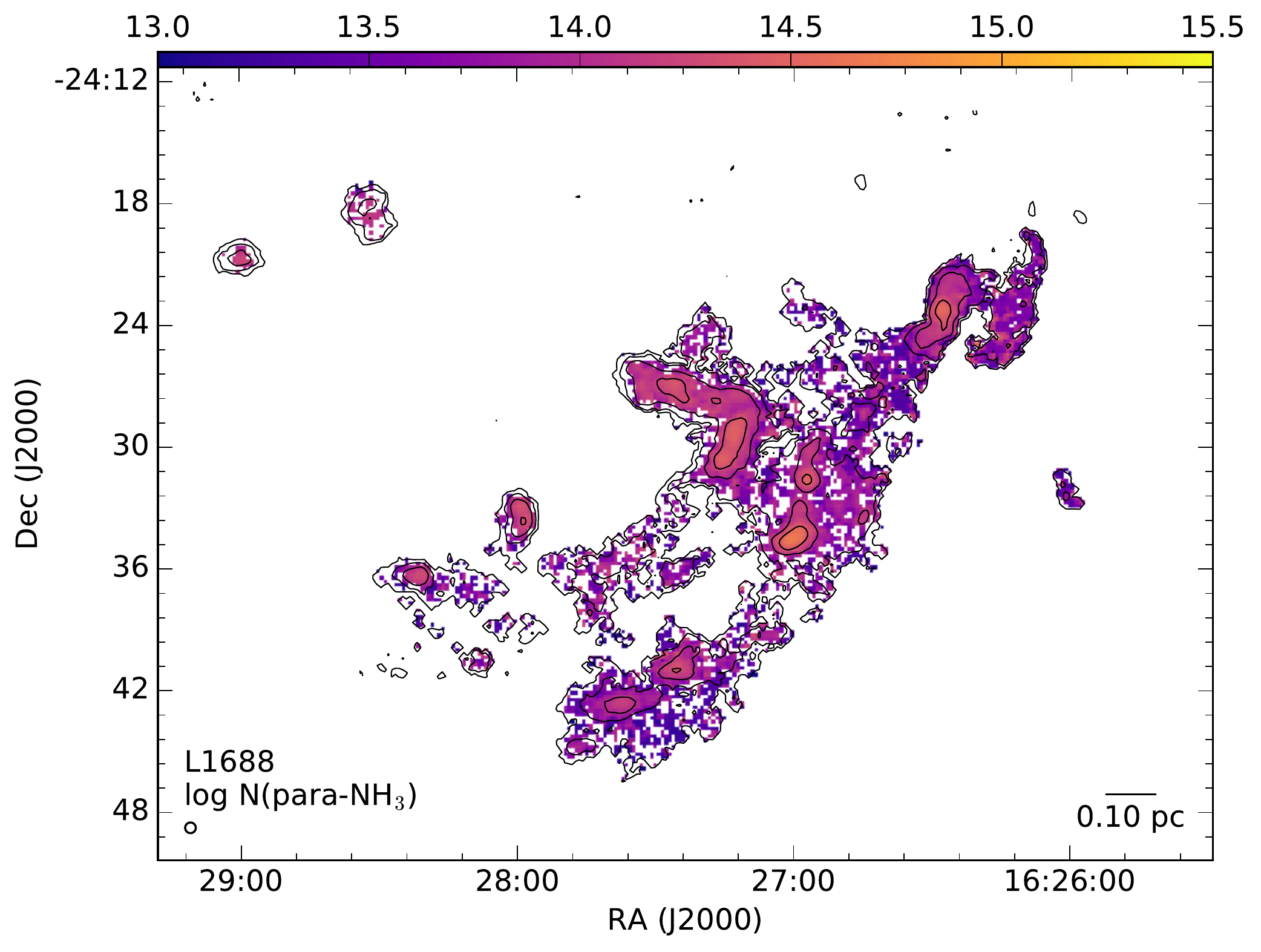}
\caption{Like Figure \ref{fig-b18-N_NH3} but for L1688. \label{fig_l1688_nnh3}
}
\end{figure*}

\begin{figure}
\includegraphics[width=0.45\textwidth]{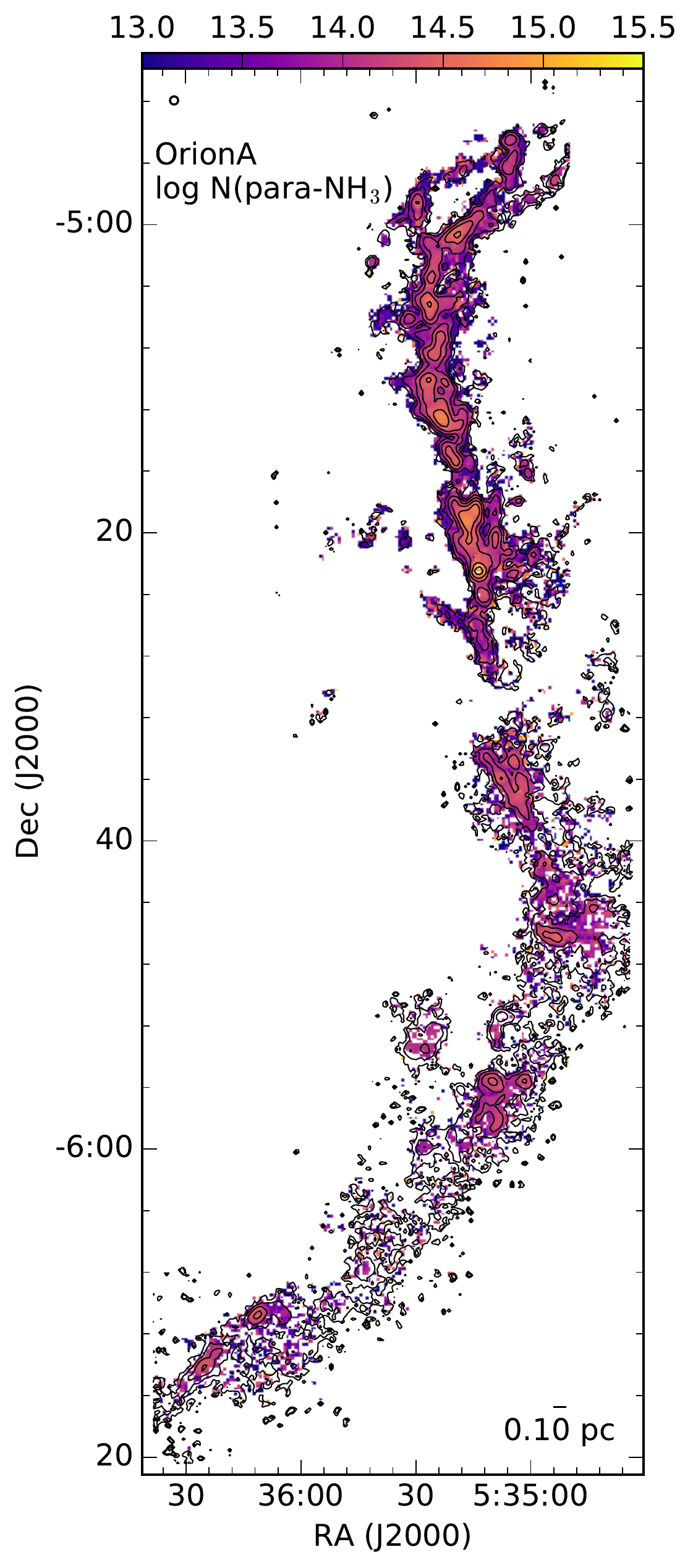}
\caption{Like Figure \ref{fig-b18-N_NH3} but for Orion A (North). \label{fig_oriona_nnh3}
}

\end{figure}
\end{document}